\documentclass[]{aa}
\usepackage{graphicx}
\usepackage[varg]{txfonts}
\usepackage{natbib}
\bibpunct{(}{)}{;}{a}{}{,} 
\begin{document}

\title{Near-infrared spectro-interferometry of Mira variables
and comparisons to 1D~dynamic model atmospheres and 
3D~convection simulations
\thanks{Based on observations made with the VLT Interferometer (VLTI)
at Paranal Observatory under program IDs 082.D-0723, 
084.D-0839, 088.D-0160, 090.D-0817, and 091.D-0765.}}
\titlerunning{Spectro-Interferometry of Mira Variables}
\author{
M.~Wittkowski\inst{1}\and
A.~Chiavassa\inst{2}\and
B.~Freytag\inst{3}\and
M.~Scholz\inst{4,5}\and
S.~H\"ofner\inst{3}\and
I.~Karovicova\inst{4}\and
P.~A.~Whitelock\inst{6,7}
}
\institute{
ESO, Karl-Schwarzschild-Str. 2,
85748 Garching bei M\"unchen, Germany,
\email{mwittkow@eso.org}
\and
Laboratoire Lagrange, UMR 7293, Universit\'e de Nice Sophia-Antipolis, CNRS,
Observatoire de la C\^ote d’Azur, BP. 4229, 06304 Nice Cedex 4, France
\and
Department of Physics and Astronomy at Uppsala University,
Regementsv{\"a}gen 1, Box 516,
SE-75120 Uppsala, Sweden
\and
Zentrum f\"ur Astronomie der Universit\"at Heidelberg (ZAH), 
Institut f\"ur Theoretische Astrophysik, Albert-Ueberle-Str. 2, 
69120 Heidelberg, Germany
\and 
Sydney Institute for Astronomy, School of Physics, University of Sydney, 
Sydney NSW 2006, Australia
\and
South African Astronomical Observatory, P.O.Box 9, 7935 Observatory, South Africa
\and
Astronomy, Cosmology and Gravity Centre, Astronomy Department, University of Cape Town, 
7701 Rondebosch, South Africa
}
\date{Received \dots; accepted \dots}
\abstract{}
{We aim at comparing spectro-interferometric observations of Mira variable
asymptotic giant branch (AGB) stars with the latest 1D dynamic model 
atmospheres based on self-excited pulsation models (CODEX models) and with 
3D dynamic model atmospheres including pulsation and convection 
(CO5BOLD models) to better understand the processes that extend 
the molecular atmosphere to radii where dust can form.}
{We obtained a total of 20 near-infrared $K$-band spectro-interferometric 
snapshot observations of the Mira variables o~Cet, R~Leo, R~Aqr, X~Hya, 
W~Vel, and R~Cnc with a spectral resolution of about 1500.
We compared observed flux and visibility spectra with predictions by
CODEX 1D dynamic model atmospheres and with azimuthally averaged 
intensities based on
CO5BOLD 3D dynamic model atmospheres.}
{Our visibility data confirm the presence of spatially extended 
molecular atmospheres located above the continuum radii with large-scale 
inhomogeneities or clumps that contribute a few percent of the total flux.
The detailed structure of the inhomogeneities or clumps show a variability 
on time scales of 3 months and above.
Both modeling attempts provided satisfactory fits to our data. 
In particular, they are both consistent with the observed decrease in the
visibility function at molecular bands of water vapor and CO, indicating
a spatially extended molecular atmosphere. Observational variability
phases are mostly consistent with those of the best-fit CODEX models, except
for near-maximum phases, where data are better described
by near-minimum models. 
Rosseland angular diameters derived from the model
fits are broadly consistent between those based on the 
1D and the 3D models and with earlier observations.
We derived fundamental parameters including absolute radii, effective
temperatures, and luminosities for our sources. 
}
{Our results provide a first observational support for theoretical
results that shocks induced by convection and pulsation in the 3D CO5BOLD
models of AGB stars are roughly spherically expanding and of similar nature to 
those of self-excited pulsations in 1D CODEX models. Unlike for red supergiants,
the pulsation- and shock-induced dynamics can levitate the molecular 
atmospheres of Mira variables to extensions that are consistent with 
observations.}
\keywords{
Techniques: interferometric --
Stars: AGB and post-AGB -- 
Stars: atmospheres -- 
Stars: mass-loss 
}
\maketitle
\section{Introduction}
Mira variable stars are evolved low or intermediate mass stars located on
the asymptotic giant branch (AGB) of the Hertzsprung-Russell (HR) diagram.
They are long-period ($\sim$\,1\,year) fundamental-mode \citep{Wood1999}
pulsators with $V$ band amplitudes of typically 5--7\,mag, near-IR $K$ band
amplitudes typically of 0.3--0.8\,mag, and bolometric amplitudes
typically of 0.5--1.2\,mag, which correspond to luminosity variations by
factors of about 1.5--3 \citep{Whitelock2000}.

AGB stars show mass-loss rates of 
$(4\times10^{-8}-8\times10^{-5})$M$_\odot/$yr \citep{deBeck2010}
and are one of the major sources of the chemical enrichment
of galaxies and of dust in the universe. 
The AGB mass loss is thought to be driven by
{\it pulsation} and {\it convection}, which levitate the atmospheres
to regions where dust can form, and subsequently by radiation pressure
on the dust grains dragging along the gas. Observationally, the presence 
of extended molecular atmospheres of Mira stars extending to a few stellar 
radii has been confirmed by spectro-interferometric observations
\citep[e.g.,][]{Mennesson2002,Perrin2004,Ohnaka2005,Wittkowski2008}. 

This scheme is theoretically constrained best
for carbon-rich Mira stars, which are large amplitude pulsators where carbon 
has been dredged up to the surface so that carbon dust can form. 
Carbon dust has absorption coefficients that are high enough
to drive the wind 
\citep[e.g.,][]{Wachter2002,Mattsson2010}.
The process is less understood for oxygen-rich Mira stars, which typically
have a lower mass and higher effective temperature than do the 
carbon-rich Miras.
This is due to uncertainties regarding the properties of silicate grains in the 
close stellar vicinity and the resulting radiative pressure
\citep{Woitke2006,Hoefner2008,Bladh2012}. The dust condensation
sequence is currently being refined, possibly
including Al$_2$O$_3$ grains closest to the surface serving as 
seed particles, followed by relatively large scattering iron-free silicates 
at small radii, which can drive the wind, and --depending on the mass-loss
rate-- subsequently by iron-coated 
silicates at larger radii \citep[e.g.,][]{Karovicova2013,Bladh2015}.

For semi-regular AGB stars with lower pulsation amplitudes 
as well as for red supergiants, it is not yet clear how the atmospheres are
levitated to regions that are favorable to dust formation and growth.
Indeed, \citet{Arroyo2015} show that current 1D pulsation 
models of red supergiants did not lead to shock fronts
that enter the inner atmospheres of red supergiants\footnote{The model atmospheres 
reach to 3--5 photospheric radii, so that shocks at larger radii cannot
be excluded.}. For 3D convection models of red supergiants, there were shocks in the 
atmospheres, mostly on smaller (i.e. granular), but not on global scales. 
However, while travelling outward, the density behind the shocks declined
drastically so that not enough material was lifted to cause noticeable 
optical depths significantly outside one stellar (photospheric) radius.
Both 1D and 3D model attempts cannot currently levitate the atmospheres 
to observed extensions of a few stellar radii where dust can form.
This suggests a missing process for levitating the atmospheres of red
supergiants.

Unlike for red supergiants,
however, recent 1D dynamic model atmospheres based on self-excited pulsation 
models of oxygen-rich Mira stars (CODEX models)
\citep{Ireland2008,Ireland2011,Scholz2014} were successful 
at describing interferometric observations of these sources, including their 
extended atmospheric molecular layers 
\citep{Woodruff2009,Wittkowski2011,Hillen2012}. However, they do not yet
include a wind.
Observations indicate surface inhomogeneities of the 
extended molecular layers corresponding to a few spots across the
stellar surface at intensity levels of a few percent
\citep[e.g.,][]{Wittkowski2011,Haubois2015}.

\citet{Freytag2008} have presented 3D radiation hydrodynamic (RHD) simulations
of the convective interior and the atmosphere of a typical AGB star.
They find that the surface of the model star is dominated by
a few giant convection cells. The interaction of irregular pulsations
with the convective cells triggers shock waves in the photospheric
layers. These shock waves were found to be roughly spherically expanding,
similar to those of 1-D pulsation models, but with certain non-radial
structures. They show that these shock waves levitate the material into 
regions where dust can form. However, these predictions of 3D CO5BOLD dynamic 
models of AGB stars have so far not been compared with spatially resolved 
observations of the extended atmospheres of AGB stars. The model
has been compared to data of a red supergiant, VX Sgr, by 
\citet{Chiavassa2010}, where it surprisingly provided a better fit
to the data than similar models of red supergiants (see also the discussion of 
VX Sgr in \citet{Arroyo2015}.

Here, we analyze the spectro-interferometric data of the Mira
variables R~Cnc, X~Hya, and W~Vel again from \citet{Wittkowski2011}, 
together with so far unpublished second epochs of these targets 
with a separation of three months. We also analyze new spectro-interferometric 
data of the Mira variables o~Cet, R~Aqr, and R~Leo in the same
way.
For the first time we compare
spectro-interferometric observations of Mira stars to predictions 
by these 3D CO5BOLD dynamic model atmospheres of AGB stars \citep{Freytag2008}, 
in addition to those by 1D CODEX dynamic model atmospheres
\citep{Ireland2008,Ireland2011,Scholz2014}.

\section{Observations and data reduction}
\label{sec:obs}
\begin{table*}
\centering
\caption{Observation log\label{tab:obs}}
\begin{tabular}{rlrrrrrrl}
\hline\hline
No. & Target    & Date          & JD\tablefootmark{a}           & Mode  & Baseline & Proj. baseline length & Proj. baseline angle & Calibrators\\
    &           &               &               &       &          & (m)                   & (deg)                &            \\\hline
1   &R Cnc      & 2008-12-29    & 4830.28       & MR23  & E0-G0-H0 & 15.90/31.78/47.69     & -73.7/-73.7/-73.7    & 81 Gem, $\iota$ Hya \\
2   &R Cnc      & 2008-12-30    & 4831.29       & MR21  & E0-G0-H0 & 16.00/31.98/47.98     & -72.7/-72.7/-72.7           & 81 Gem, $\iota$ Hya \\
3   &R Cnc      & 2009-03-01    & 4892.17       & MR23  & E0-G0-H0 & 15.57/31.12/46.69     & -68.9/-68.9/-68.9    & 31 Ori, 31 Leo     \\
4   &R Cnc      & 2009-03-03    & 4894.13       & MR21  & E0-G0-H0 & 15.98/31.94/47.92     & -71.7/-71.7/-71.7    & $\iota$ Hya                \\
5   &R Cnc      & 2010-02-11    & 5239.17       & MR23  & E0-G0-H0 & 16.00/31.97/47.97     & -72.8/-72.8/-72.8    & $\iota$ Hya                \\
6   &R Cnc      & 2010-02-11    & 5239.20       & MR23  & E0-G0-H0 & 15.88/31.74/47.63     & -70.7/-70.7/-70.7    & $\iota$ Hya                \\[1ex]

7   &W Vel      & 2008-12-29    & 4830.39       & MR23  & E0-G0-H0 & 15.29/30.55/45.83     & -84.4/-84.4/-84.4    & 31 Leo, $\gamma$ Pyx \\
8   &W Vel      & 2009-03-01    & 4892.21       & MR23  & E0-G0-H0 & 15.43/30.83/46.26     & -82.0/-82.0/-82.0    & 31 Ori, $\iota$ Hya        \\[1ex]

9   &X Hya      & 2008-12-29    & 4830.32       & MR23  & E0-G0-H0 & 15.97/31.92/47.89     & -71.1/-71.1/-71.1    & $\iota$ Hya, 31 Leo        \\ 
10  &X Hya      & 2008-12-30    & 4831.37       & MR21  & E0-G0-H0 & 15.46/30.89/46.55     & -75.8/-75.8/-75.8    & $\iota$ Hya, 31 Leo        \\
11  &X Hya      & 2009-03-01    & 4892.15       & MR23  & E0-G0-H0 & 15.95/31.88/47.83     & -70.8/-70.8/-70.8    & $\iota$ Hya, 31 Leo        \\[1ex]

12  &R Aqr      & 2012-10-18    & 6219.03       & MR21  & A1-B2-C1 & 10.51/10.76/14.07     &  -70.4/12.4/60.2     & $\phi$ Aqr           \\
13  &R Aqr      & 2012-10-18    & 6219.07       & MR21  & A1-B2-C1 & 10.98/11.18/15.38     &  20.0/-67.9/66.6     & $\phi$ Aqr           \\[1ex]

14  &o Cet      & 2012-10-18    & 6219.22       & MR21  & A1-B2-C1 & 10.70/11.12/15.91     &  28.1/65.5/72.3      & $\alpha$ Cet \\
15  &o Cet      & 2012-10-18    & 6219.25       & MR21  & A1-B2-C1 & 10.53/11.03/15.91     &  63.4/31.7/73.0      & $\alpha$ Cet \\
16  &o Cet      & 2013-09-13    & 6549.25       & MR21  & A1-B2-C1 & 9.97/11.07/14.05      &  17.0/-66.7/68.5     & $\alpha$ Cet \\
17  &o Cet      & 2013-09-13    & 6549.28       & MR21  & A1-B2-C1 & 10.3/11.31/15.32      &  23.4/66.6/71.0      & $\alpha$ Cet \\[1ex]

18  &R Leo      & 2012-12-10    & 6272.27       & MR21  & A1-B2-C1 & 7.87/10.63/10.64      &  8.5/76.9/-59.7      & $\beta$ Cnc          \\
19  &R Leo      & 2012-12-10    & 6272.32       & MR21  & A1-B2-C1 & 8.39/11.24/13.34      & 19.8/-64.5/76.8      & $\beta$ Cnc          \\
20  &R Leo      & 2012-12-10    & 6272.33       & MR21  & A1-B2-C1 & 8.73/11.31/14.29      & 24.0/-66.1 /76.3     & $\beta$ Cnc          \\
\hline
\end{tabular}
\tablefoot{
\tablefootmark{a}{JD-2450000}}
\end{table*}

\begin{figure*}
\centering
  \includegraphics[width=0.32\textwidth]{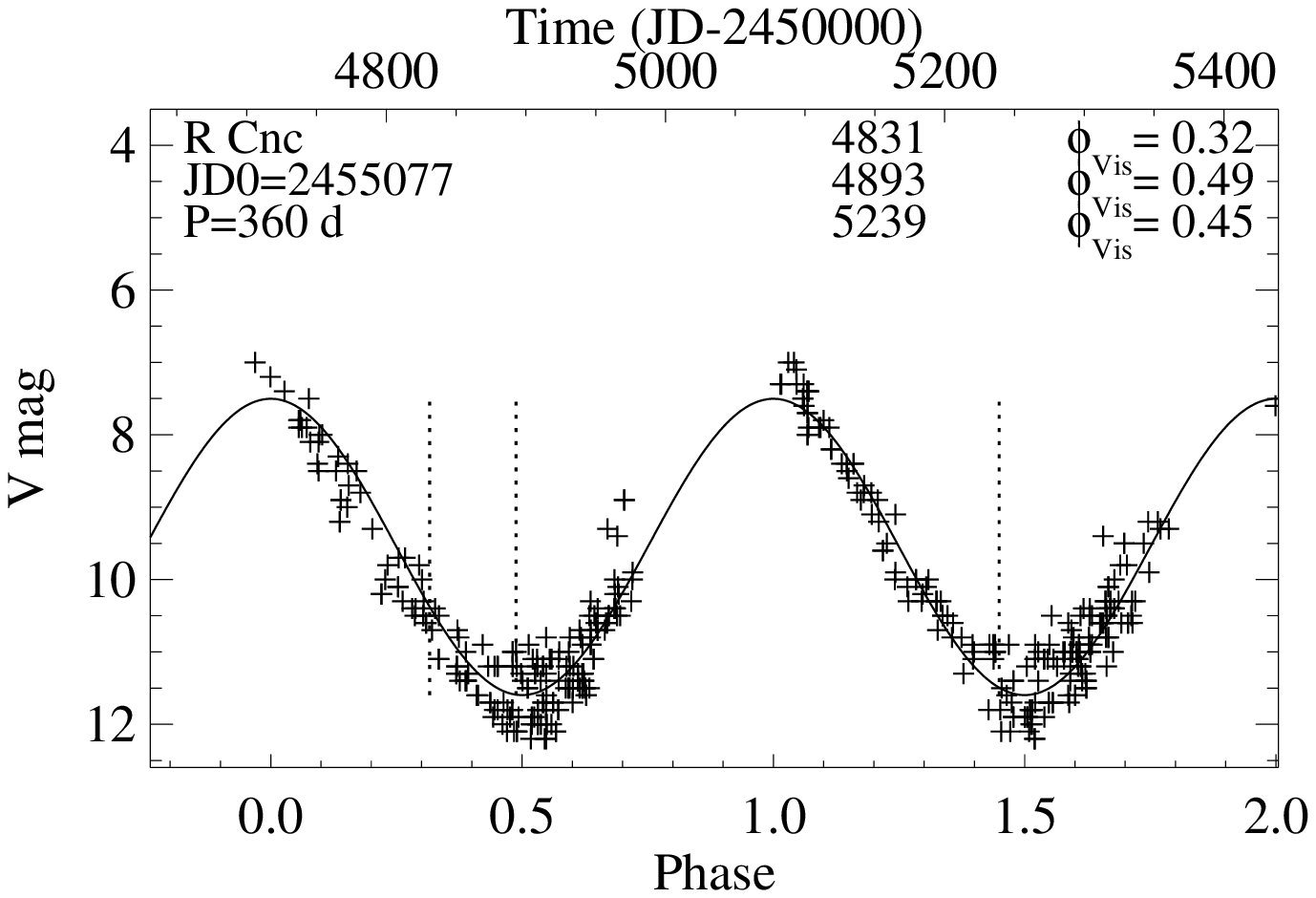}
  \includegraphics[width=0.32\textwidth]{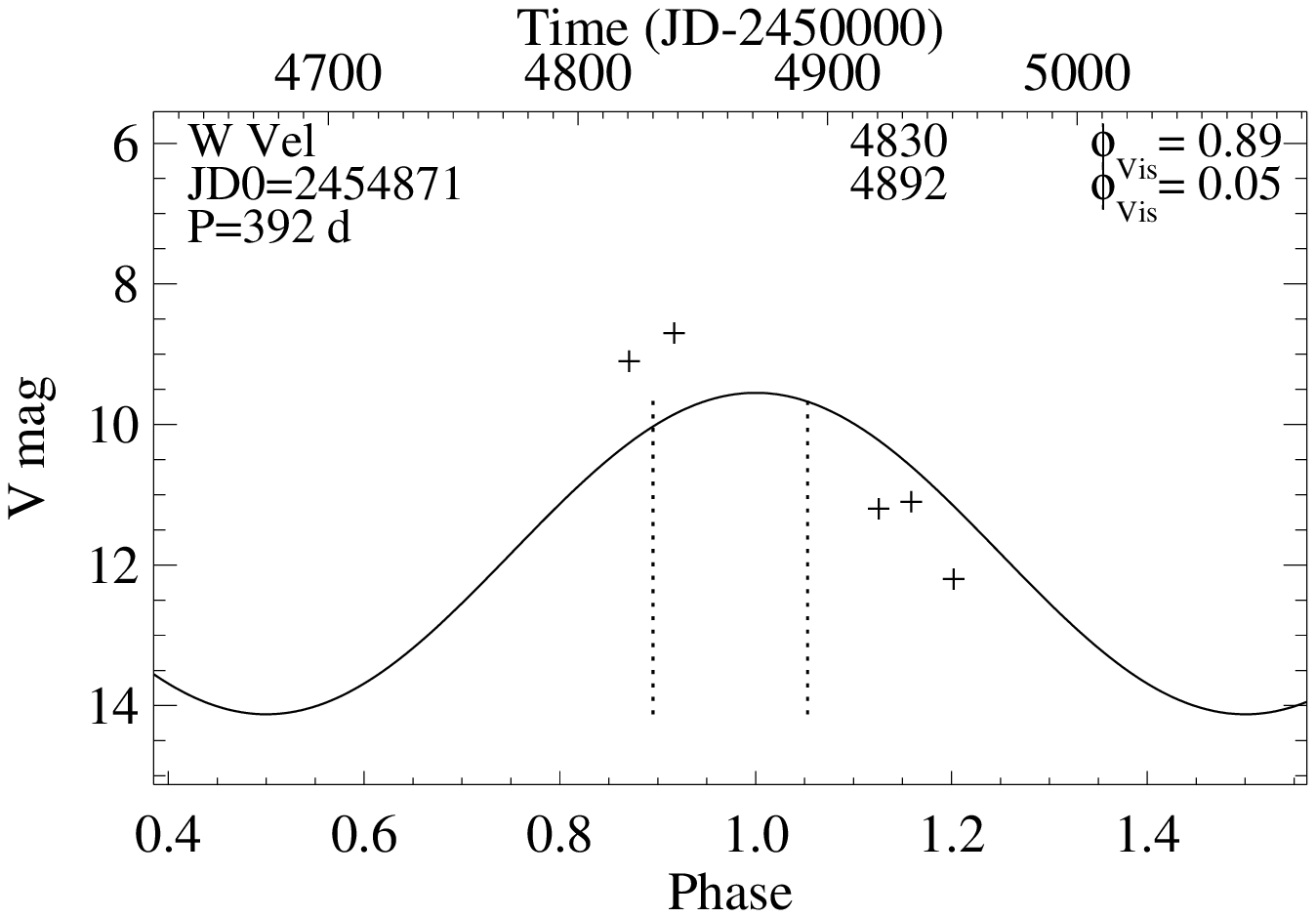}
  \includegraphics[width=0.32\textwidth]{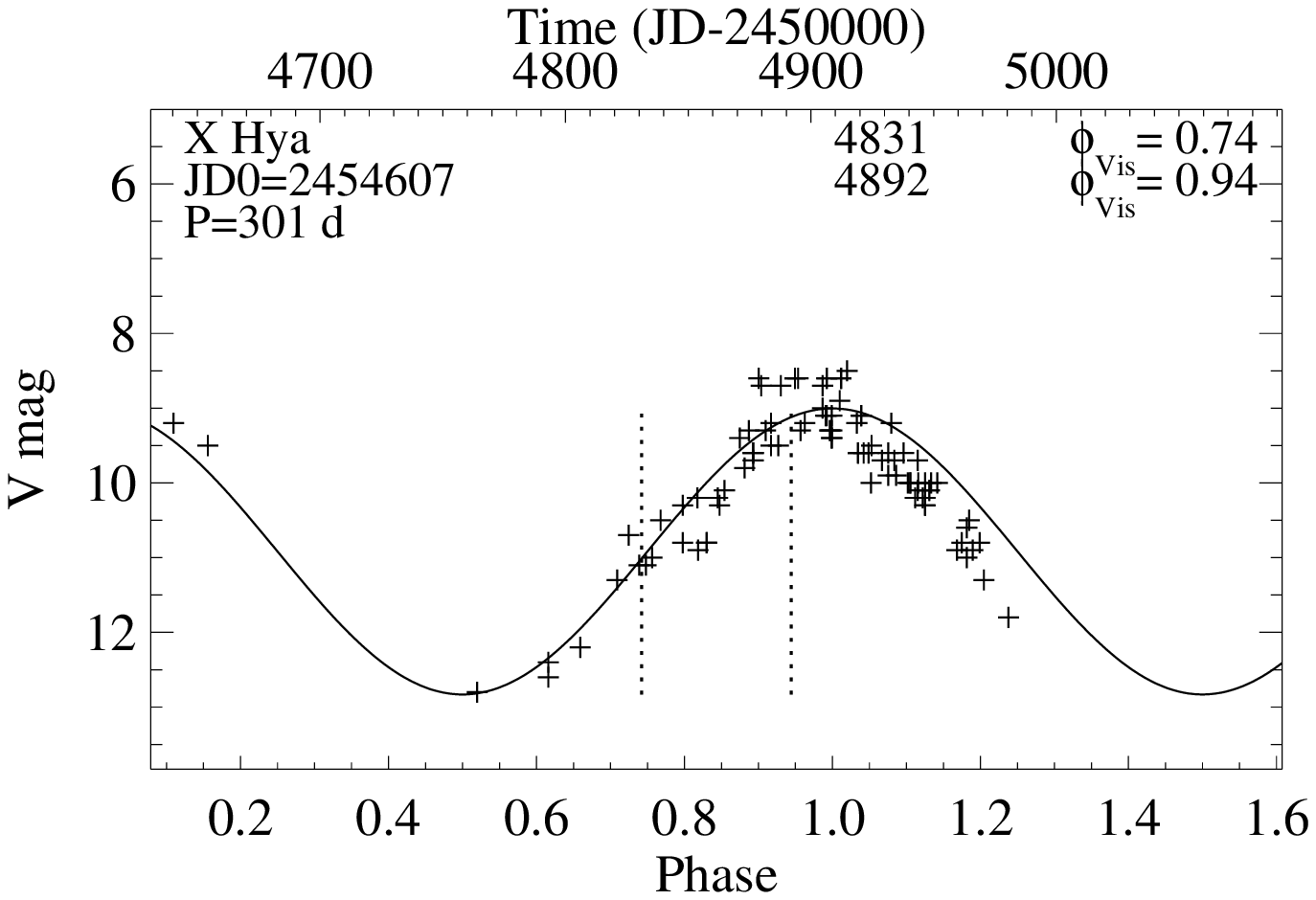}

  \includegraphics[width=0.32\textwidth]{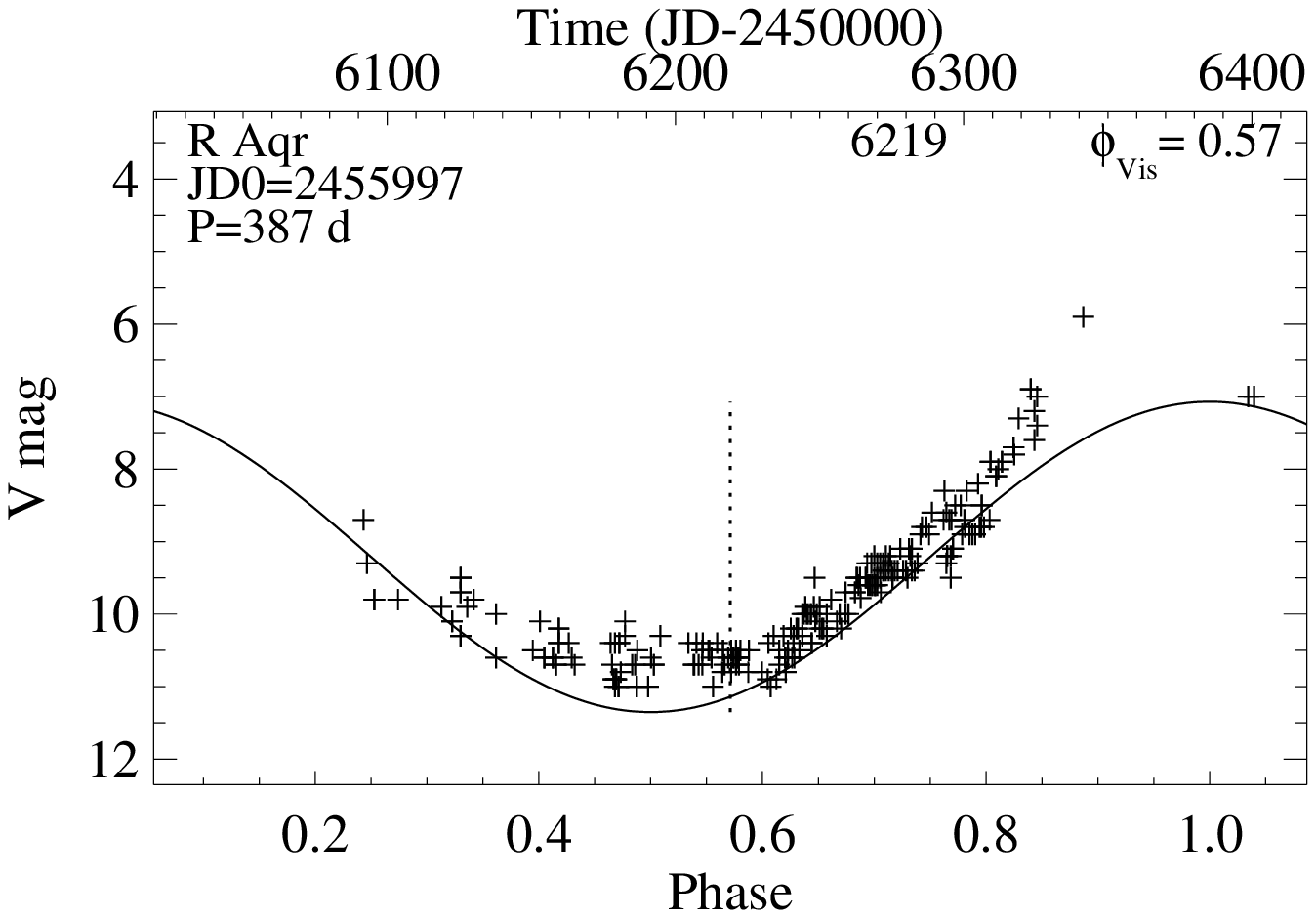}
  \includegraphics[width=0.32\textwidth]{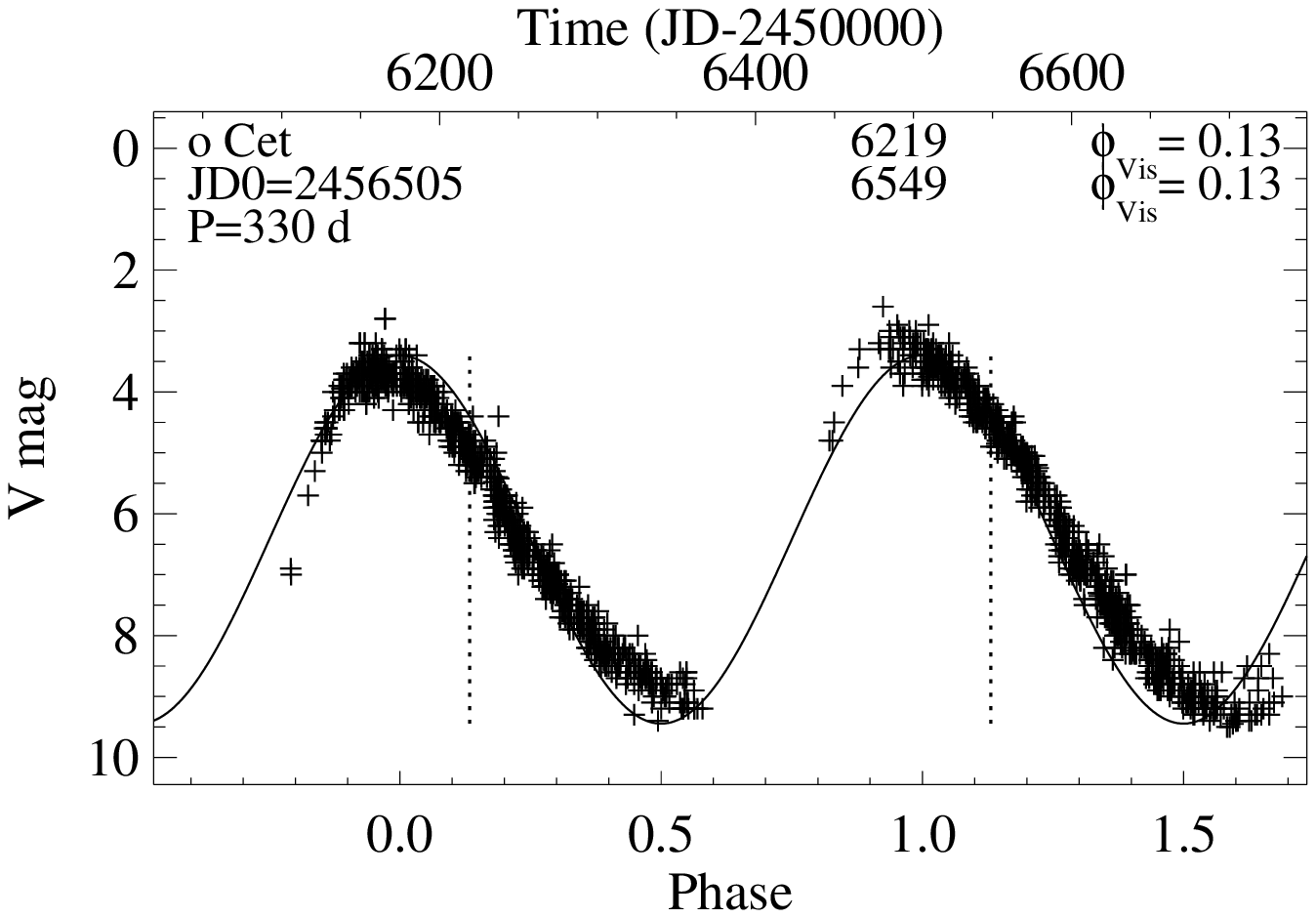}
  \includegraphics[width=0.32\textwidth]{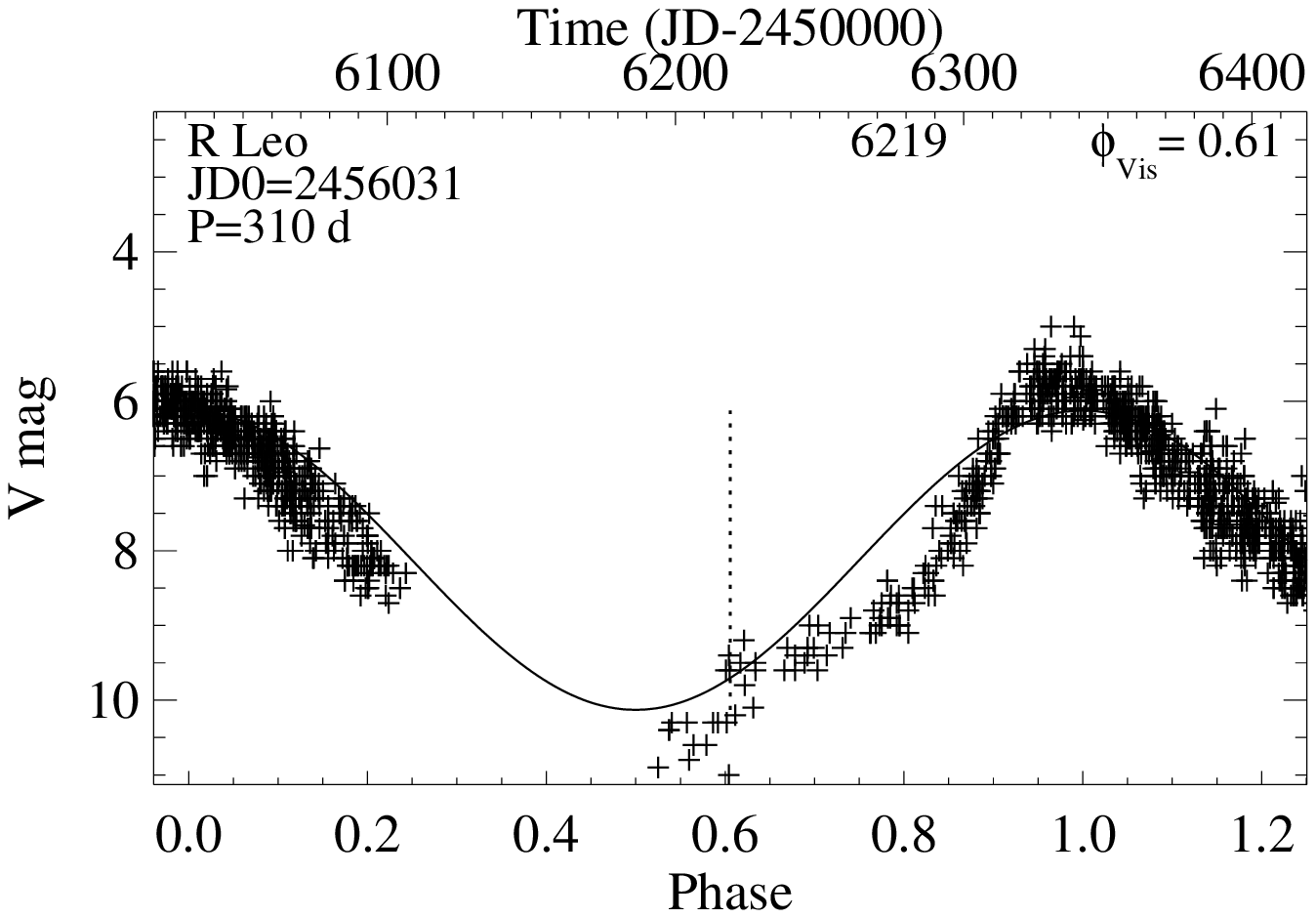}
\caption{Visual lightcurves of our sources as a function of Julian Date (top axis)
and variability phase (bottom axis) around the dates of observation. The solid
line denotes a sinusoidal fit to the data (crosses) obtained from the AAVSO and 
the AFOEV databases. The vertical dashed lines indicate the epochs of our 
observations.\label{fig:lightcurves} }
\end{figure*}

We obtained a total of 20 spectro-interferometric snapshot observations
of the Mira variables R~Cnc, W~Vel, X~Hya, R~Aqr, o~Cet, and R~Leo 
with the AMBER instrument \citep{Petrov2007} 
in the $K$-band medium resolution modes 
(MR-K 2.1\,$\mu$m, MR-K 2.3\,$\mu$m, $R\sim1500$),
employing external FINITO fringe tracking.
Table~\ref{tab:obs} shows the details of our observations, including
the dates, the baselines, the modes, and the calibrators used. 
Figure~\ref{fig:lightcurves} shows the visual lightcurves of our sources
as obtained from the AAVSO and AFOEV databases around the dates of our 
observations. 
R~Cnc was observed at three epochs;
W~Vel, X~Hya, and o~Cet at two epochs each; and R~Aqr and R~Leo at one epoch
each. Observations of these targets were interleaved with observations
of interferometric calibrators. Calibrators were selected with the 
ESO Calvin tool, and their angular $K$-band diameters were adopted from 
\citet{Lafrasse2010} (81 Gem: 2.93 mas $\pm$ 0.03 mas; 
$\iota$ Hya: 3.58 mas $\pm$ 0.28 mas; 31 Ori: 4.02 mas $\pm$ 0.39 mas; 
31 Leo: 3.33 mas $\pm$ 0.34 mas; $\gamma$ Pyx: 3.00 mas $\pm$ 0.27 mas;
$\phi$ Aqr: 6.16 mas $\pm$ 0.50 mas; $\beta$ Cnc: 5.21 mas $\pm$ 0.36 mas).
For $\alpha$ Cet, we instead adopted the measured K-band angular 
diameter of 11.05 mas $\pm$ 0.55 mas from \citet{Wittkowski2006}.
The first epochs of R Cnc, X Hya, and W Vel were previously analyzed and
published by \citet{Wittkowski2011}. Here, we reduced and analyzed
the complete data set in a uniform way with the aims of studying the
visibility variability and comparing the observations not only to 
1D CODEX dynamic model atmospheres based on self-excited pulsation models,
but also for the first time to 3D CO5BOLD dynamic model atmospheres including 
pulsation and convection.

Each data set of Table~\ref{tab:obs} typically consists of a calibrator
observation taken before the science target observation, the 
science target observation, and another calibrator
observation taken afterward. Each of these observations typically
consist of five files of 170 scans each, accompanied by a dark and a sky file.
It is essential for a good calibration of the interferometric transfer function
that the performance of the FINITO fringe tracking is comparable between
corresponding calibrator and science target observations. To this purpose,
we inspected the FINITO lock ratios and the FINITO phase rms values
as reported in the fits headers of each file. In a few cases we deselected 
individual files for which these values deviated by more than about 20\%. 
In rare cases we had to discard a complete data set because the FINITO phase 
rms values were systematically very different by clearly more than 
about 20\% between science target observation and calibrator observations.
This concerned one additional data set of R Cnc and one additional data set
of X Hya obtained in MR21 mode on 2008 December 30, as well as data of
o Cet obtained in MR23 mode on 2011 September 30.

We obtained averaged visibility and closure phase values from the raw data
using the latest version, version 3.0.8, of the {\tt amdlib} 
data reduction package \citep{Tatulli2007,Chelli2009}. The absolute wavelength
calibration and the calibration of the interferometric transfer function
were performed using IDL scripts developed in house. For the absolute wavelength
calibration, we correlated the AMBER flux spectra with a
reference spectrum that included the AMBER transmission curve, the telluric
spectrum estimated with ATRAN \citep{Lord1992}, and the expected stellar
spectrum using the spectrum of the K giant BS~4432 from 
\citet{Lancon2000}. For calibrating the interferometric transfer
function, we used an average transfer function of the calibrator observations
obtained before and after that of the science target. In some cases, when
this sequence was not available, we used the two closest calibrators in time.
The error of the transfer function was estimated by the difference between
the two calibrator measurements.
The error of the final visibility spectra includes the statistical noise
of the raw data and the systematic error of the transfer function. 

All obtained flux, squared visibility, and closure phase data are shown 
below in Figs.~\ref{fig:rcnc}--\ref{fig:rleo} (online version), together 
with model predictions as outlined in Sect.~\ref{sec:modelcomp}.
These figures include one row of flux, squared visibility, and closure phase 
panels for each row of Table~\ref{tab:obs}, so they show the results of each
dataset individually. The middle panels showing the squared visibility data
include three lines corresponding to the three baselines of each AMBER dataset,
where the shortest baseline corresponds to the highest visibility curve 
and the longest baseline to the lowest visibility curve.
\onlfig{
\begin{figure*}[p]
\centering
  \includegraphics[width=0.32\textwidth]{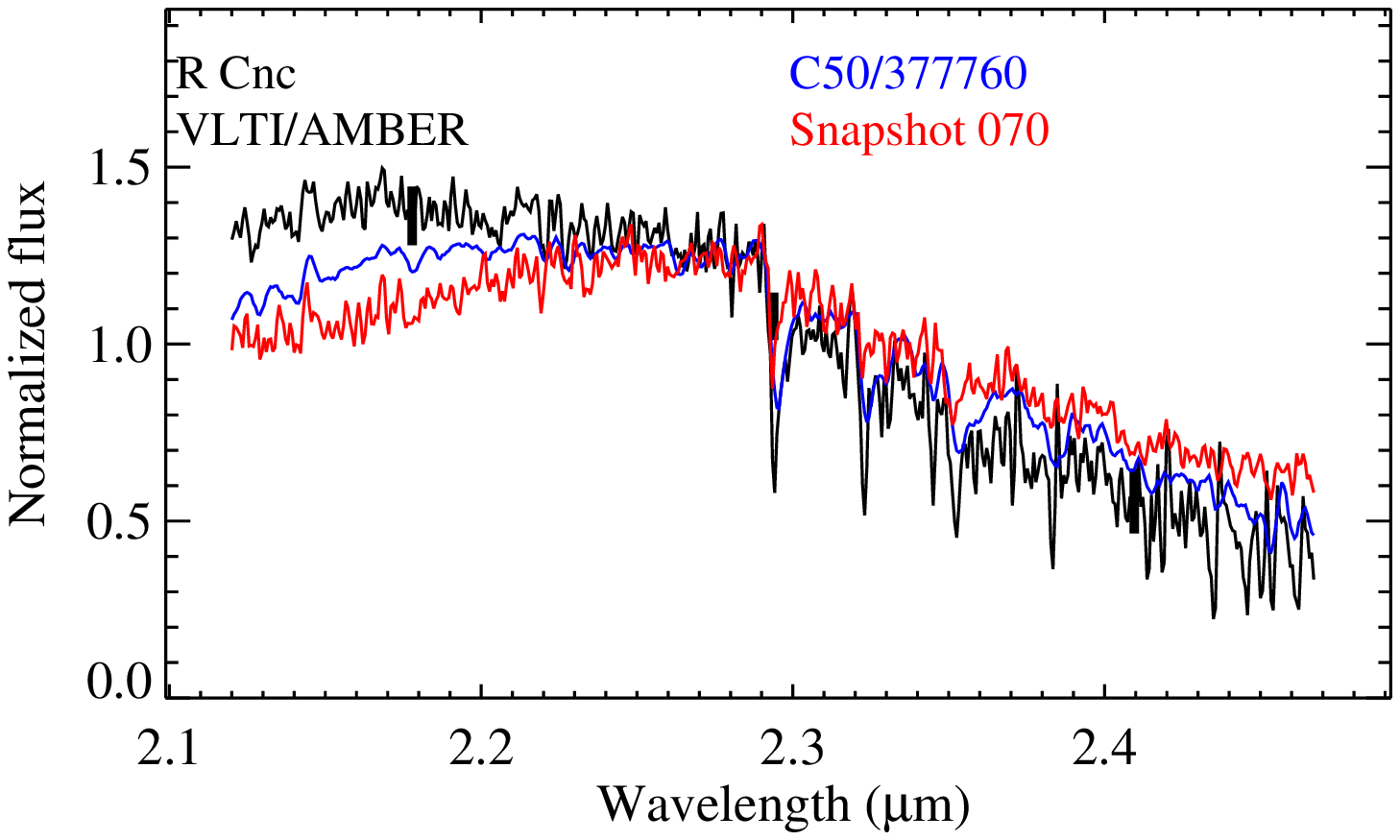}
  \includegraphics[width=0.32\textwidth]{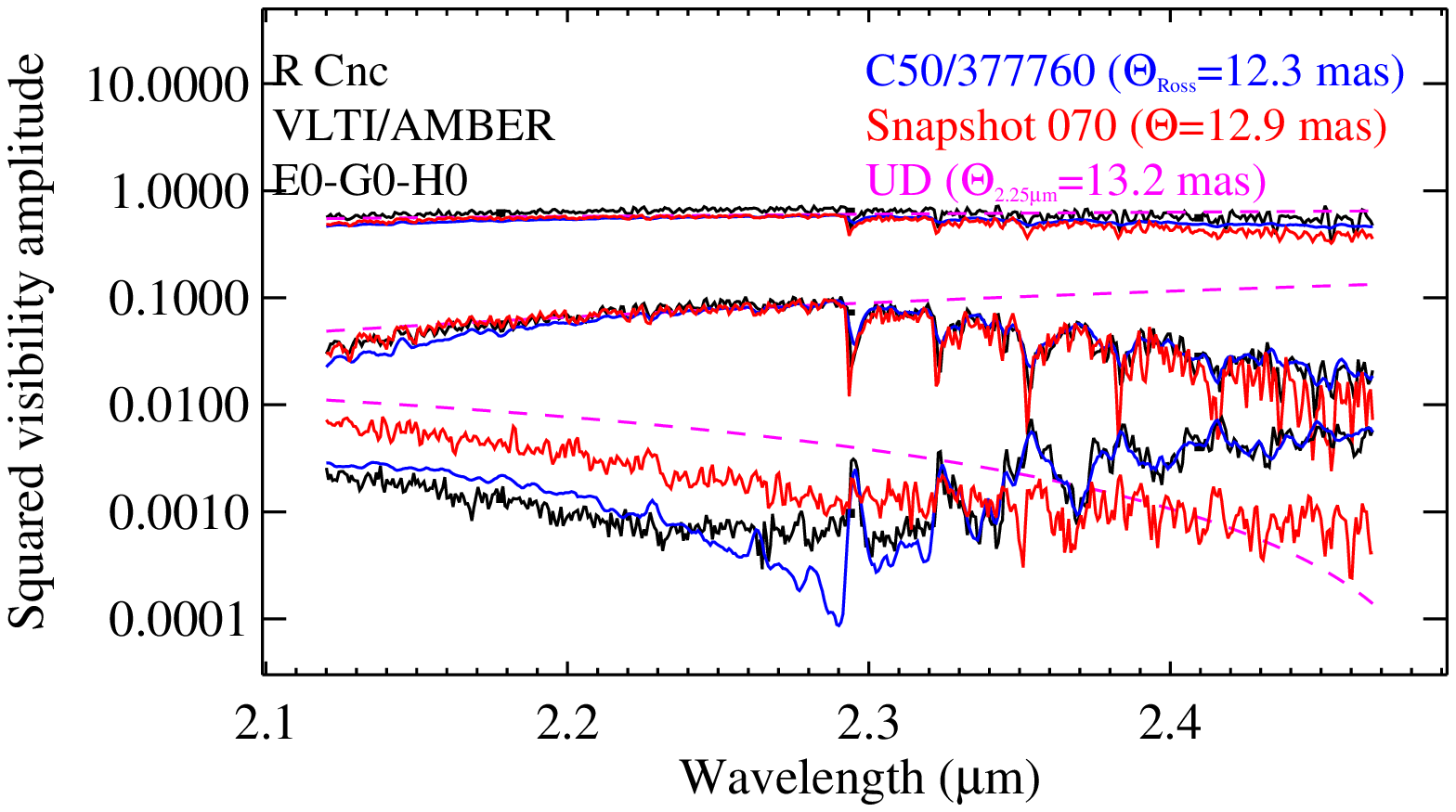}
  \includegraphics[width=0.32\textwidth]{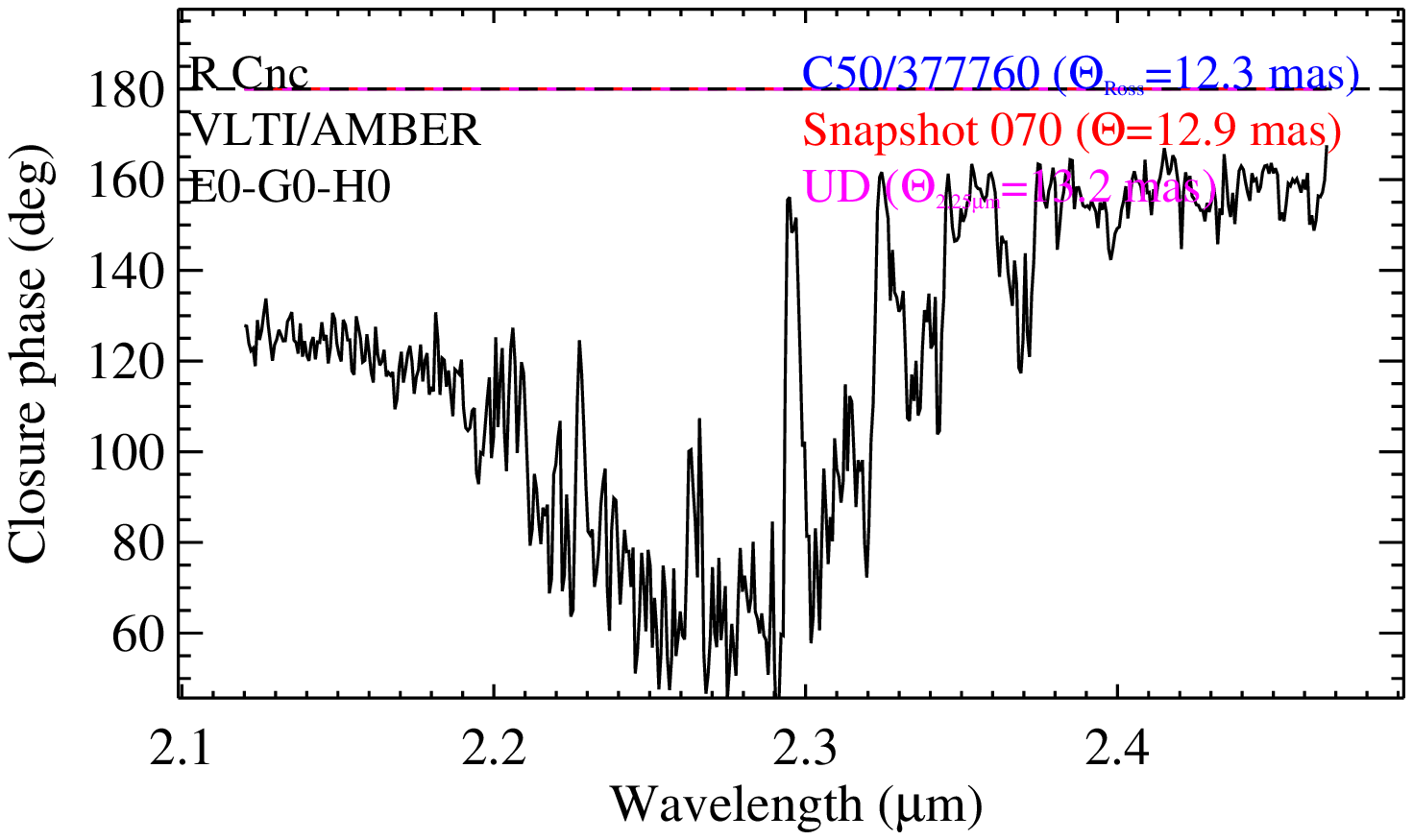}

  \includegraphics[width=0.32\textwidth]{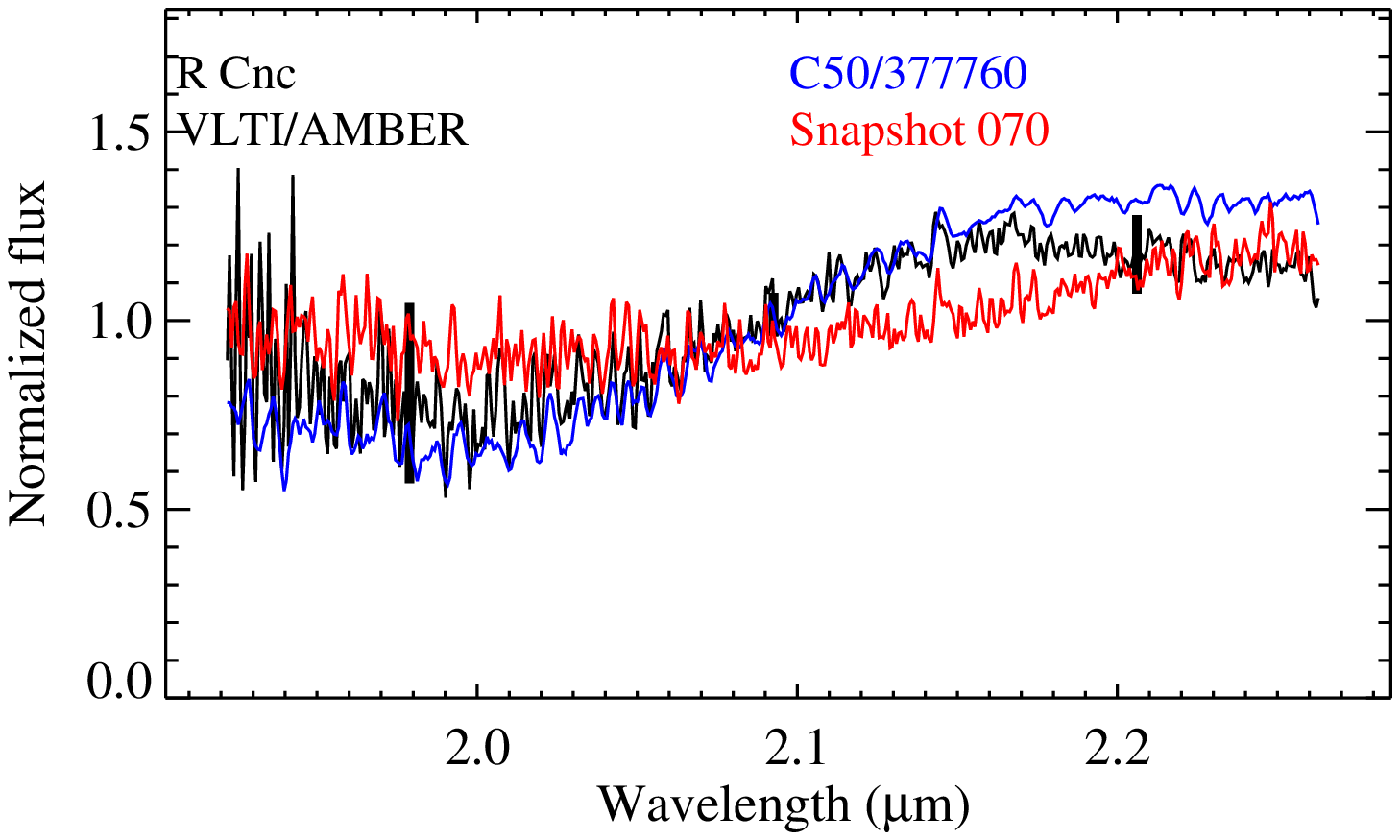}
  \includegraphics[width=0.32\textwidth]{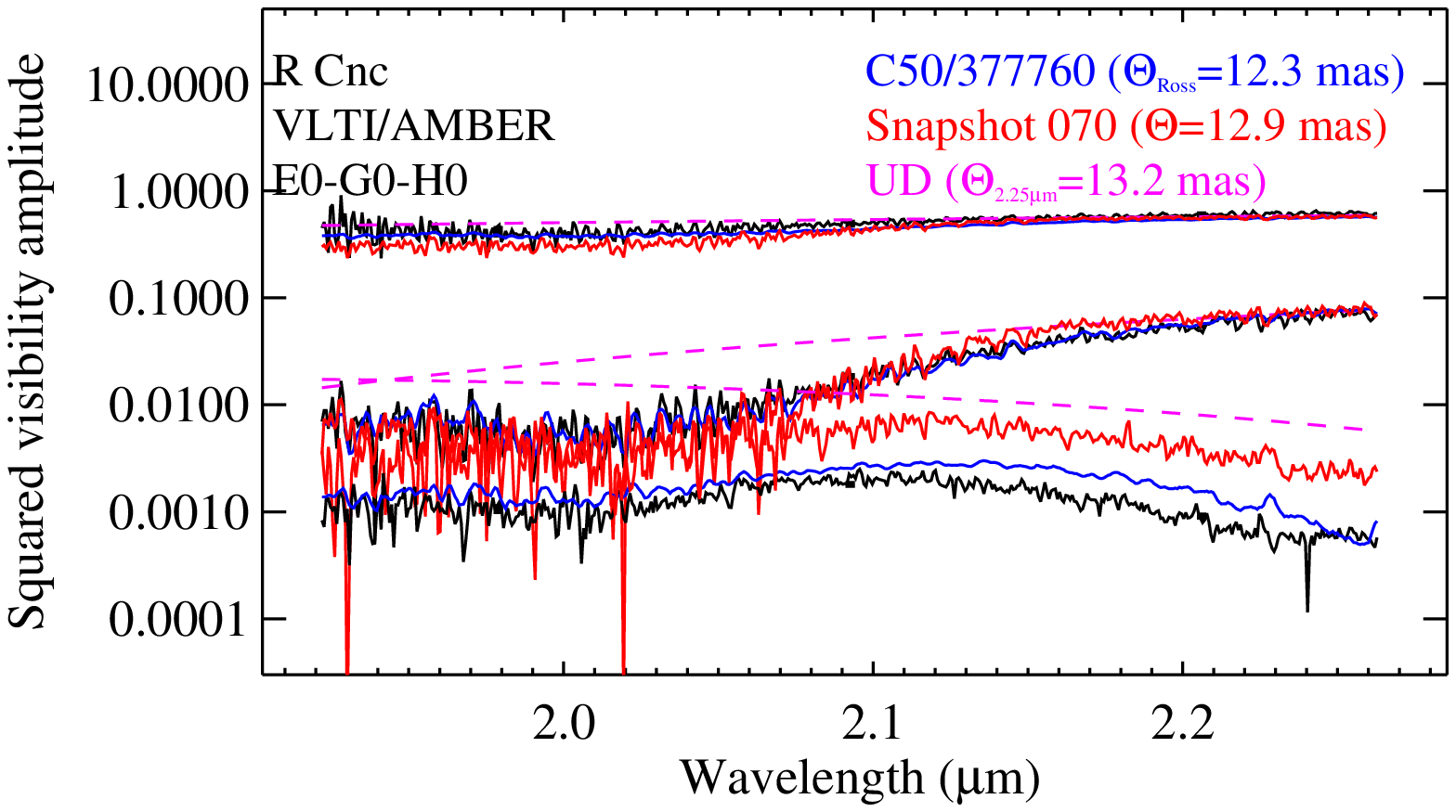}
  \includegraphics[width=0.32\textwidth]{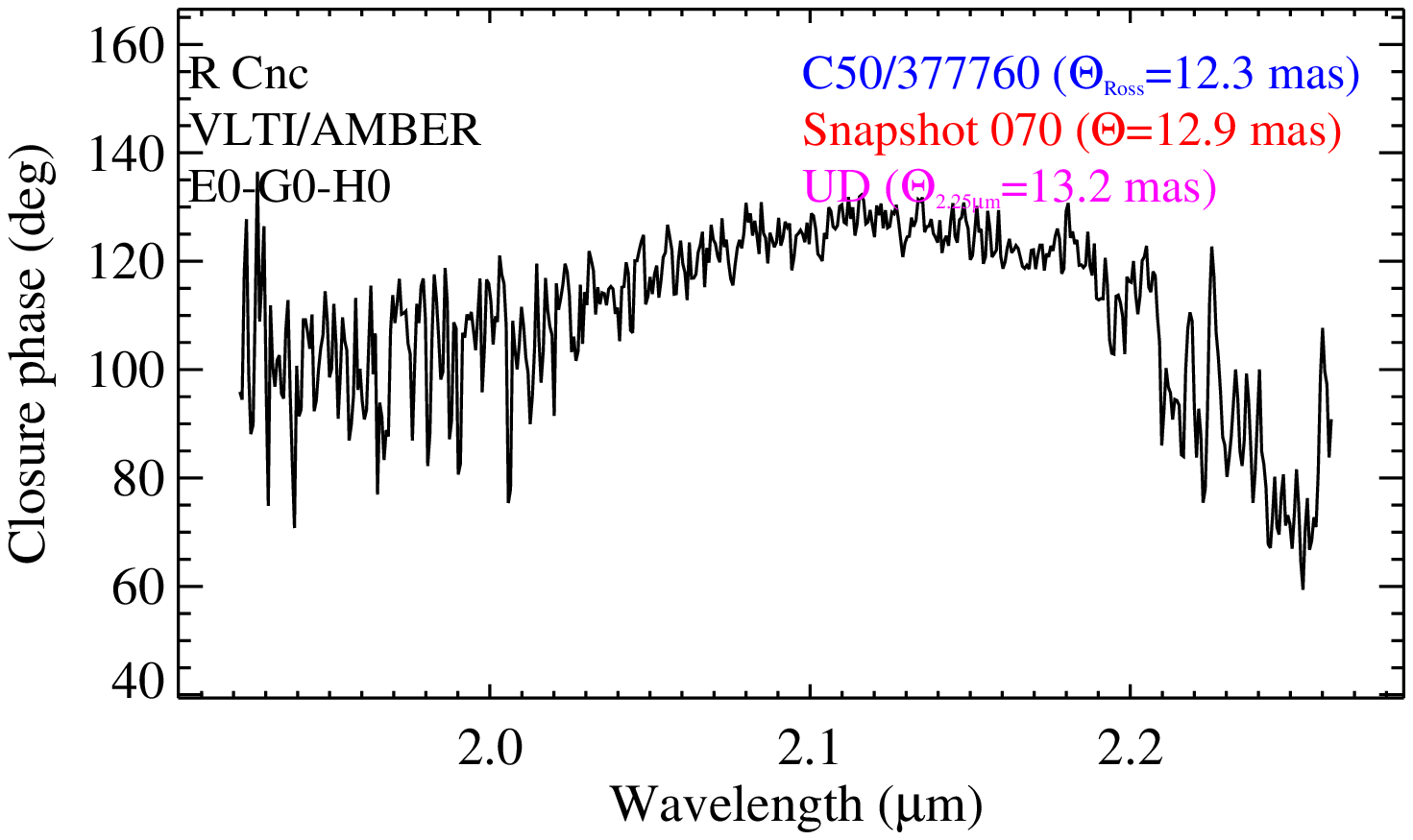}

  \includegraphics[width=0.32\textwidth]{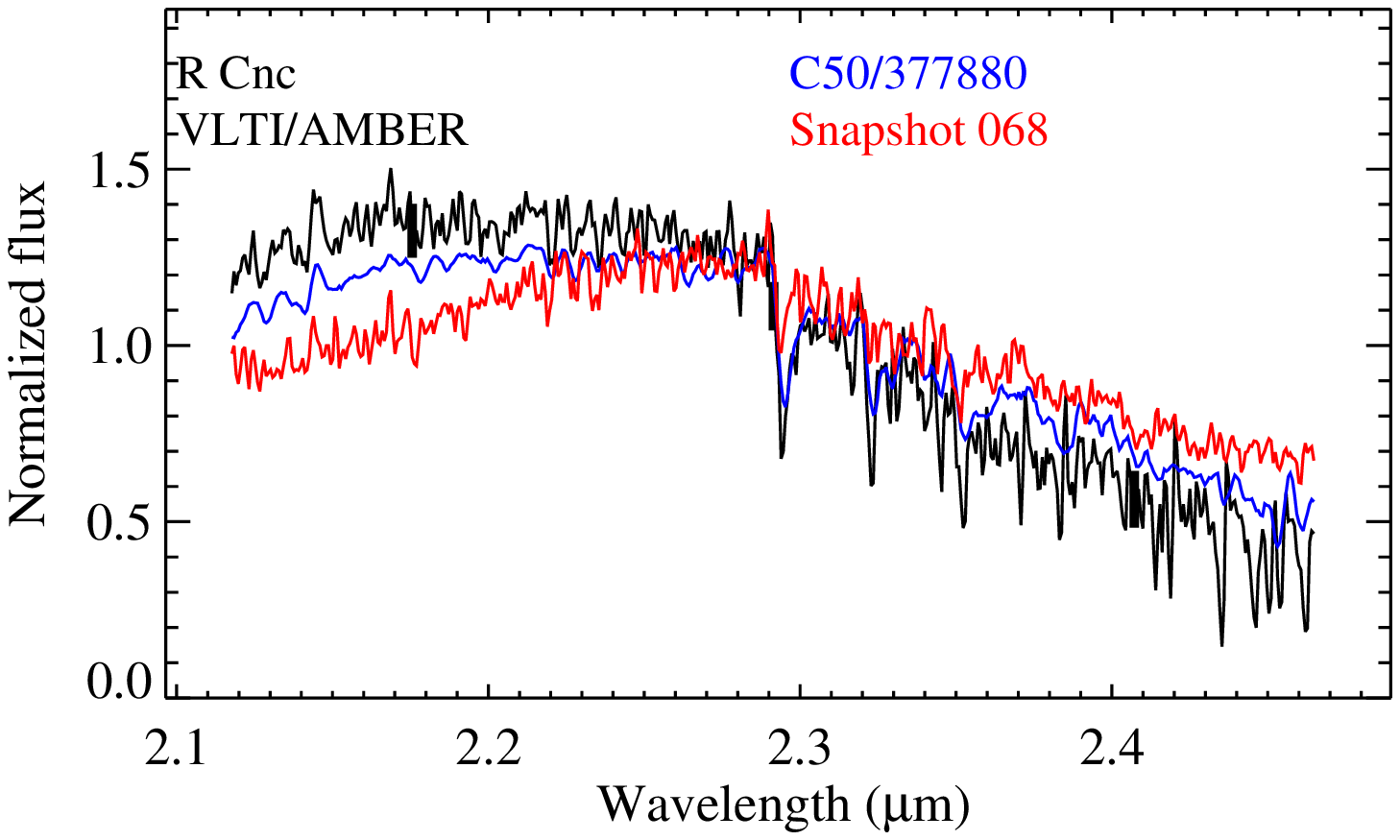}
  \includegraphics[width=0.32\textwidth]{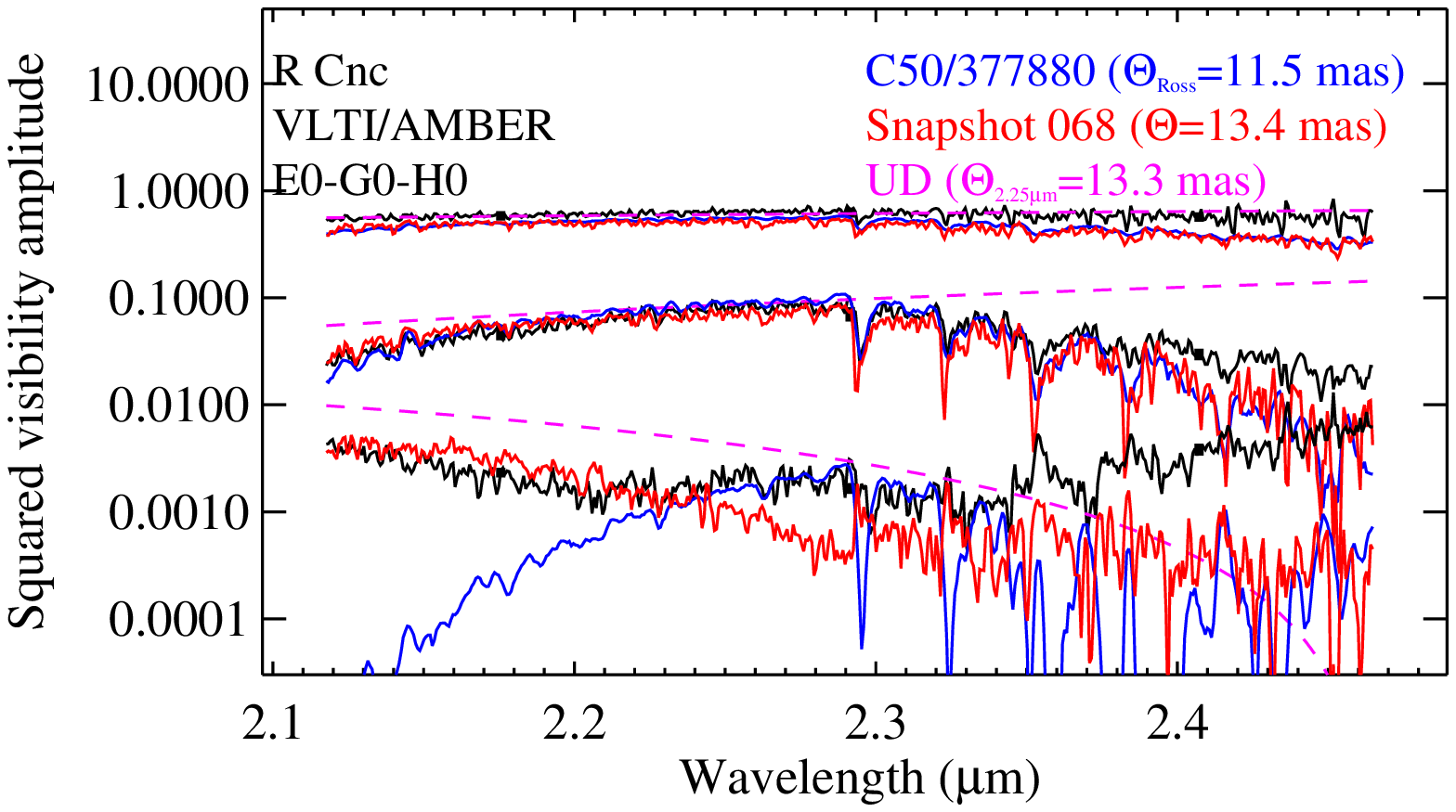}
  \includegraphics[width=0.32\textwidth]{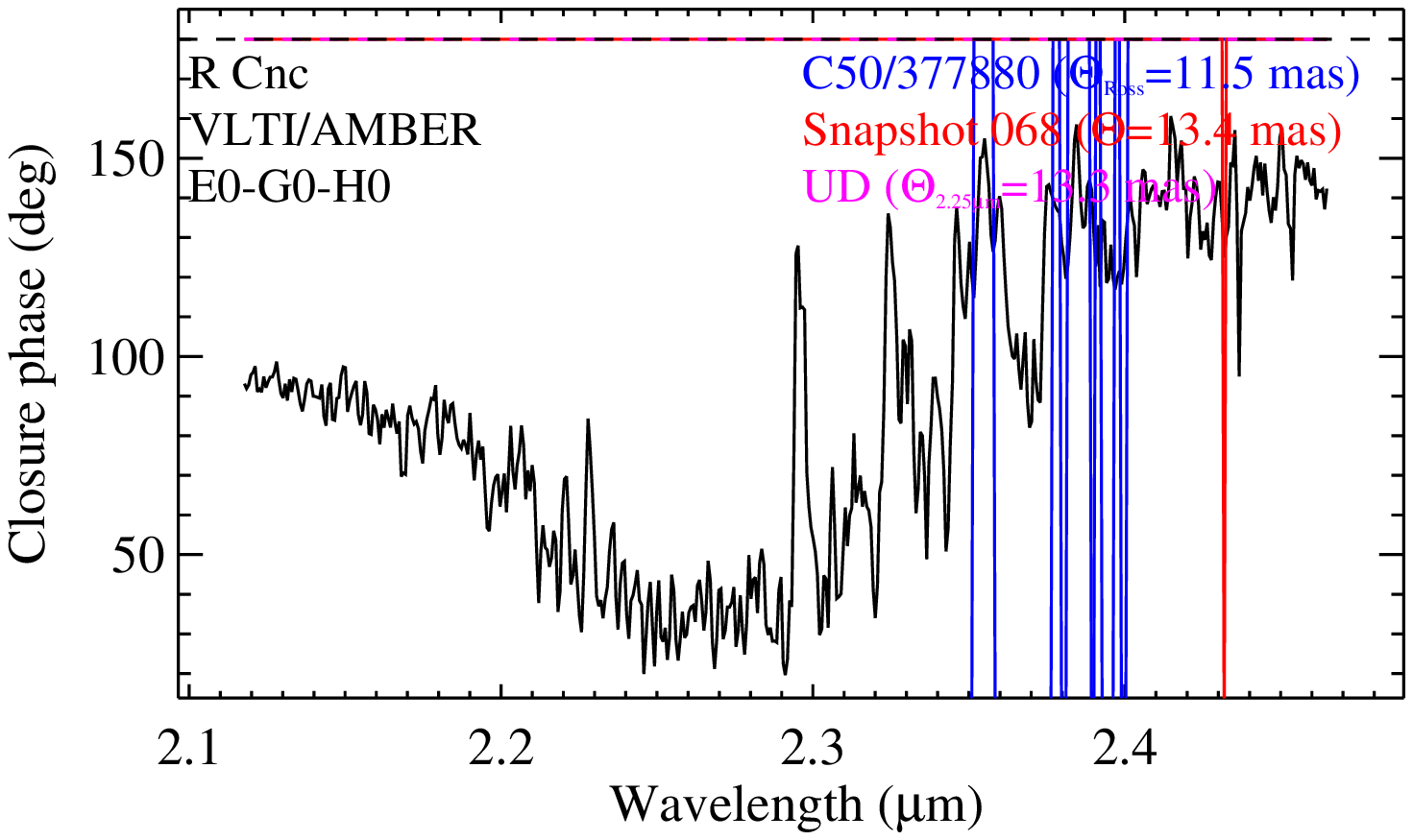}

  \includegraphics[width=0.32\textwidth]{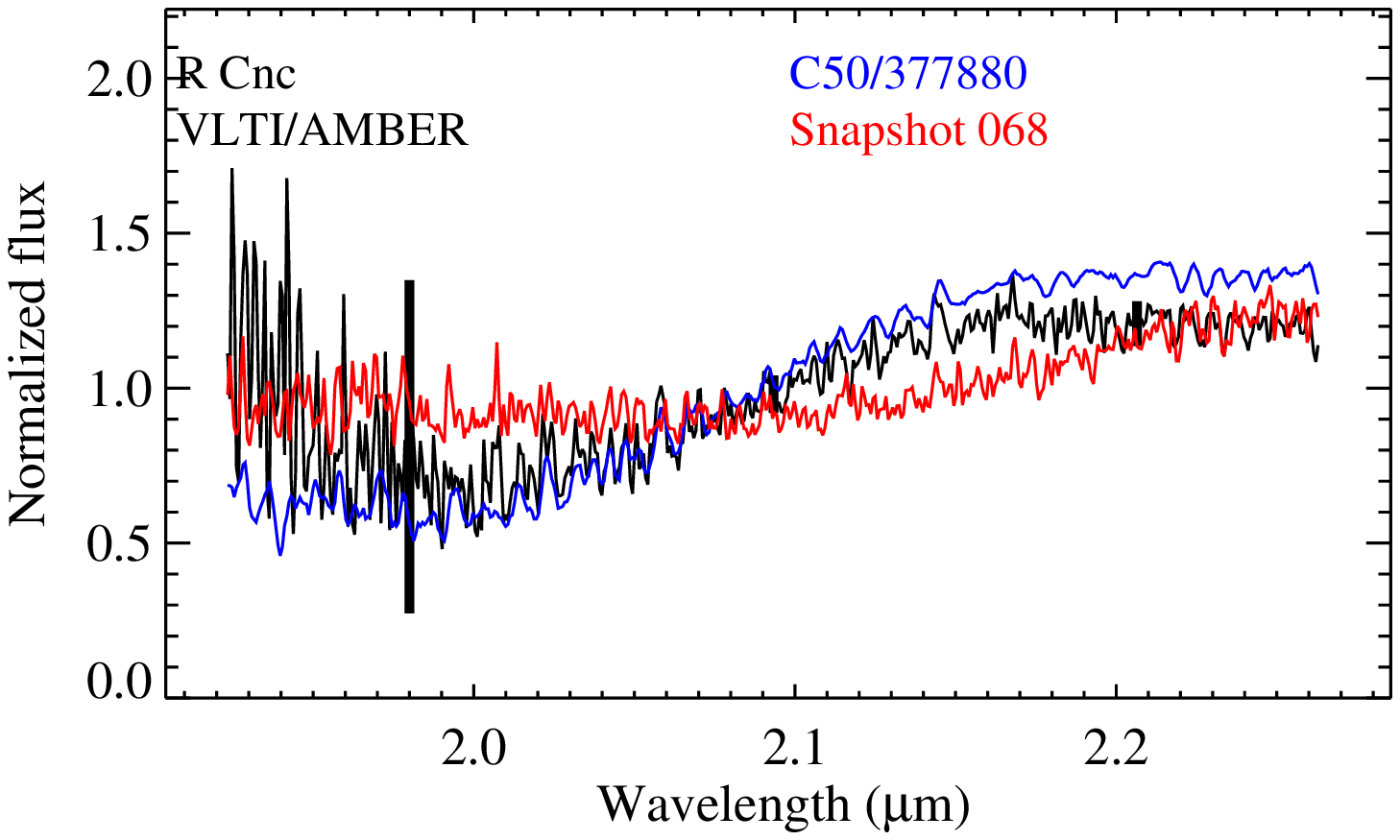}
  \includegraphics[width=0.32\textwidth]{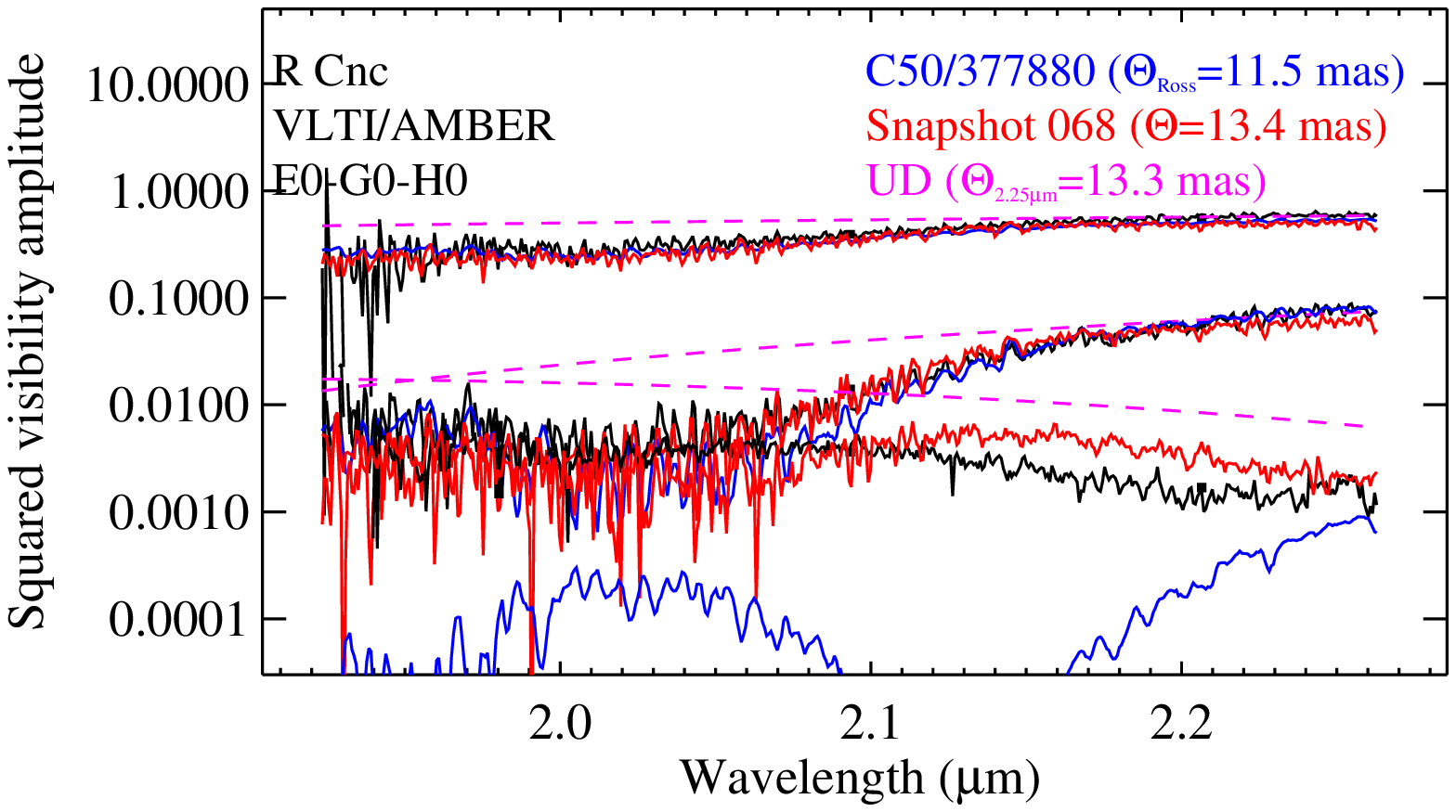}
  \includegraphics[width=0.32\textwidth]{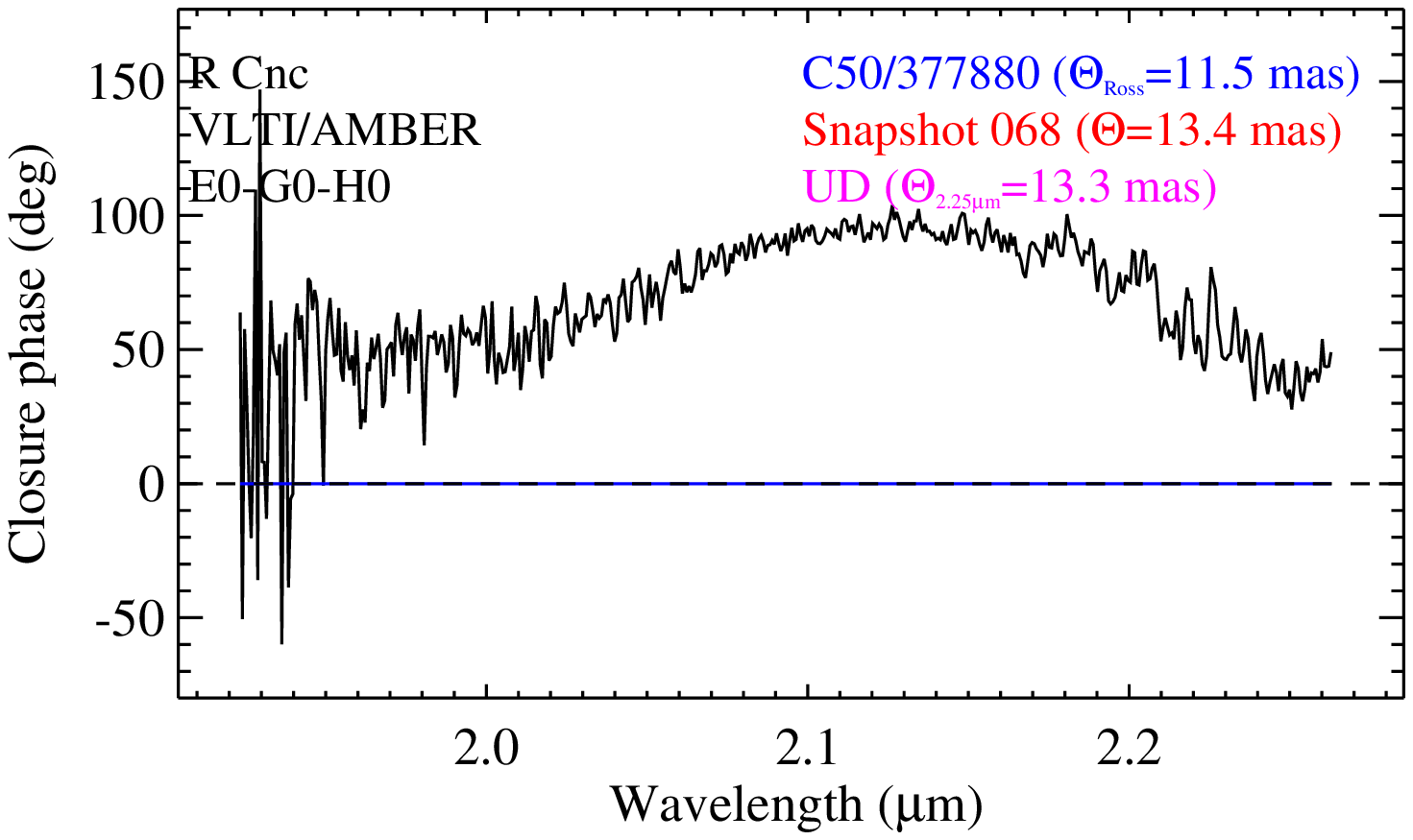}

  \includegraphics[width=0.32\textwidth]{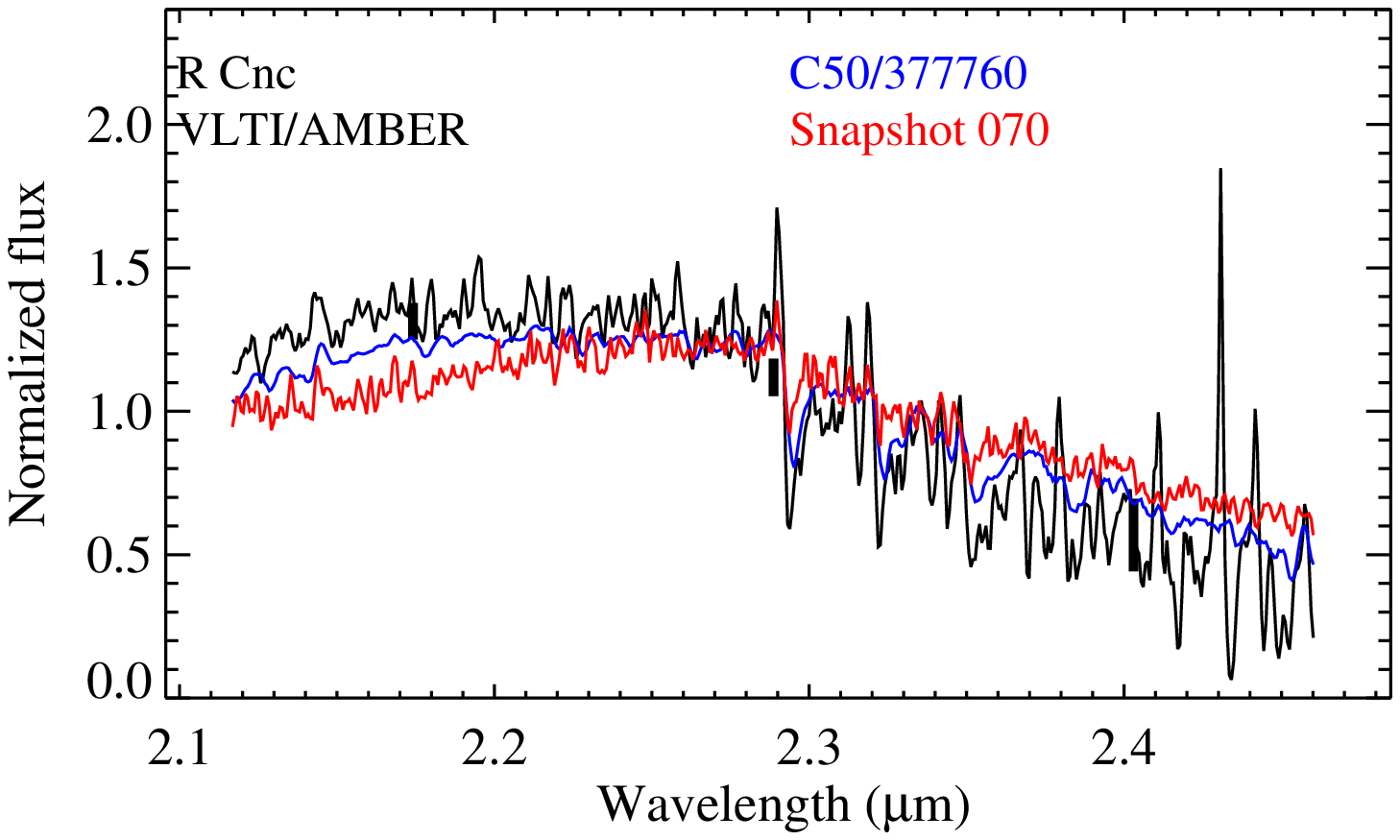}
  \includegraphics[width=0.32\textwidth]{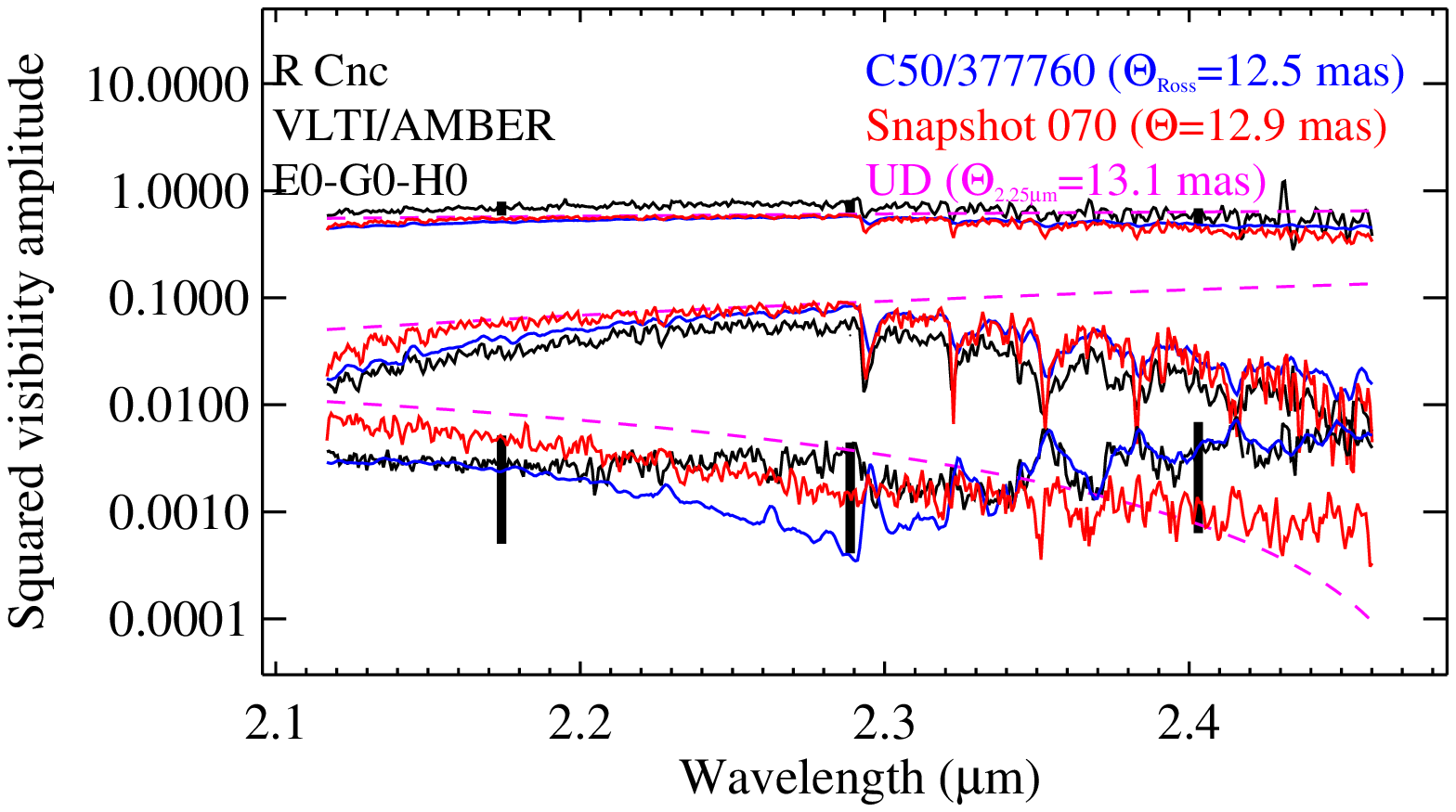}
  \includegraphics[width=0.32\textwidth]{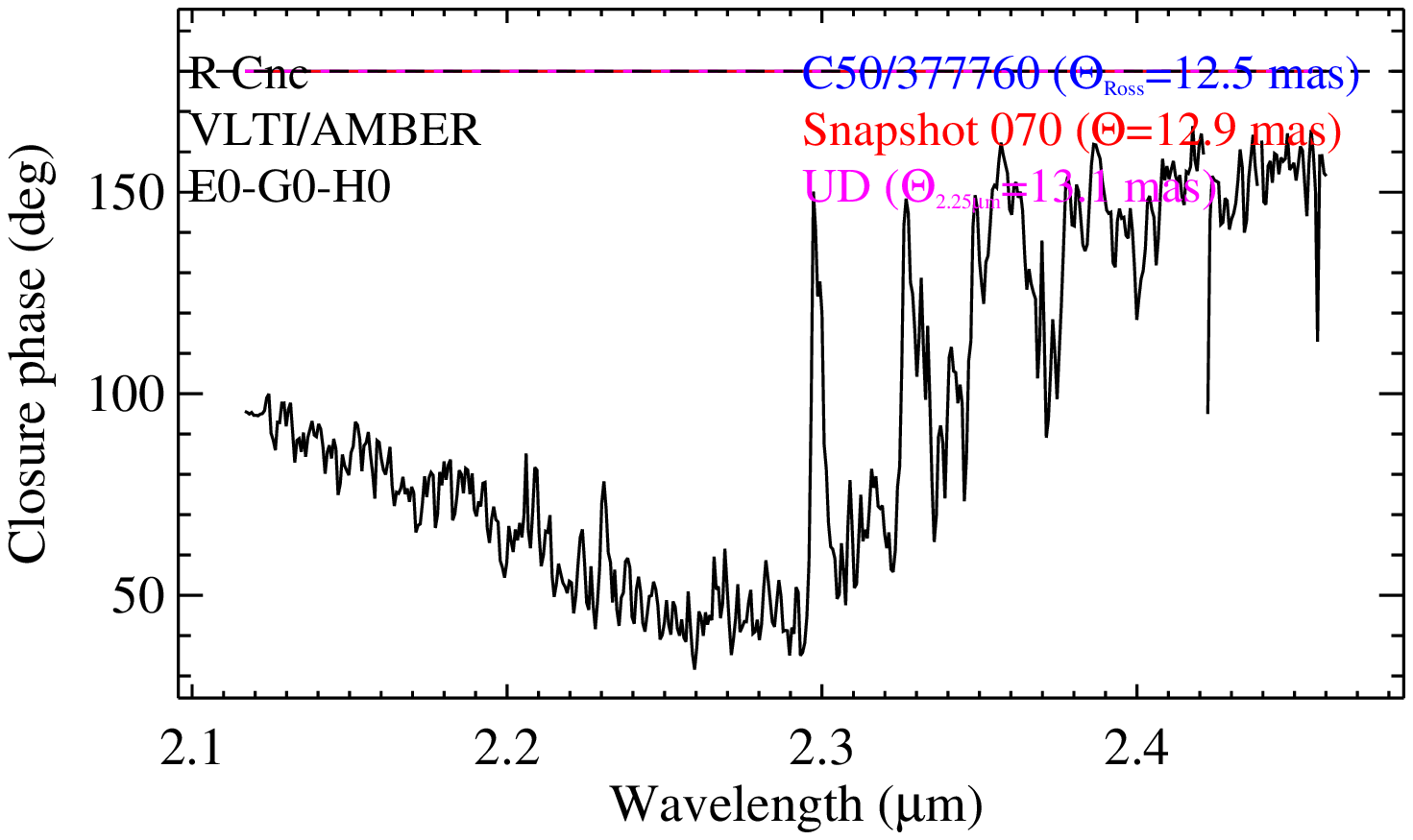}

  \includegraphics[width=0.32\textwidth]{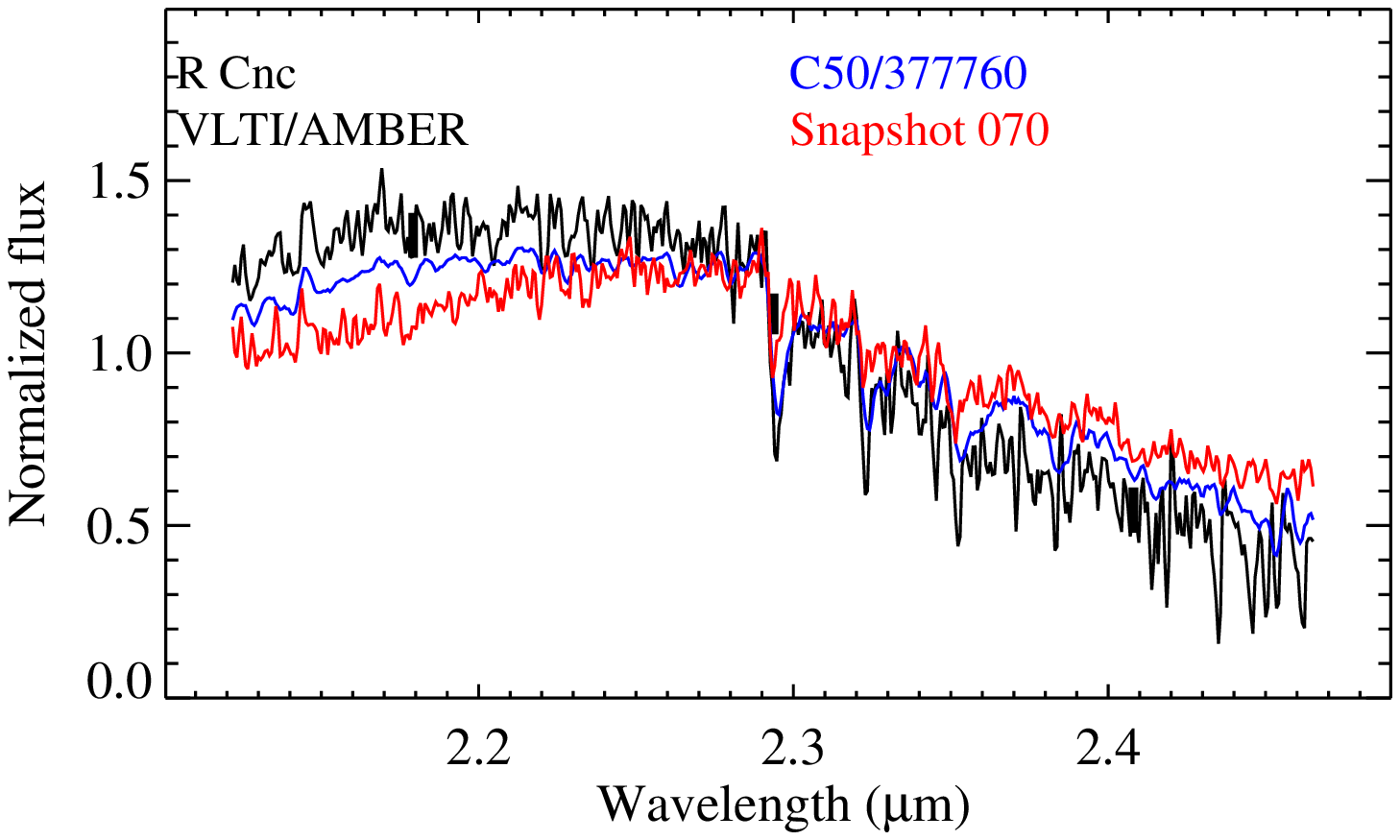}
  \includegraphics[width=0.32\textwidth]{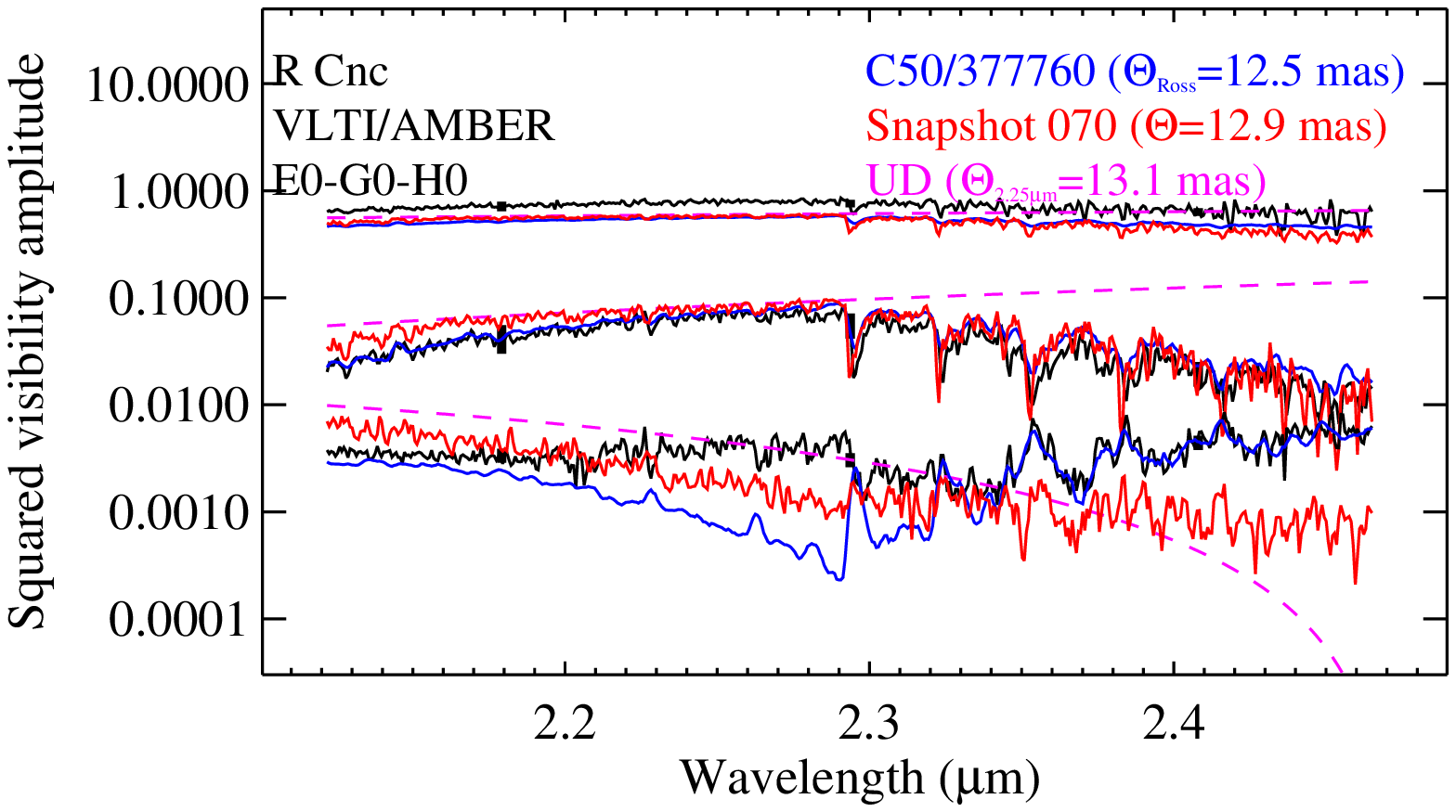}
  \includegraphics[width=0.32\textwidth]{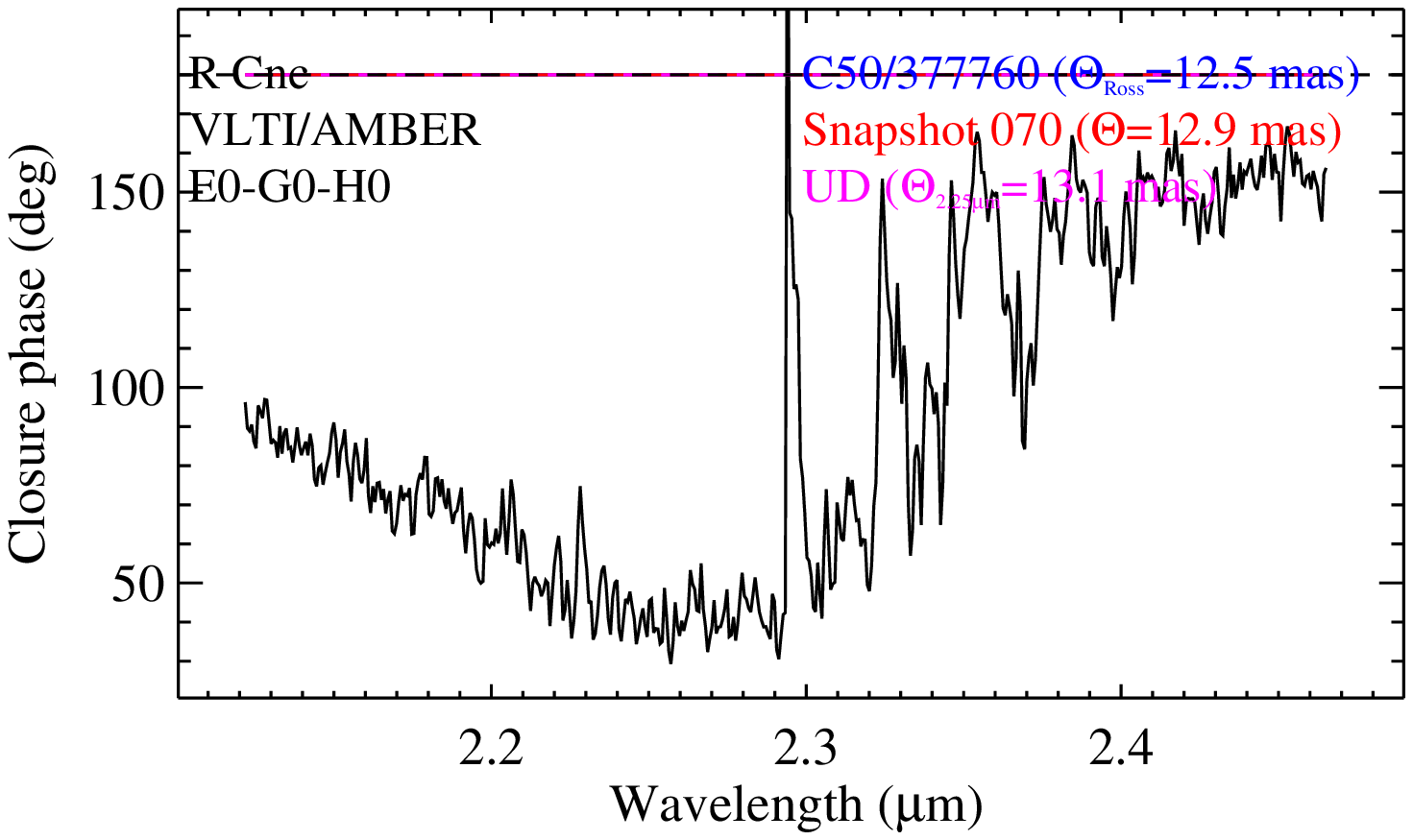}
\caption{Flux (left panels), squared visibility amplitudes (middle panels), 
and closure phases (right panels) of R Cnc from data sets no. 1 to 6 
(top to bottom) of Table~\protect\ref{tab:obs} (black lines).
Also shown are model predictions 
from the best-fit 1D CODEX (blue lines) and 3D CO5BOLD (red lines) models
from Table~\protect\ref{tab:fitresults}. The squared visibility amplitudes 
include three lines that correspond to the three baselines of each 
AMBER dataset. Mean error bars are drawn in the center of 
each third of the total wavelength range. In particular, for the logarithmic scales, 
these are sometimes too small to be easily visible. They are typically 
5--20\% for the fluxes, 5--15\% for
the squared visibility amplitudes, and 5--15$\deg$ for the closure phases.
}
\label{fig:rcnc}
\end{figure*}
}
\onlfig{
\begin{figure*}[p]
\centering
  \includegraphics[width=0.32\textwidth]{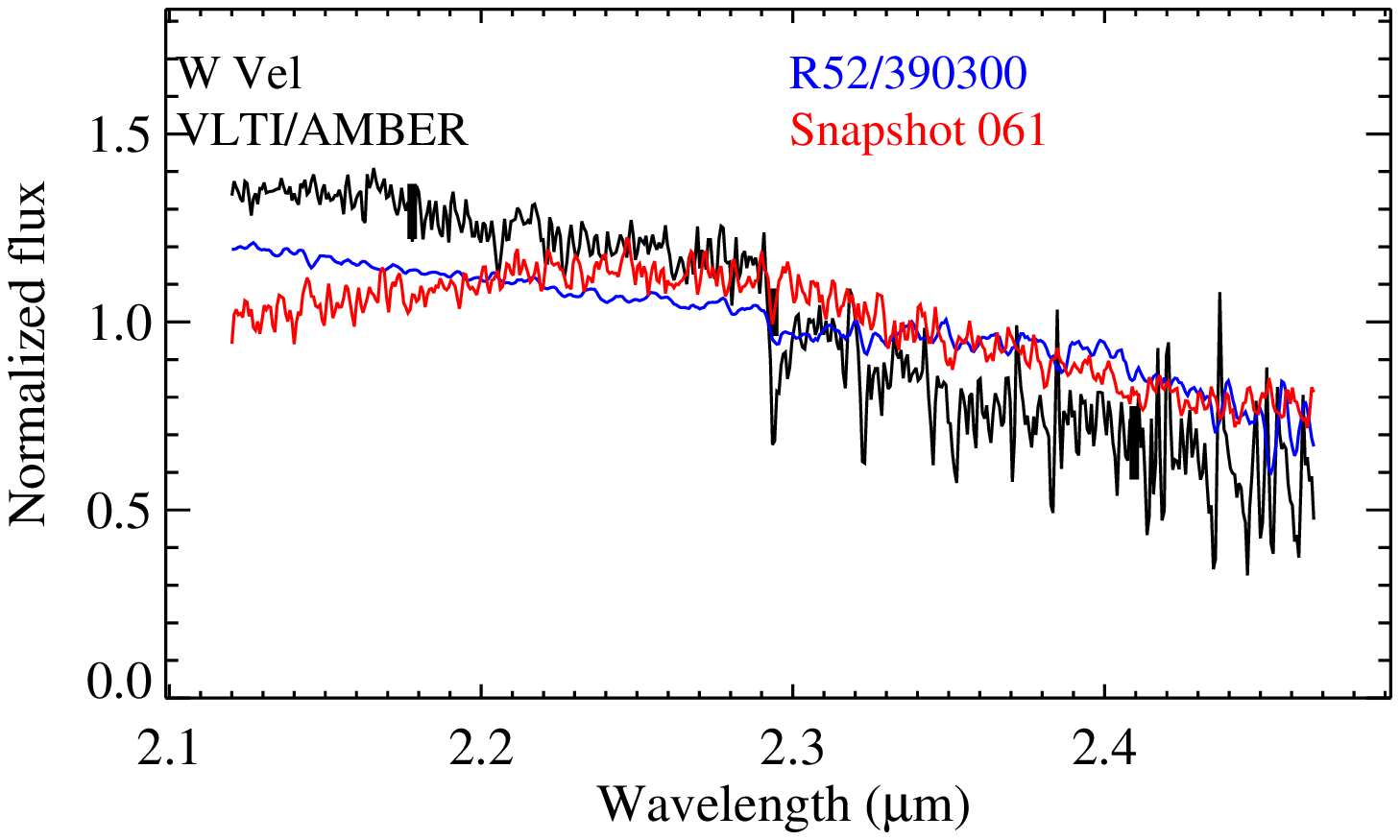}
  \includegraphics[width=0.32\textwidth]{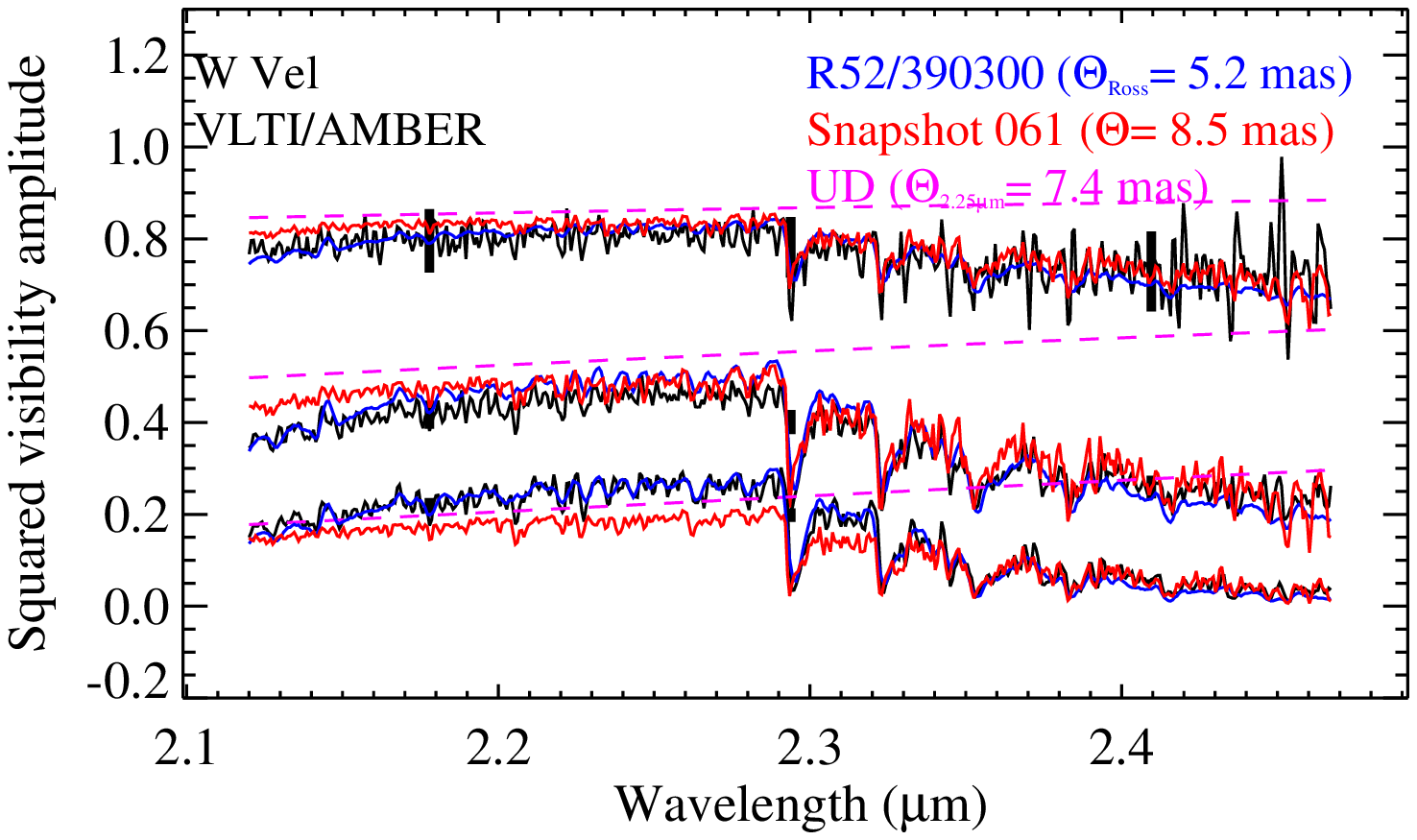}
  \includegraphics[width=0.32\textwidth]{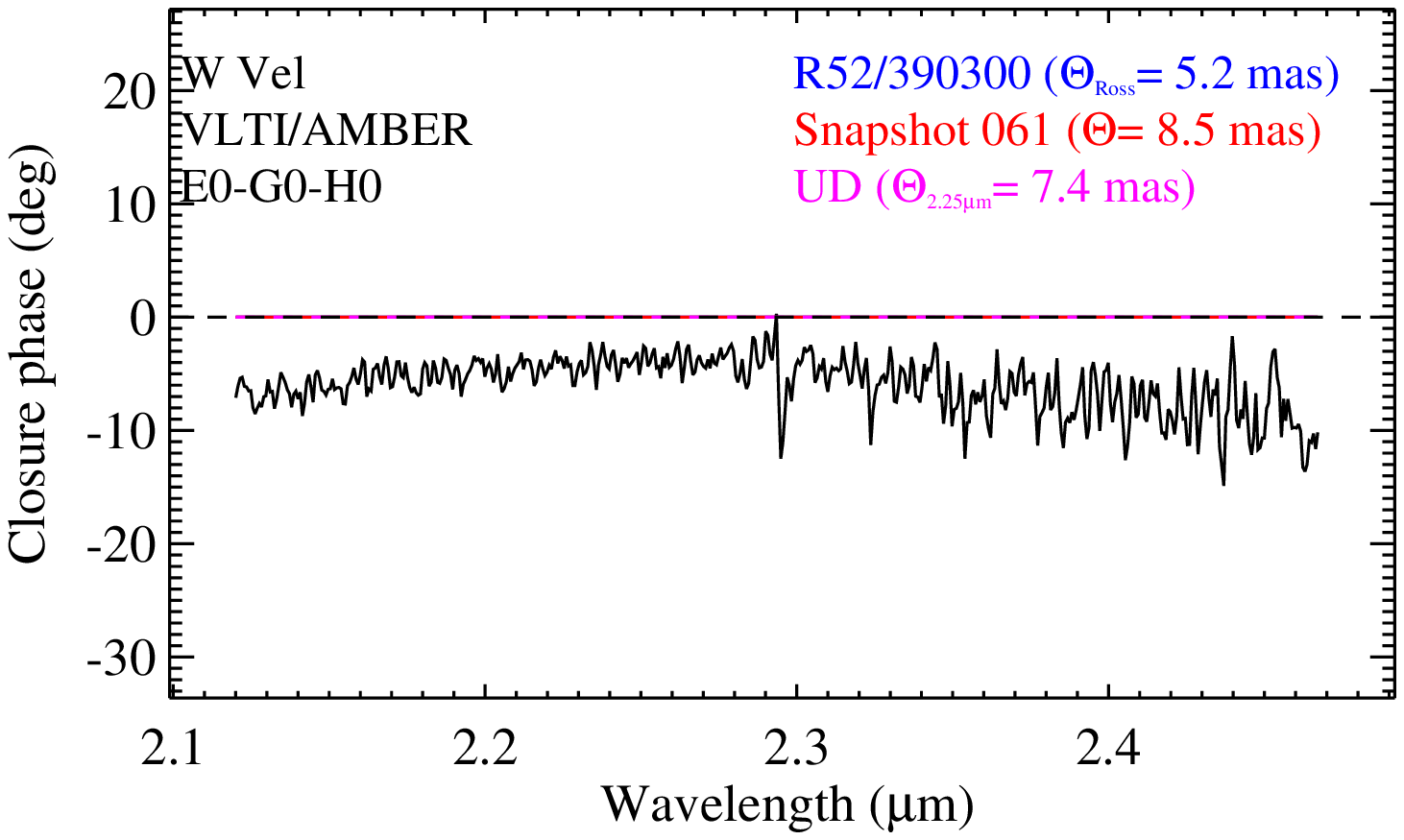}

  \includegraphics[width=0.32\textwidth]{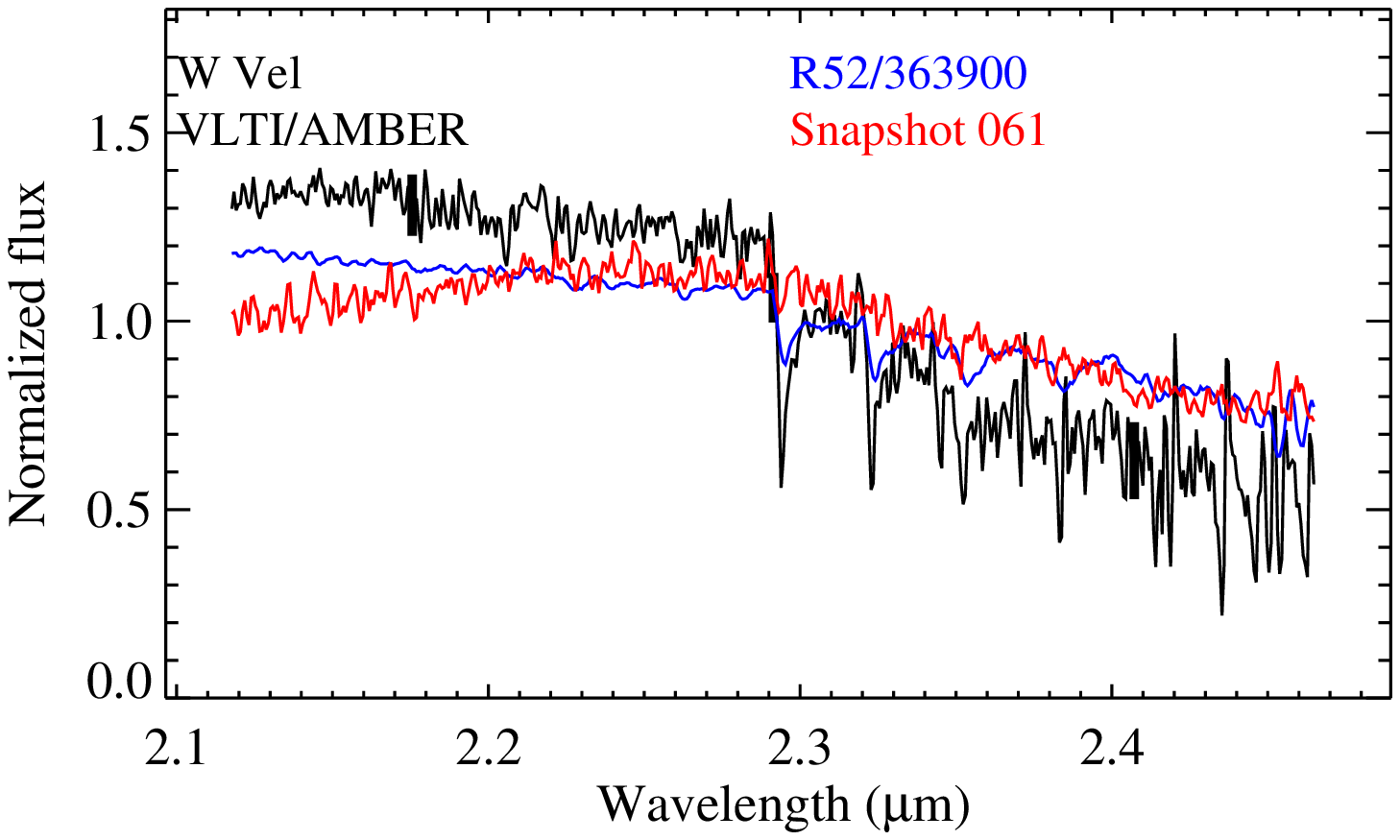}
  \includegraphics[width=0.32\textwidth]{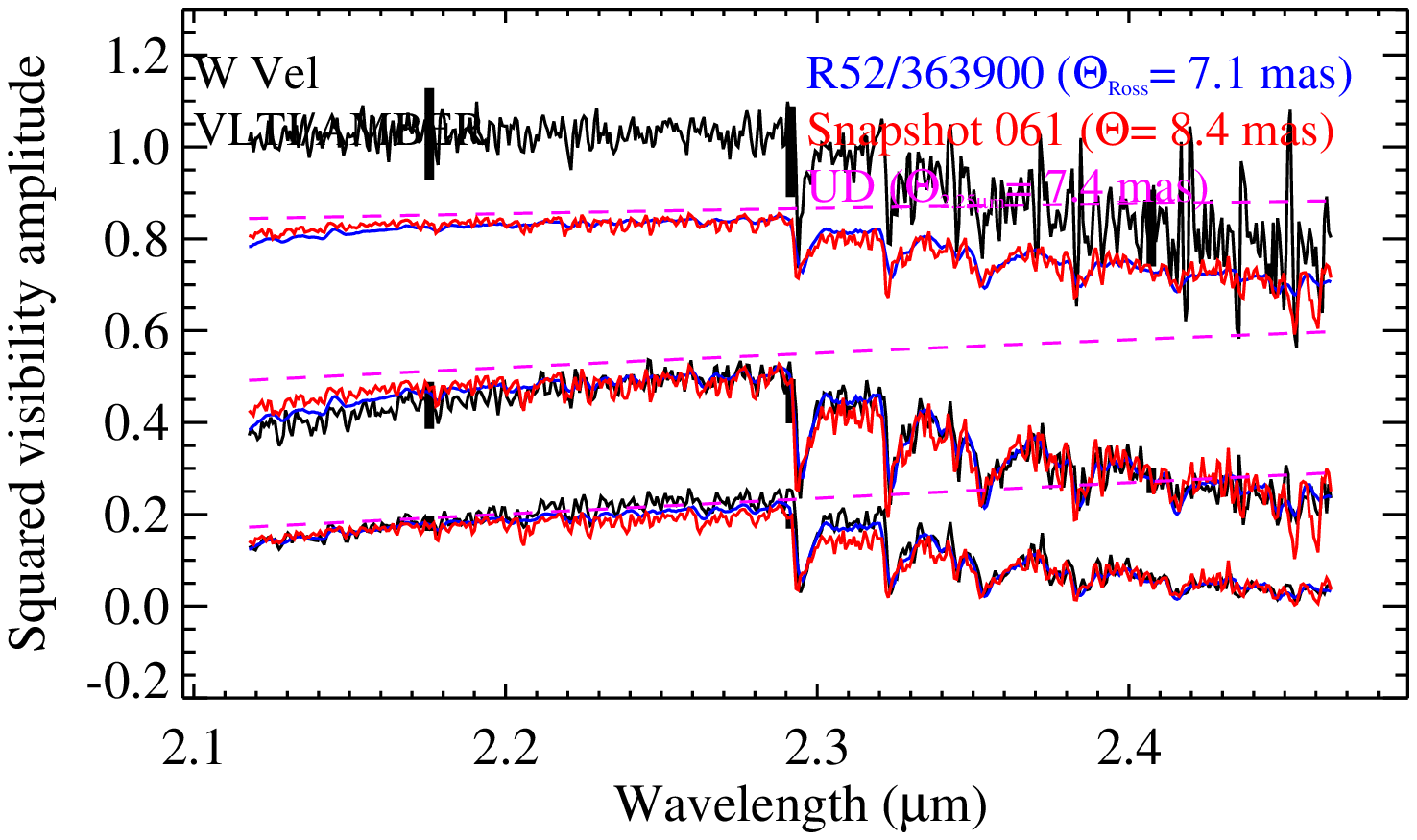}
  \includegraphics[width=0.32\textwidth]{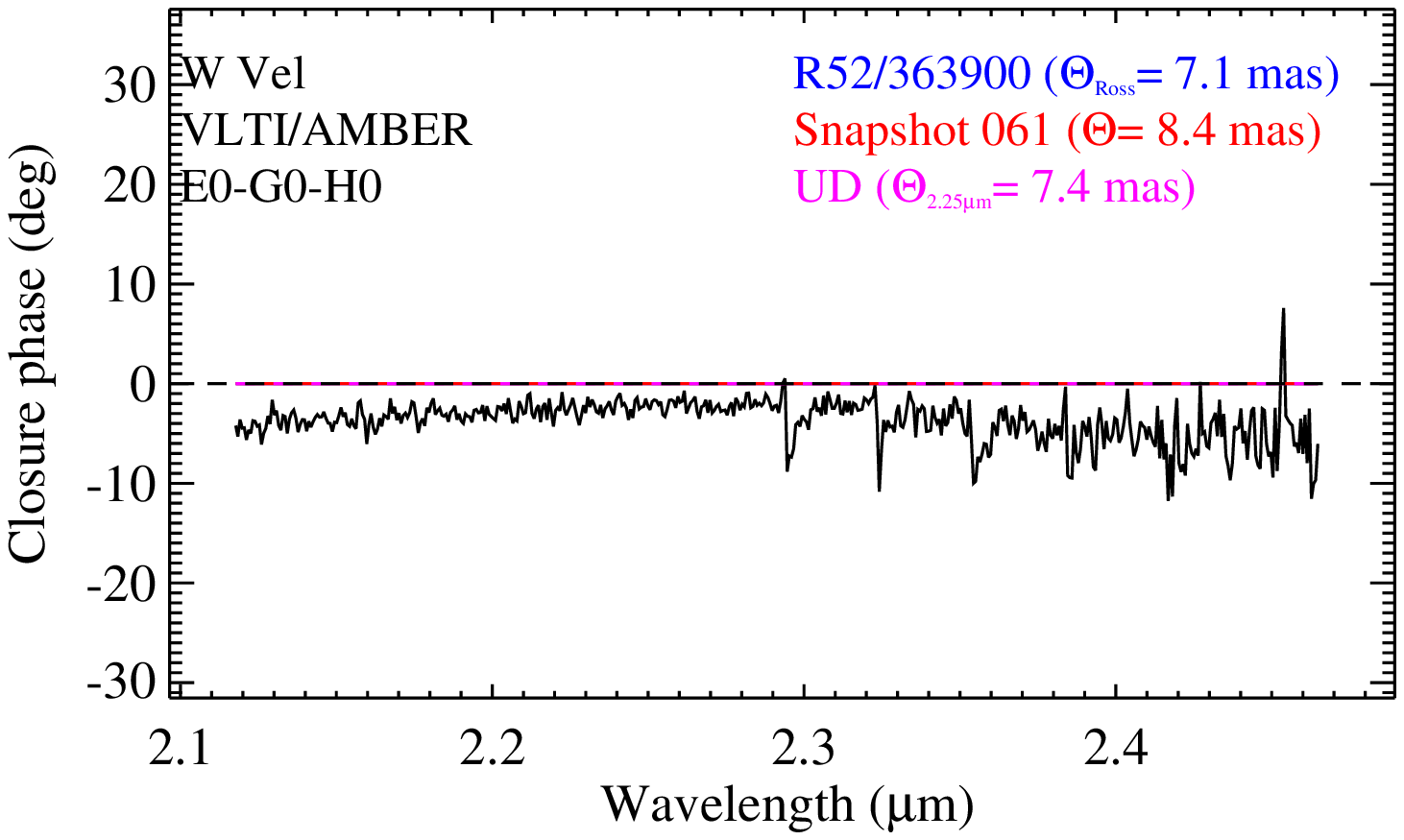}
\caption{As Fig.~\protect\ref{fig:rcnc}, but for W Vel, data sets no. 
7 \& 8 from Table~\protect\ref{tab:obs}.}
\label{fig:wvel}
\end{figure*}
}
\onlfig{
\begin{figure*}[p]
\centering
  \includegraphics[width=0.32\textwidth]{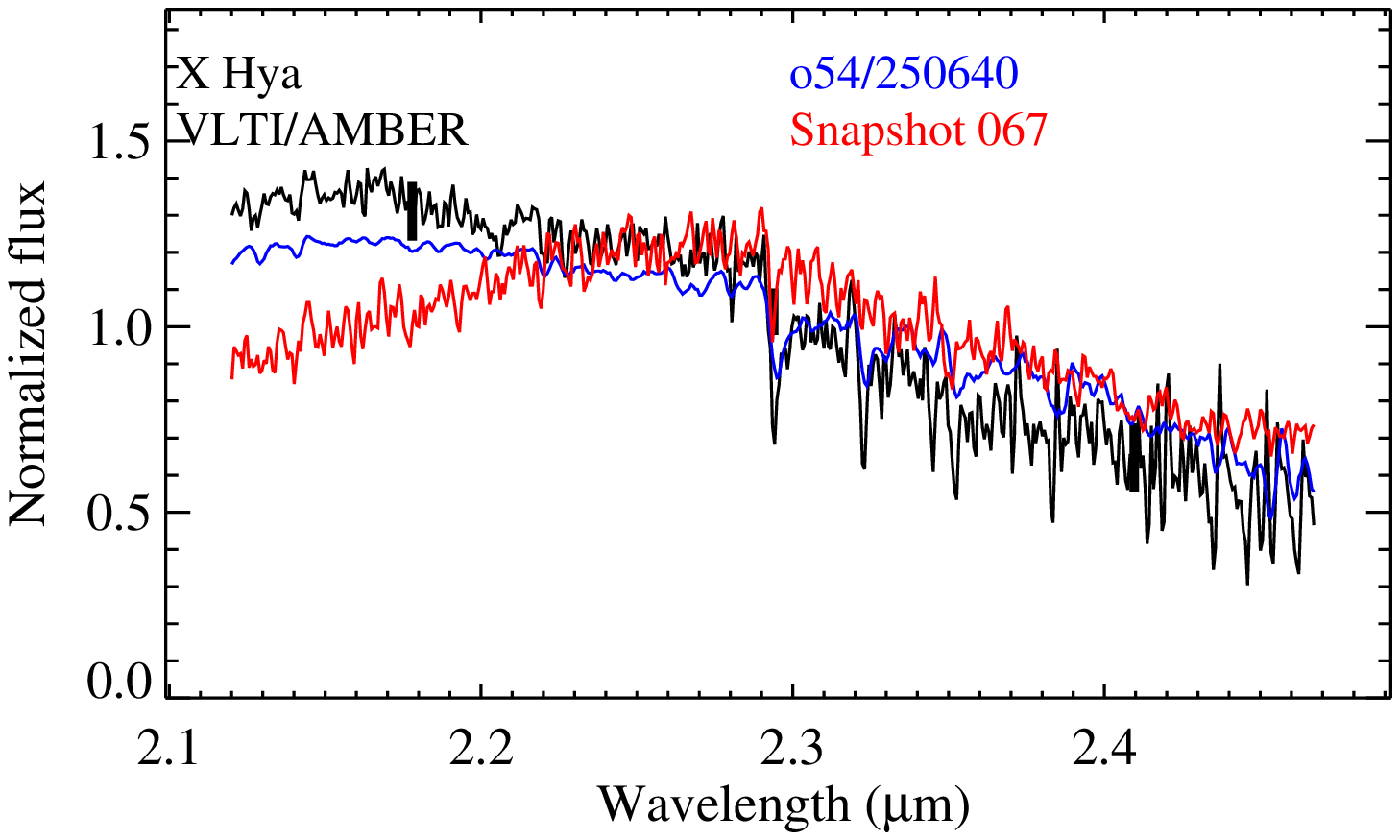}
  \includegraphics[width=0.32\textwidth]{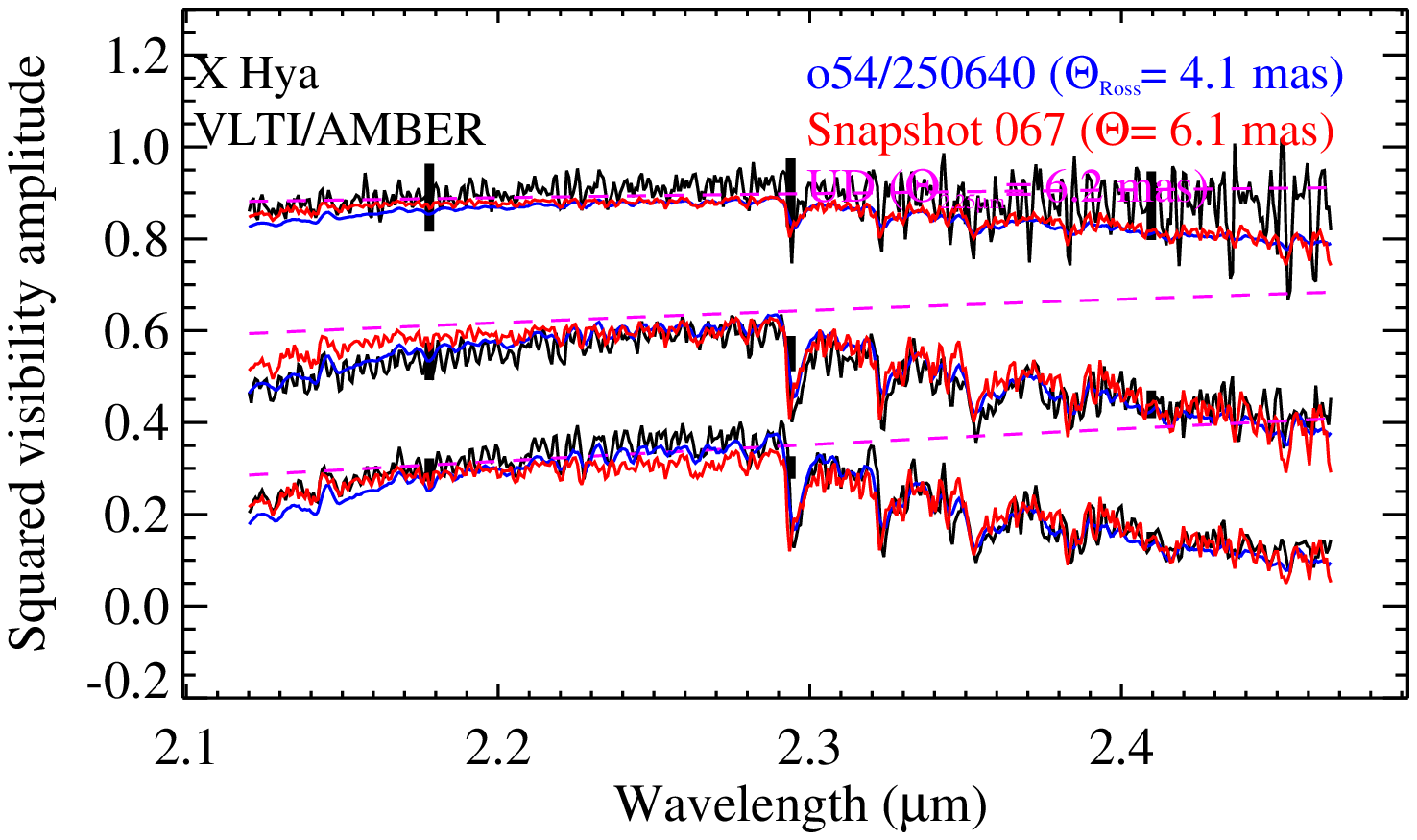}
  \includegraphics[width=0.32\textwidth]{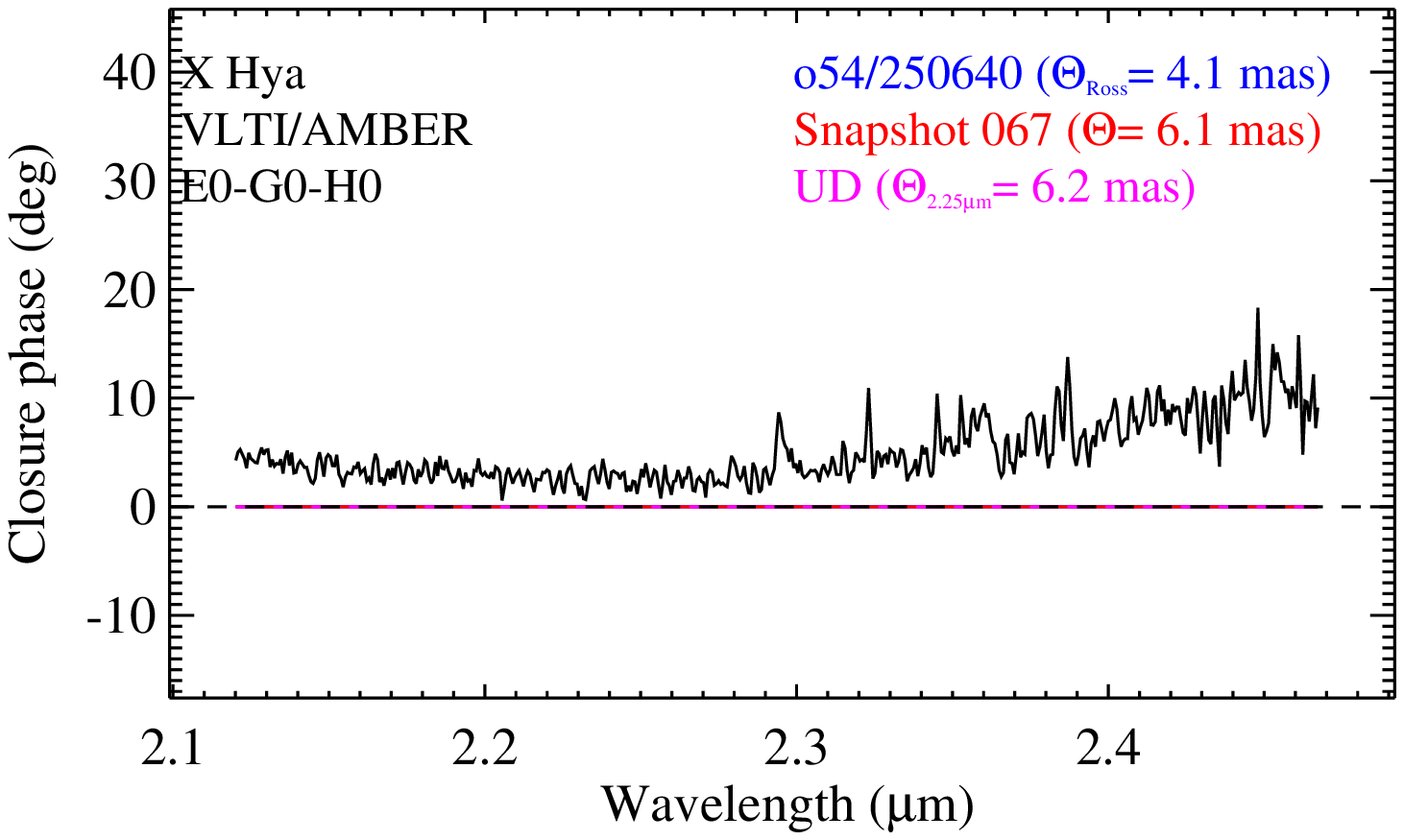}

  \includegraphics[width=0.32\textwidth]{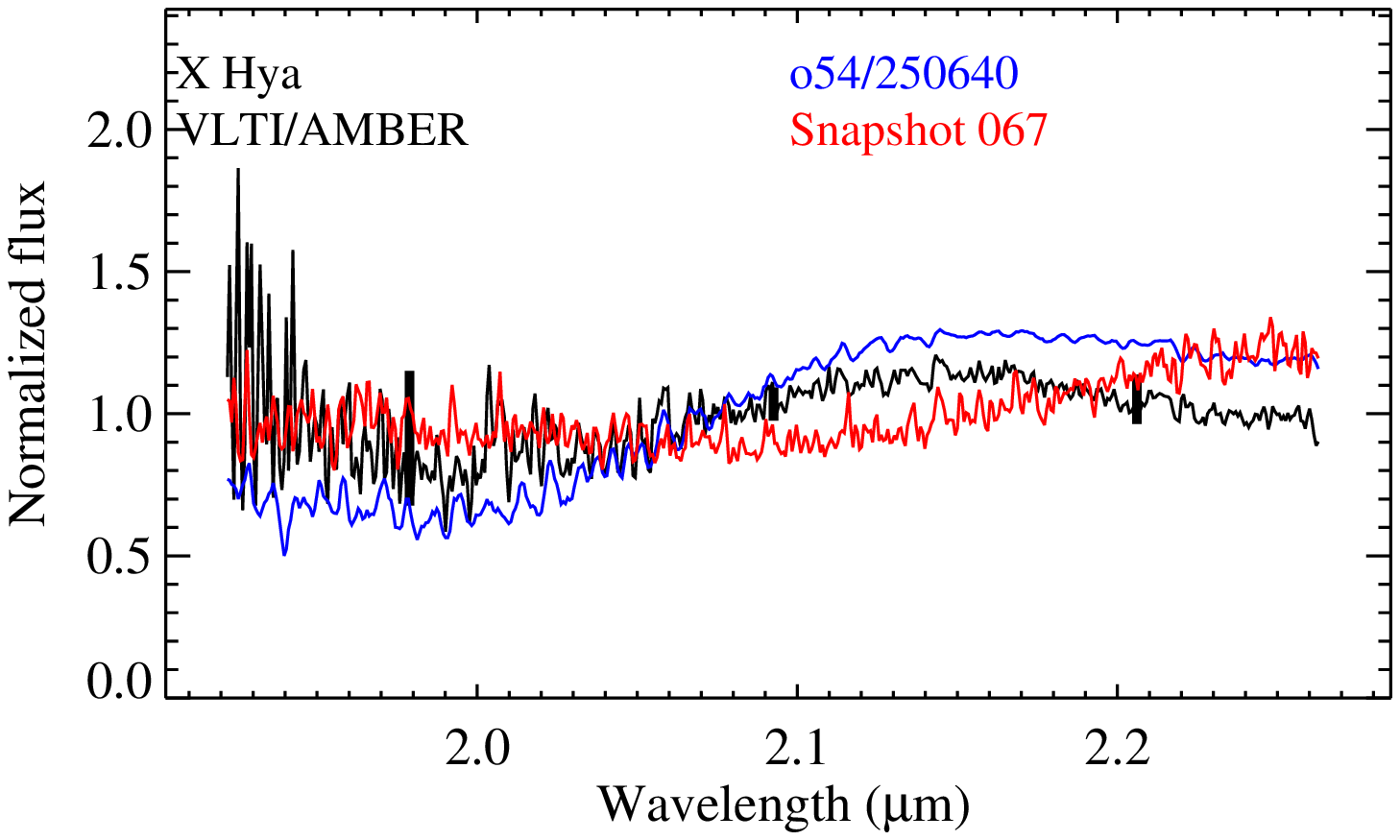}
  \includegraphics[width=0.32\textwidth]{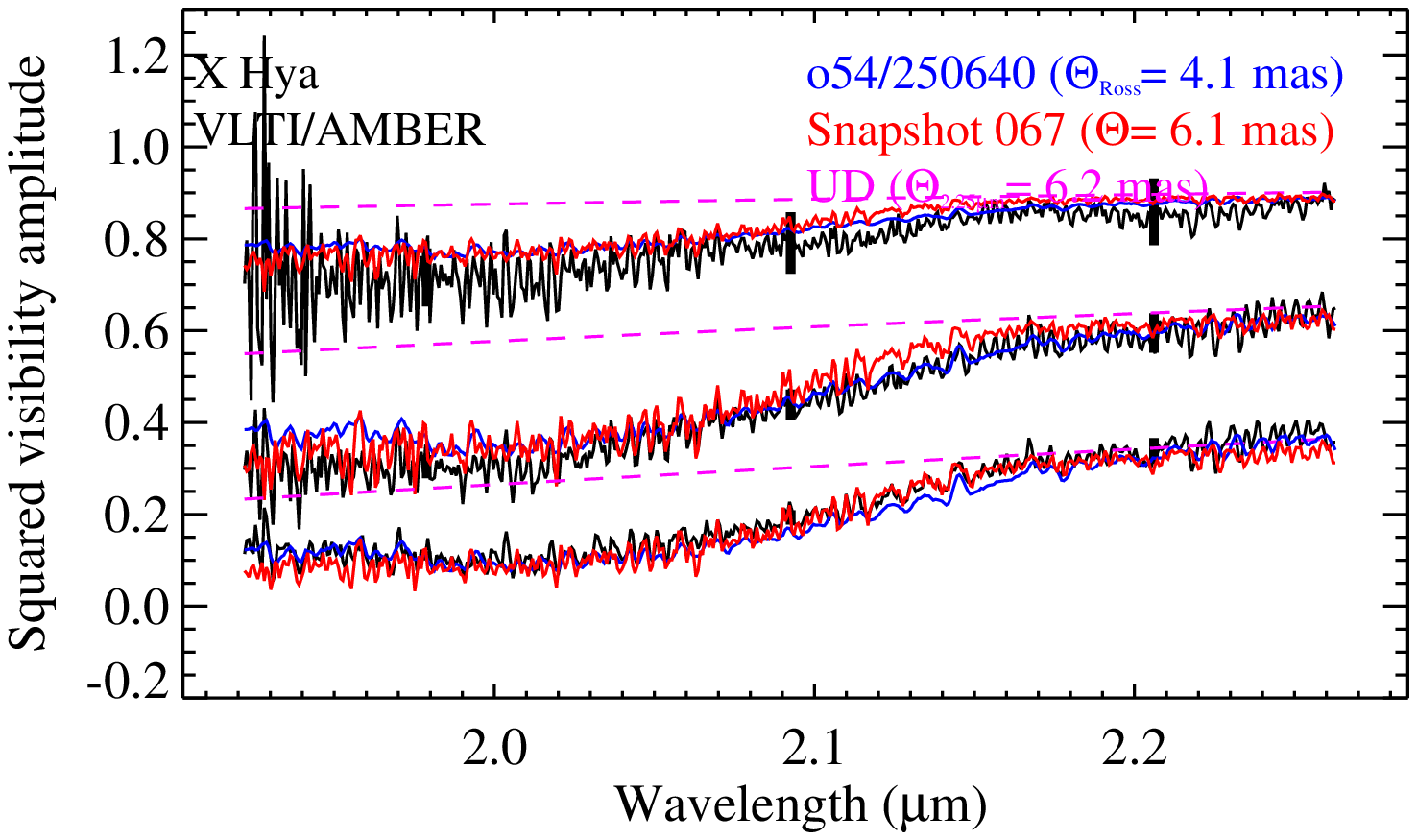}
  \includegraphics[width=0.32\textwidth]{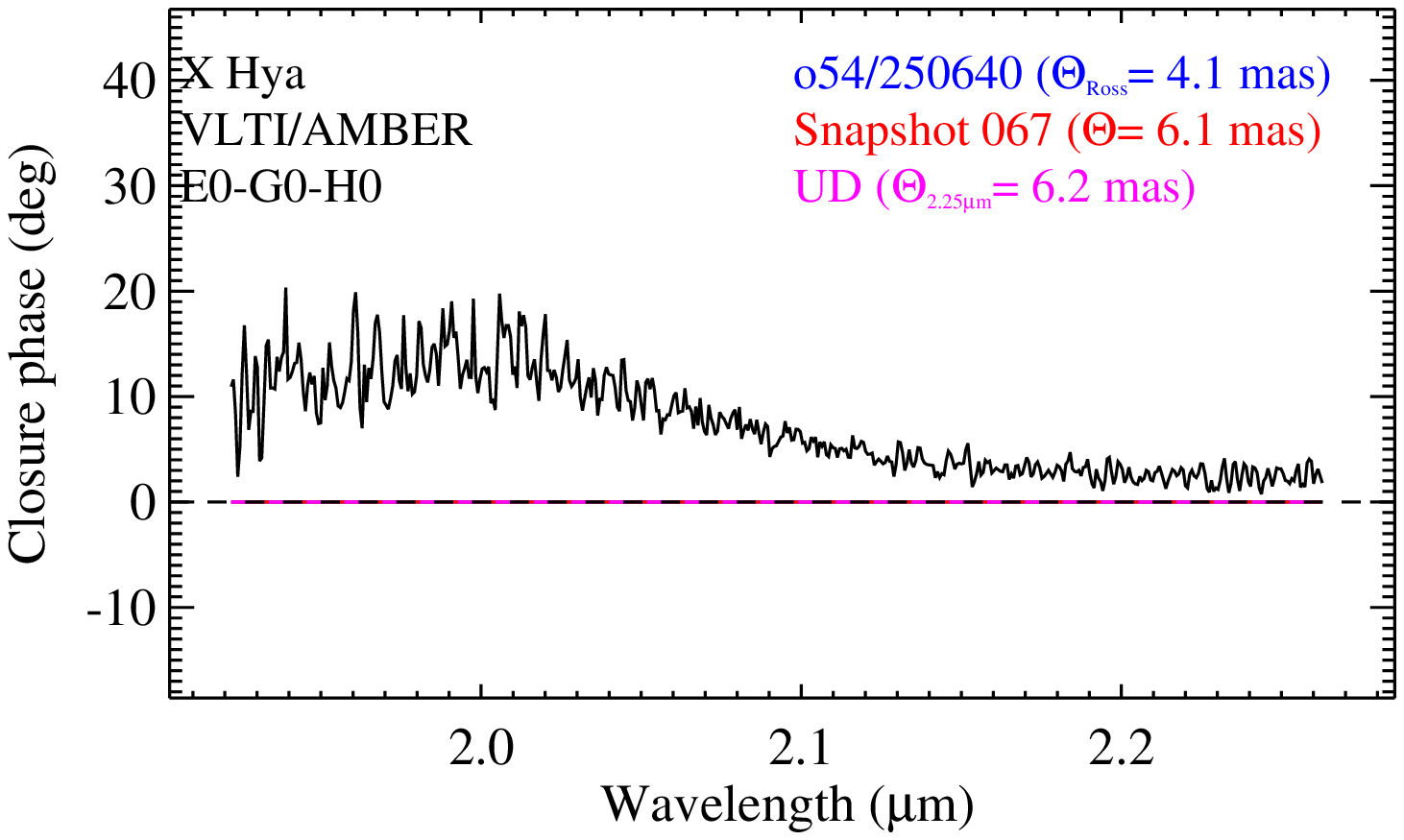}

  \includegraphics[width=0.32\textwidth]{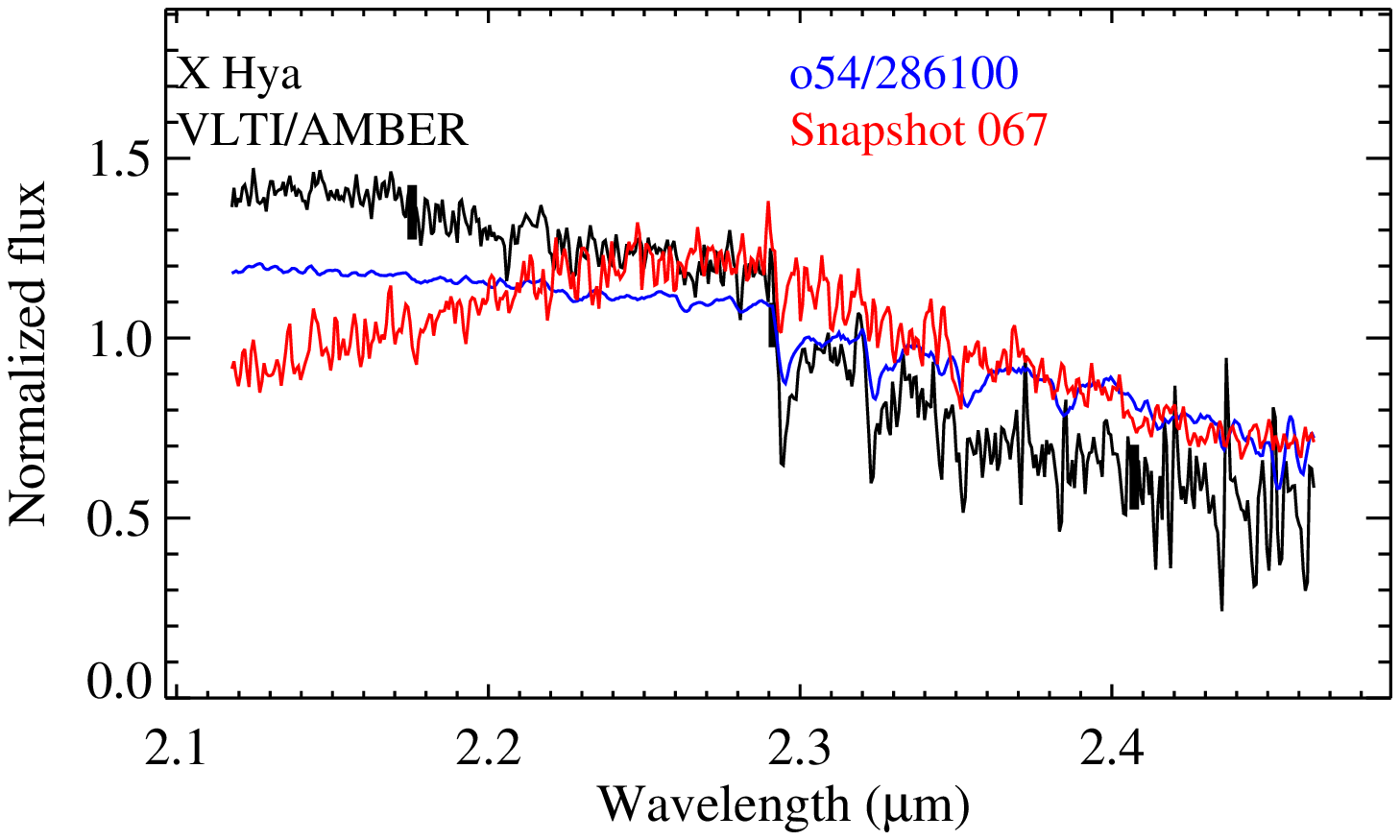}
  \includegraphics[width=0.32\textwidth]{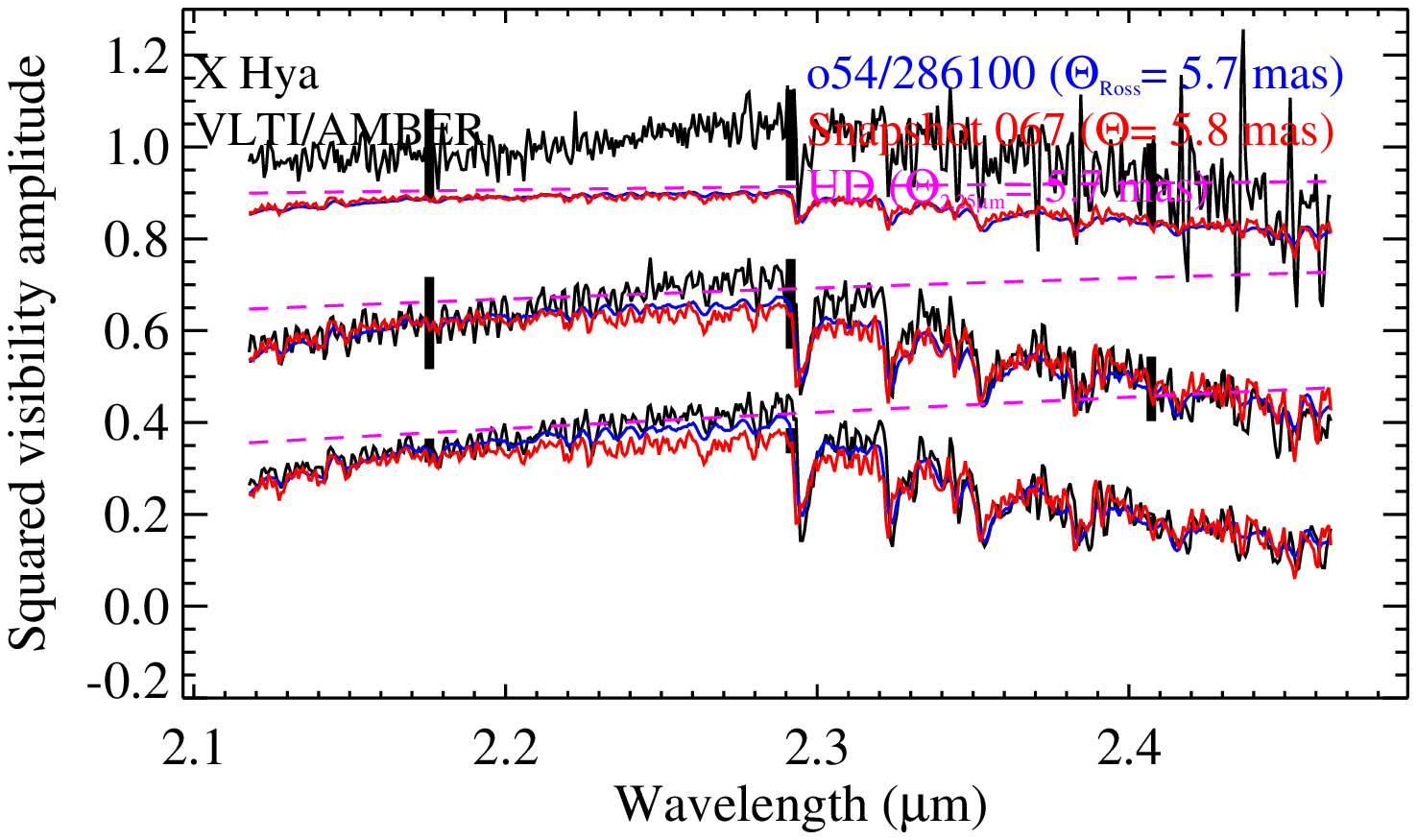}
  \includegraphics[width=0.32\textwidth]{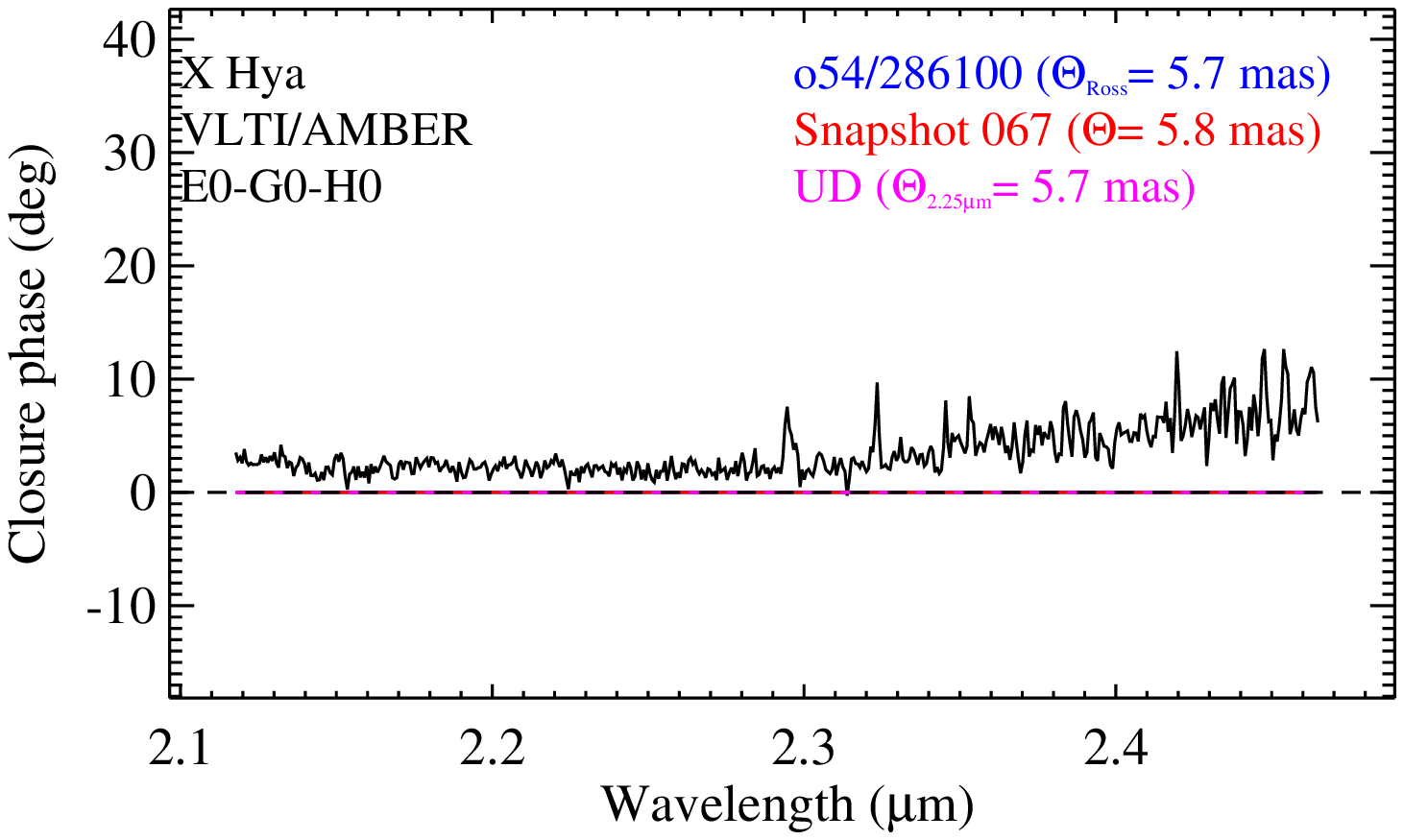}
\caption{As Fig.~\protect\ref{fig:rcnc}, but for X Hya, data sets 9--11
from Table~\protect\ref{tab:obs}.}
\label{fig:xhya}
\end{figure*}
}
\onlfig{
\begin{figure*}[p]
\centering
  \includegraphics[width=0.32\textwidth]{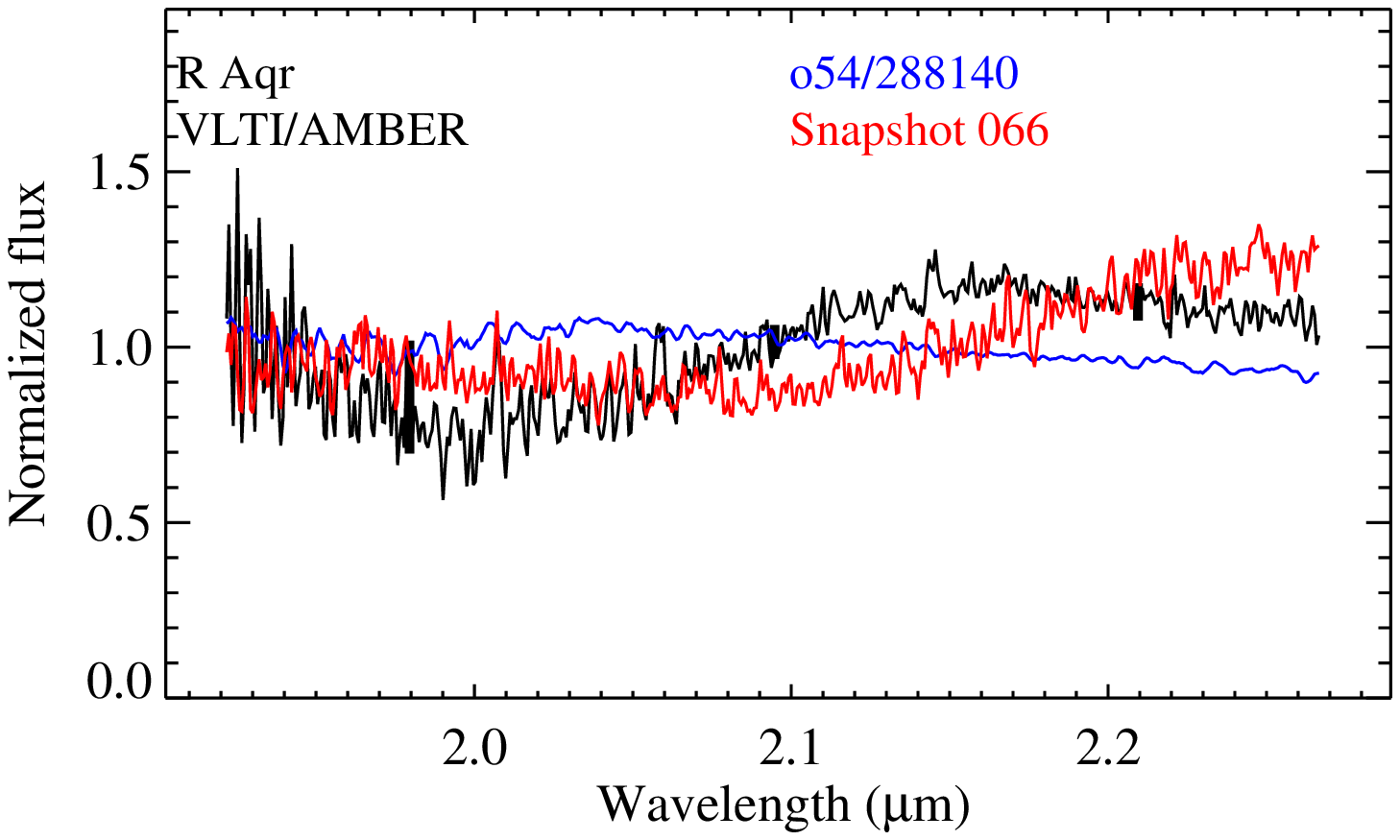}
  \includegraphics[width=0.32\textwidth]{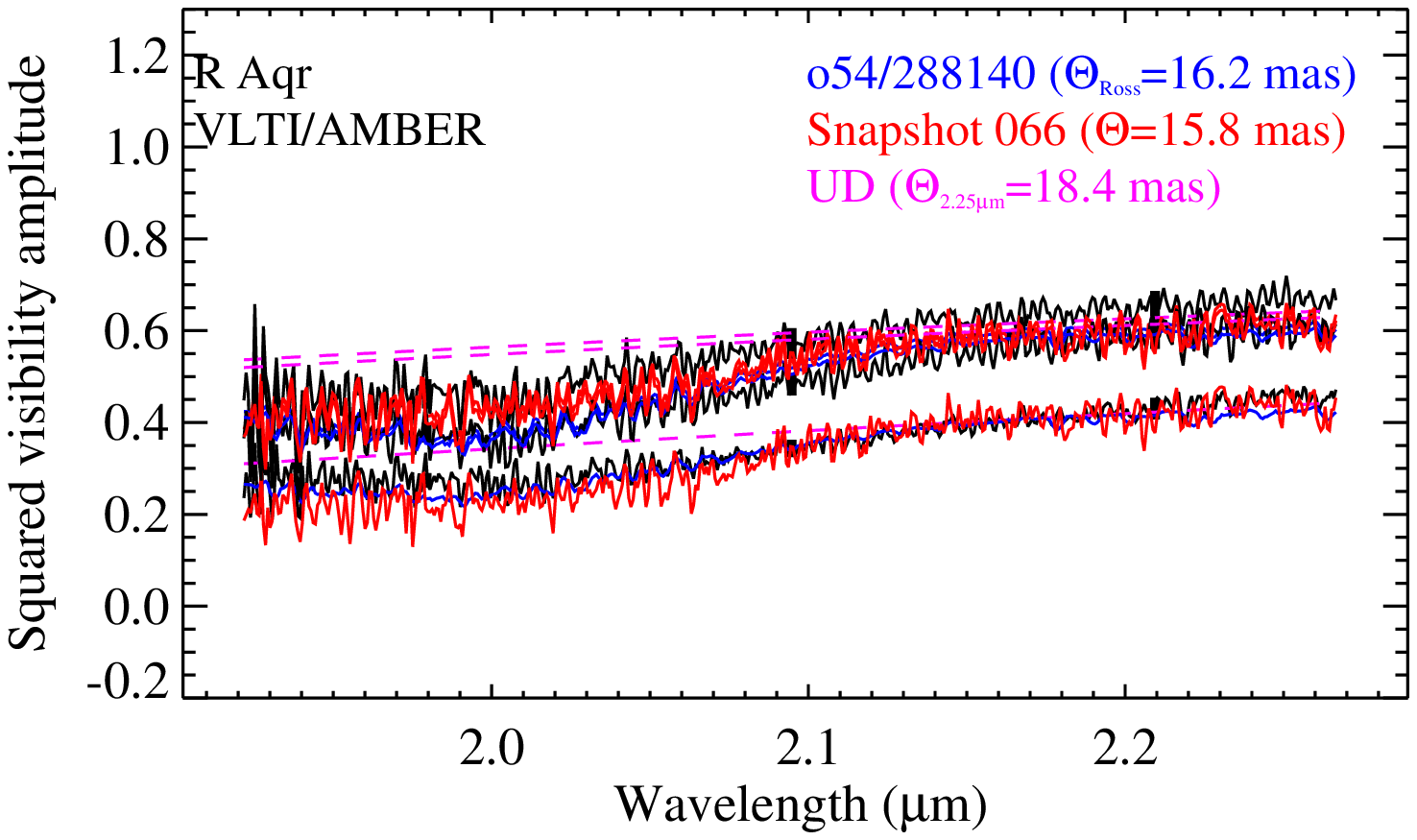}
  \includegraphics[width=0.32\textwidth]{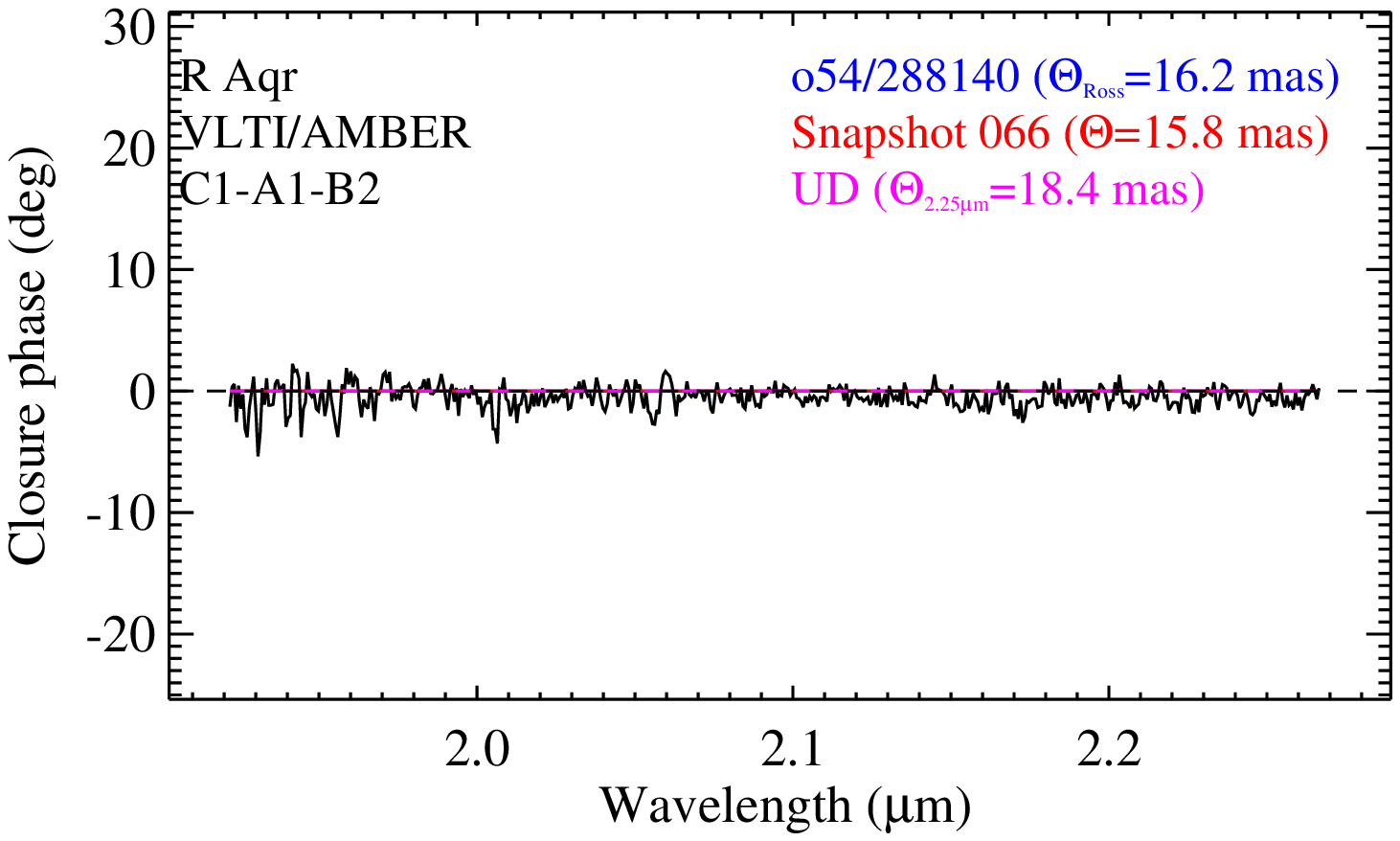}

  \includegraphics[width=0.32\textwidth]{Oct2012raqr-MR21-1.spec.ps}
  \includegraphics[width=0.32\textwidth]{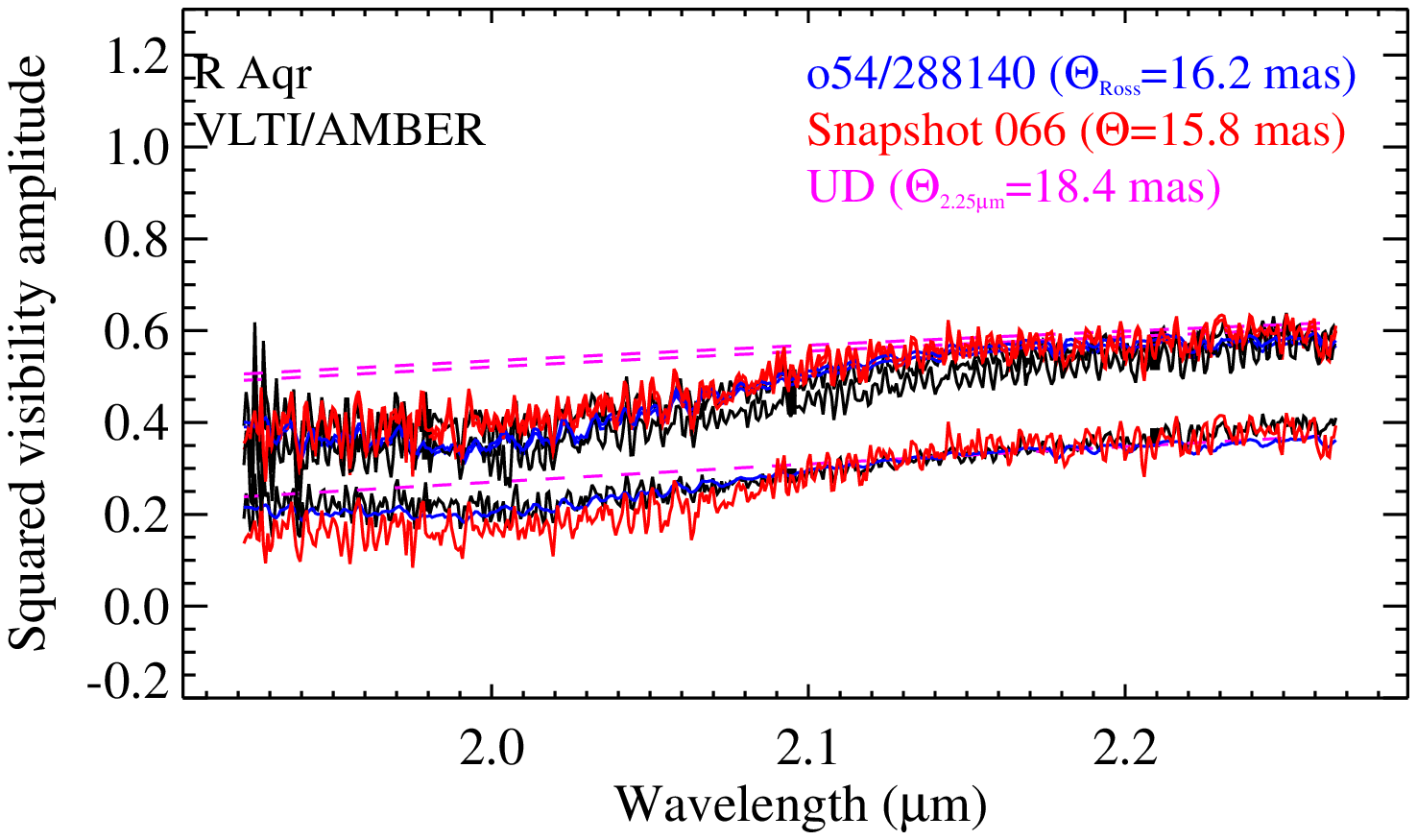}
  \includegraphics[width=0.32\textwidth]{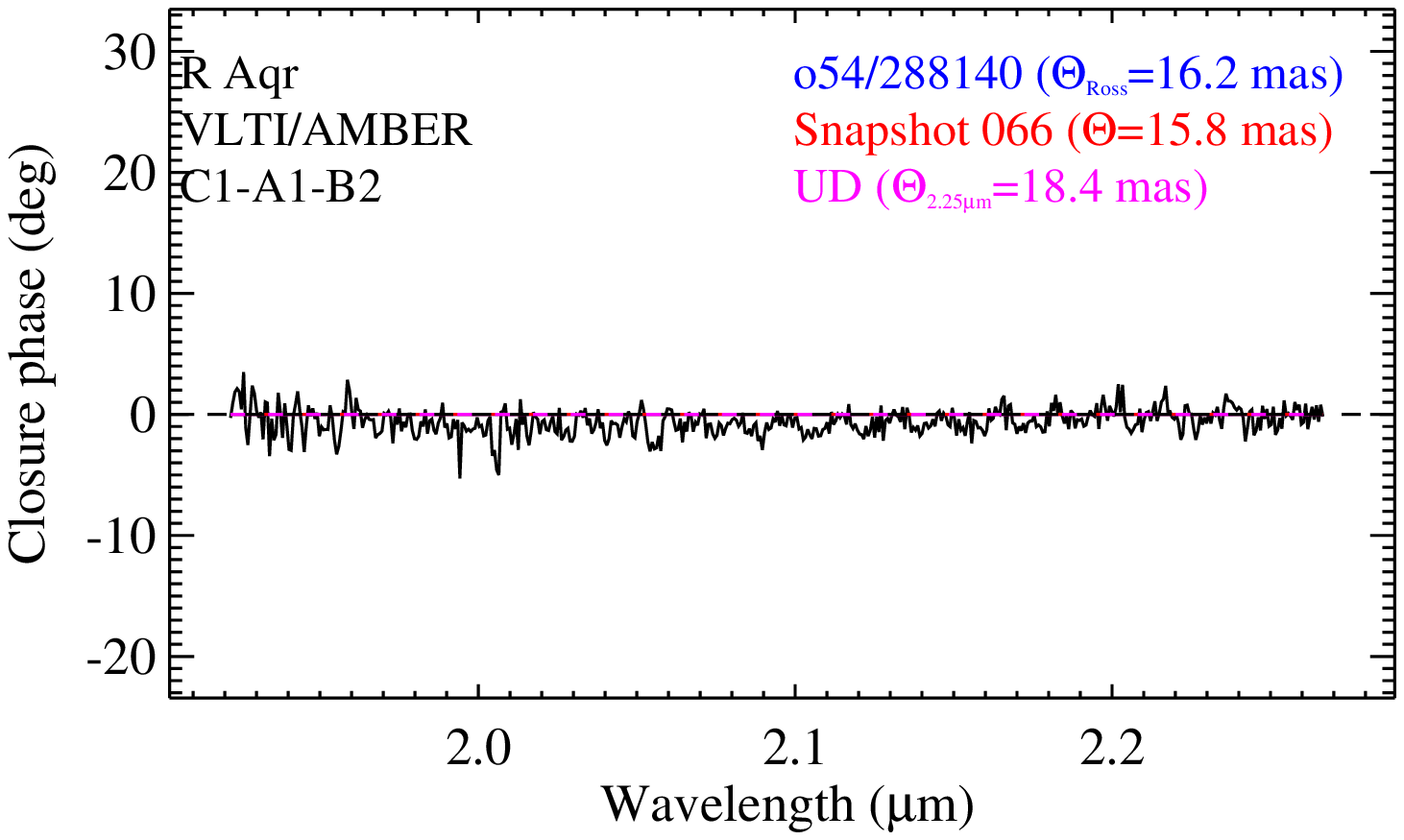}
\caption{As Fig.~\protect\ref{fig:rcnc}, but for R Aqr, data set no. 12
\& 13 from Table~\protect\ref{tab:obs}.}
\label{fig:raqr}
\end{figure*}
}
\onlfig{
\begin{figure*}[p]
\centering
  \includegraphics[width=0.32\textwidth]{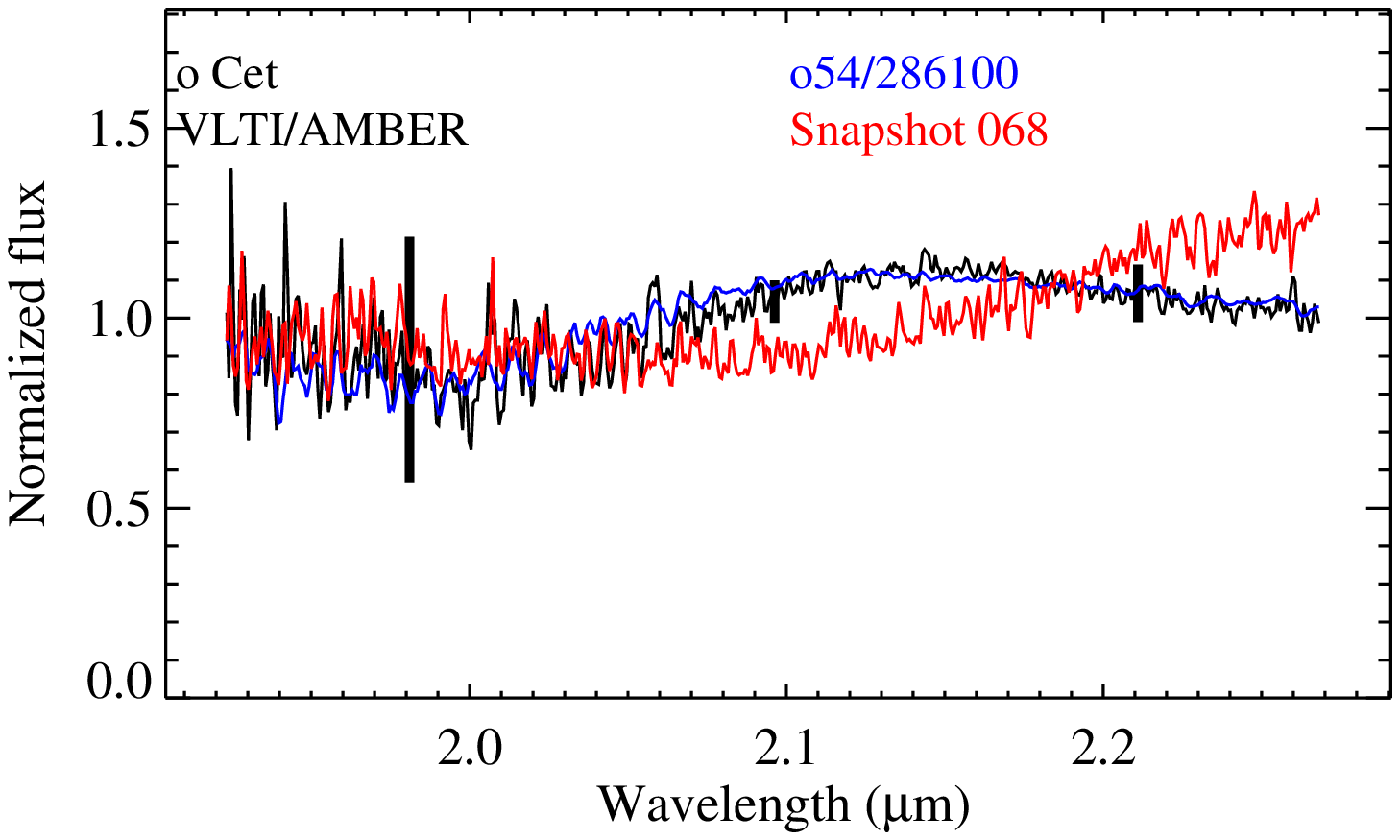}
  \includegraphics[width=0.32\textwidth]{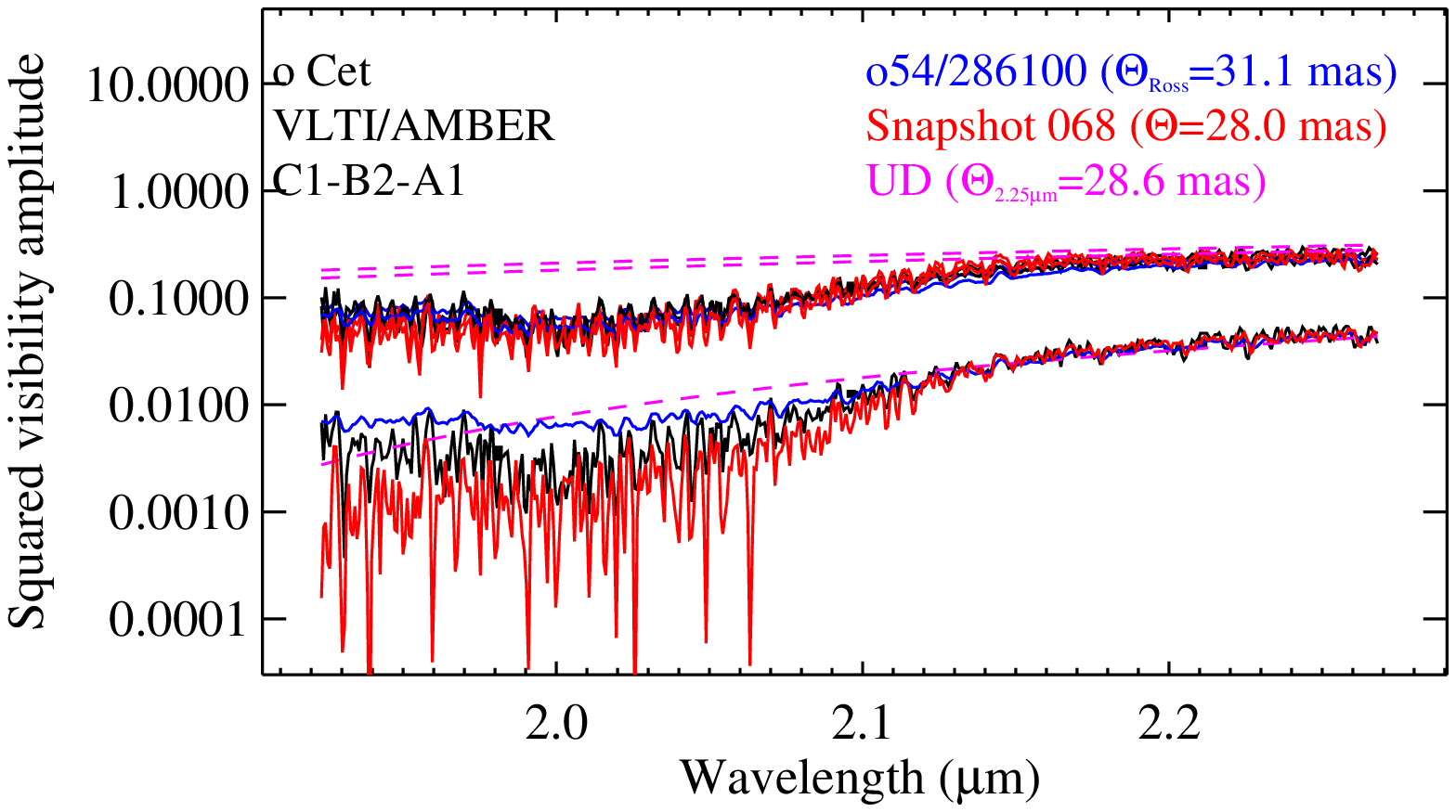}
  \includegraphics[width=0.32\textwidth]{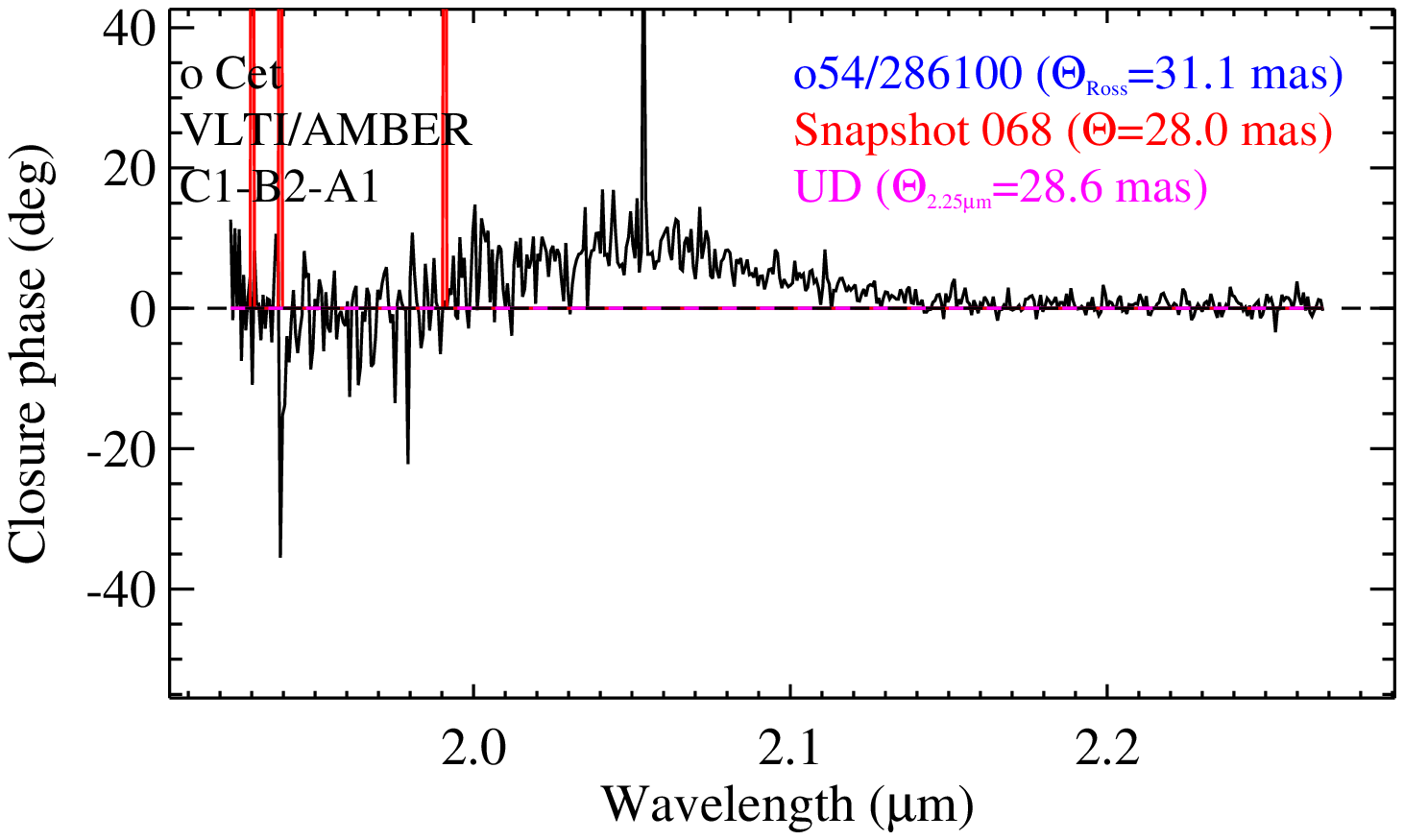}

  \includegraphics[width=0.32\textwidth]{Oct2012omicet-MR21-1.spec.ps}
  \includegraphics[width=0.32\textwidth]{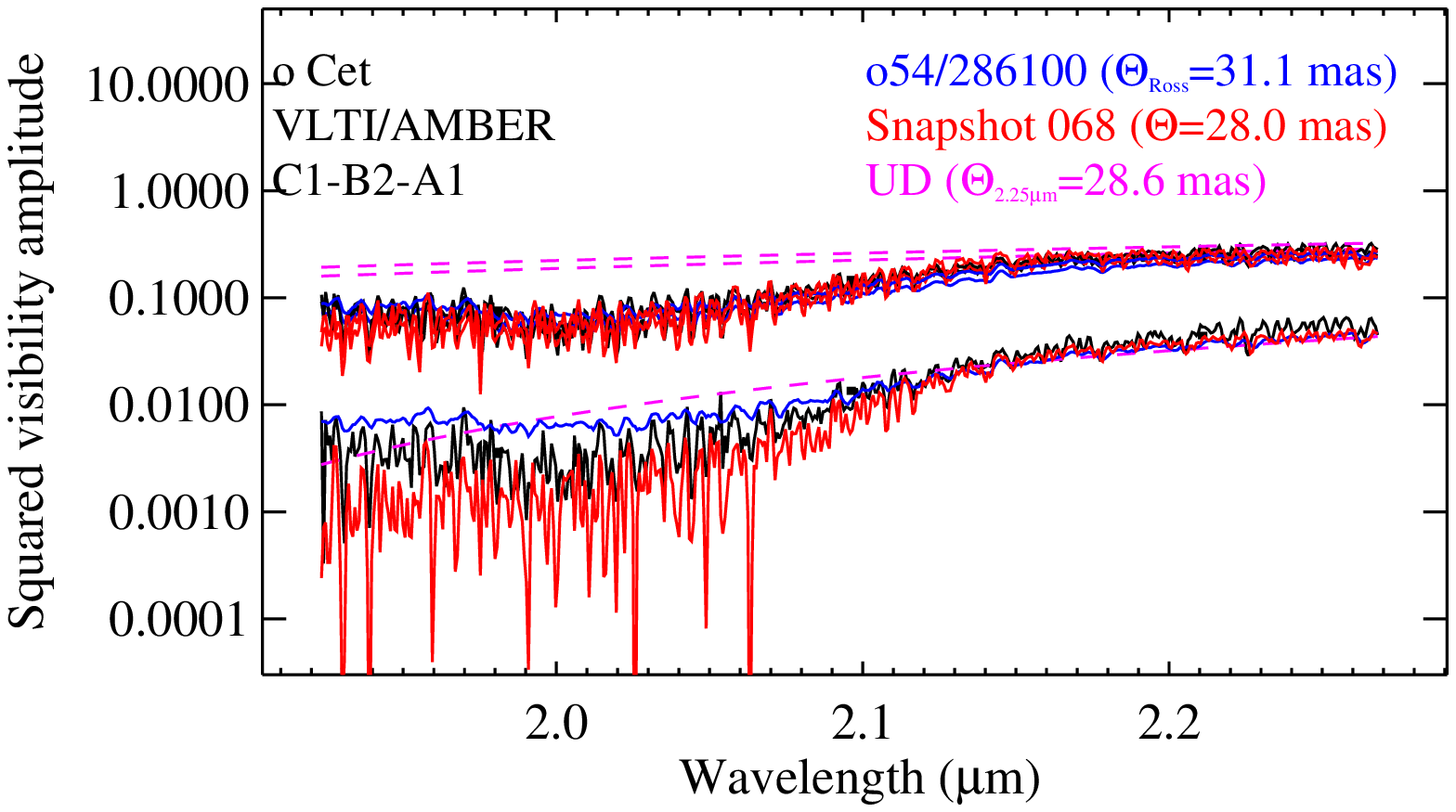}
  \includegraphics[width=0.32\textwidth]{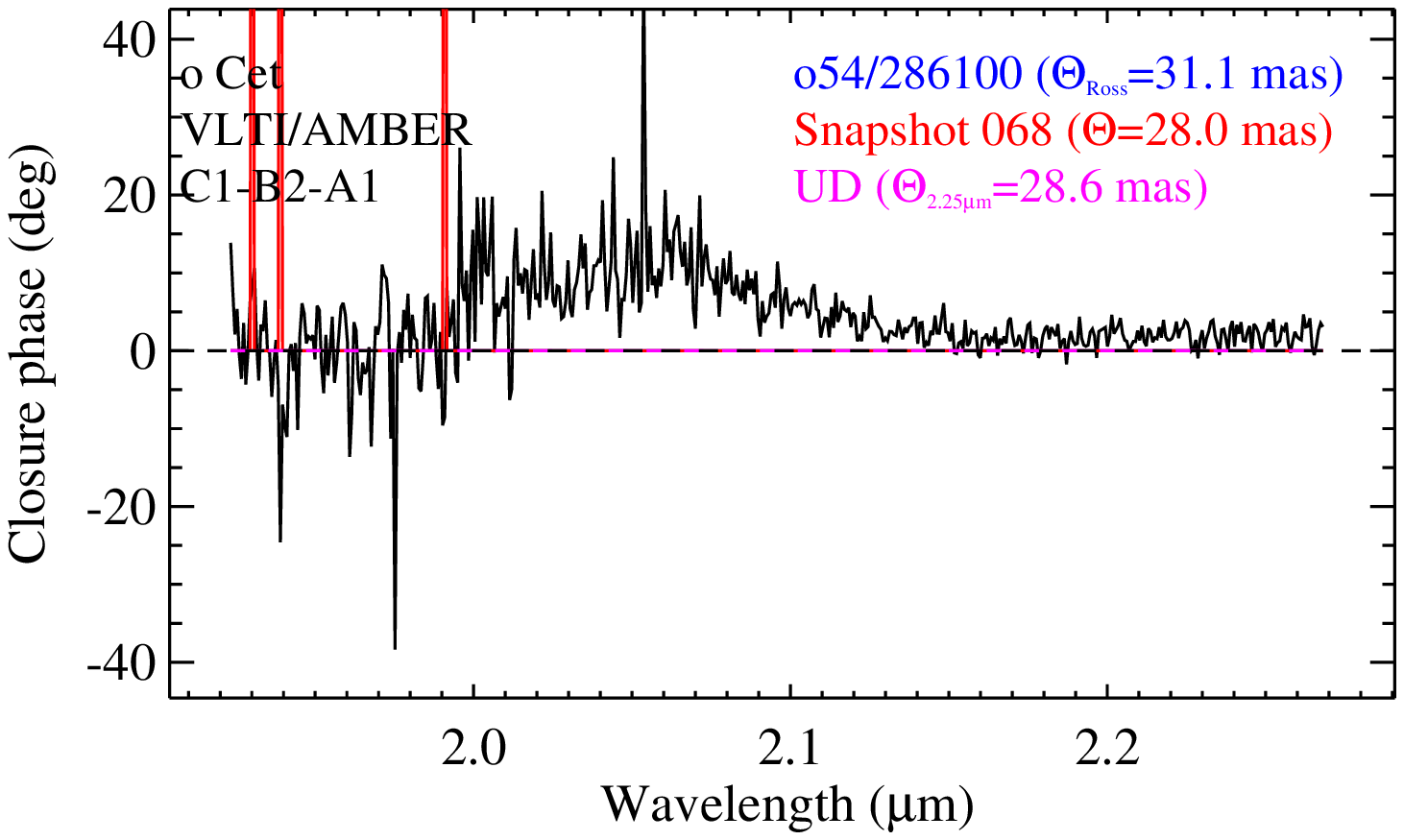}

  \includegraphics[width=0.32\textwidth]{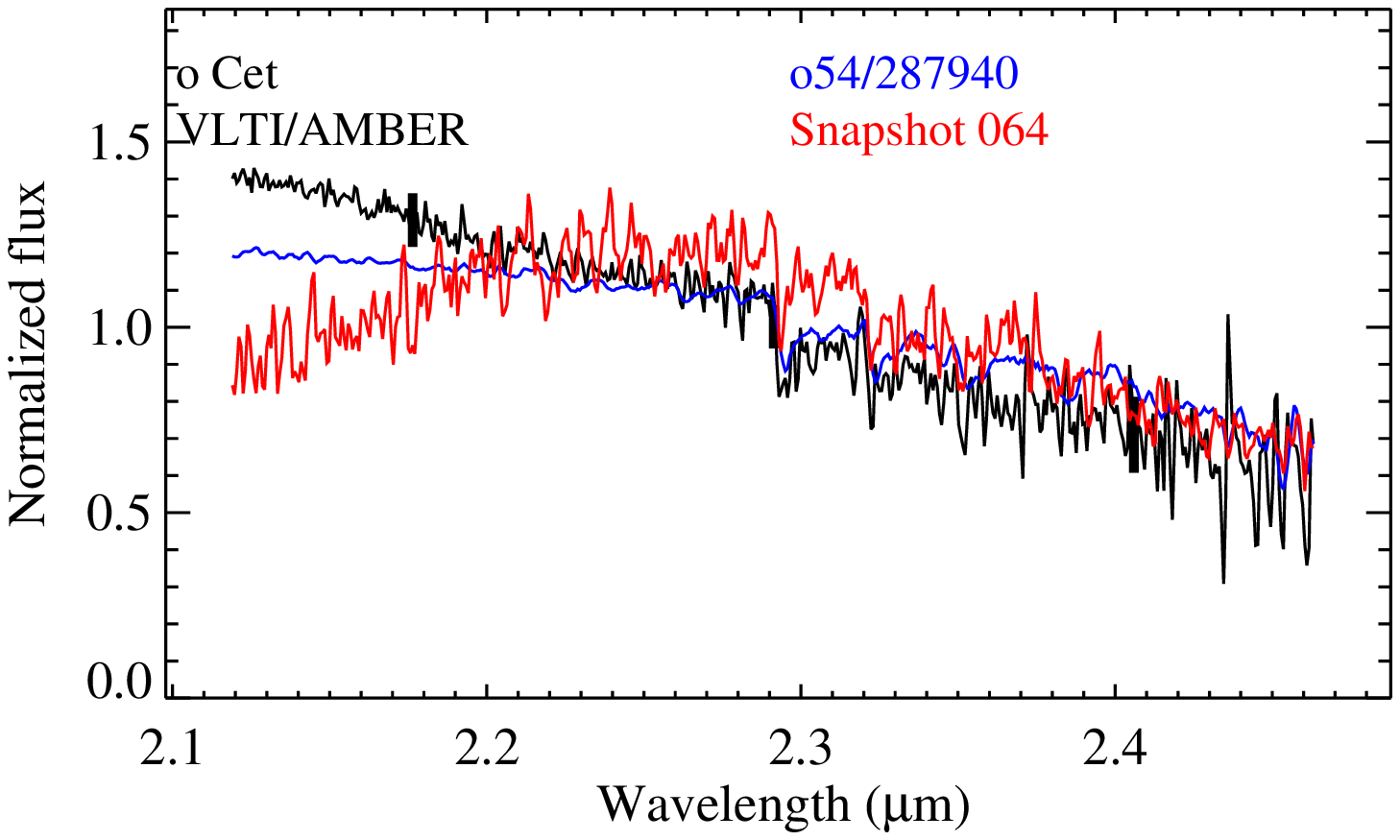}
  \includegraphics[width=0.32\textwidth]{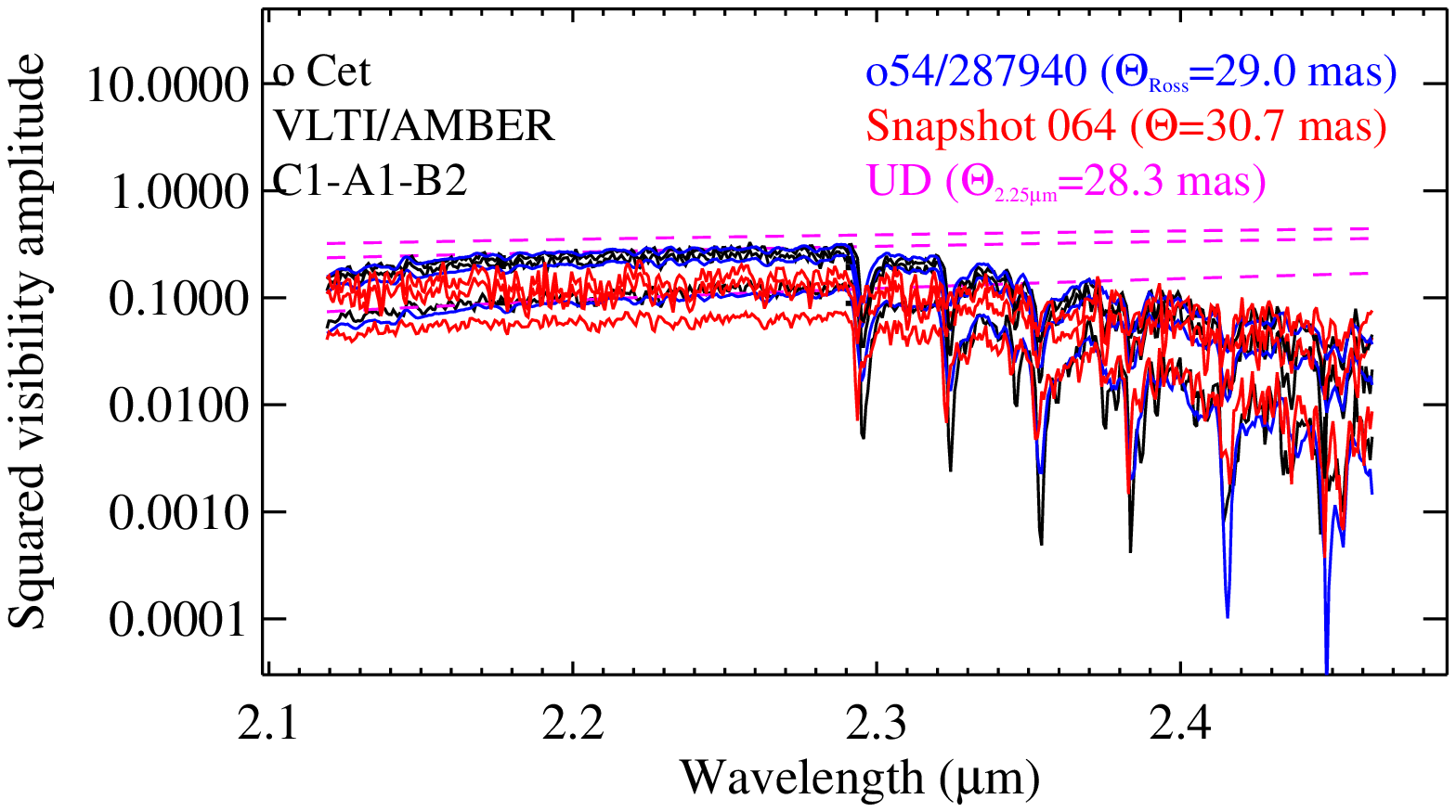}
  \includegraphics[width=0.32\textwidth]{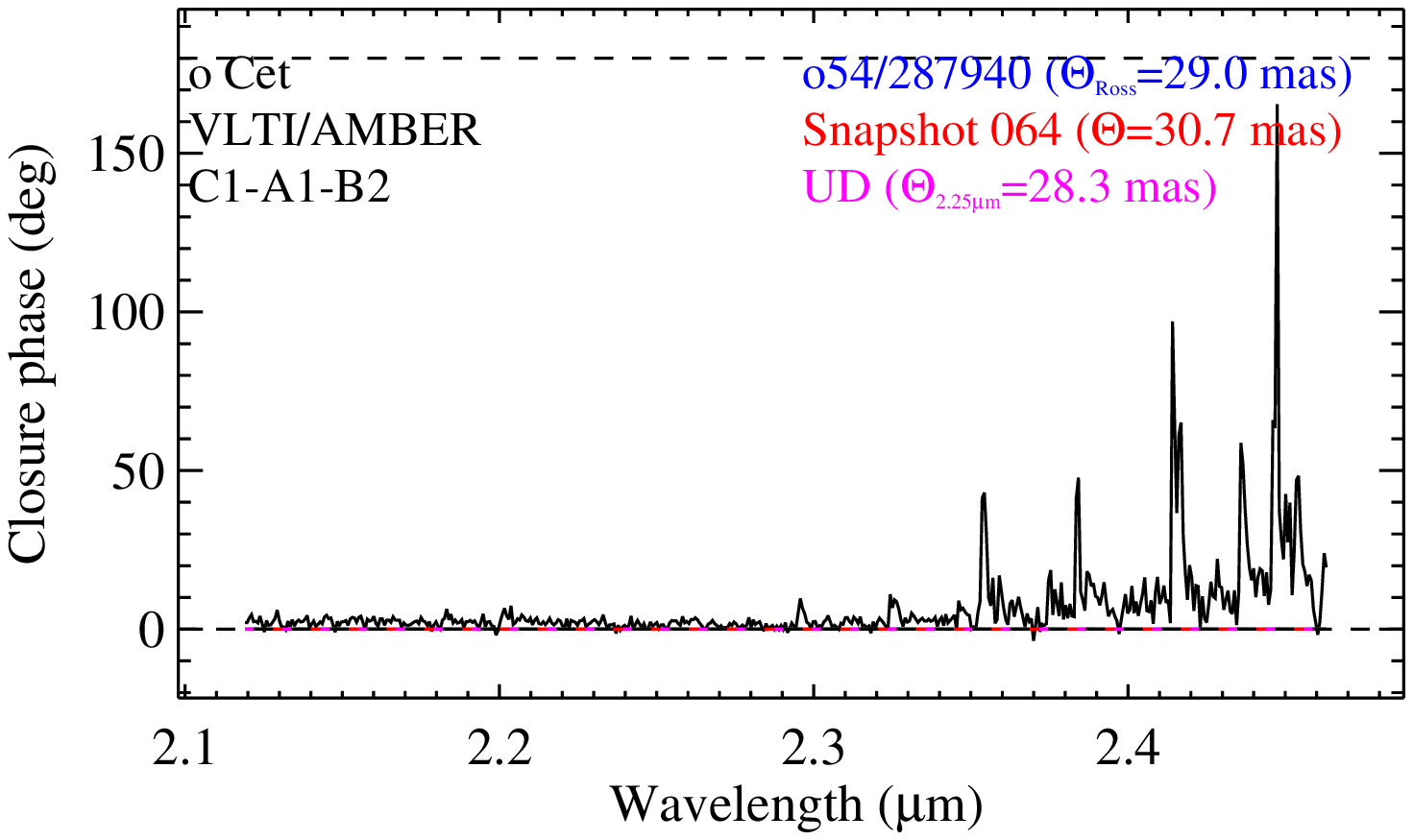}

  \includegraphics[width=0.32\textwidth]{Sep2013omicet-MR21-1.spec.ps}
  \includegraphics[width=0.32\textwidth]{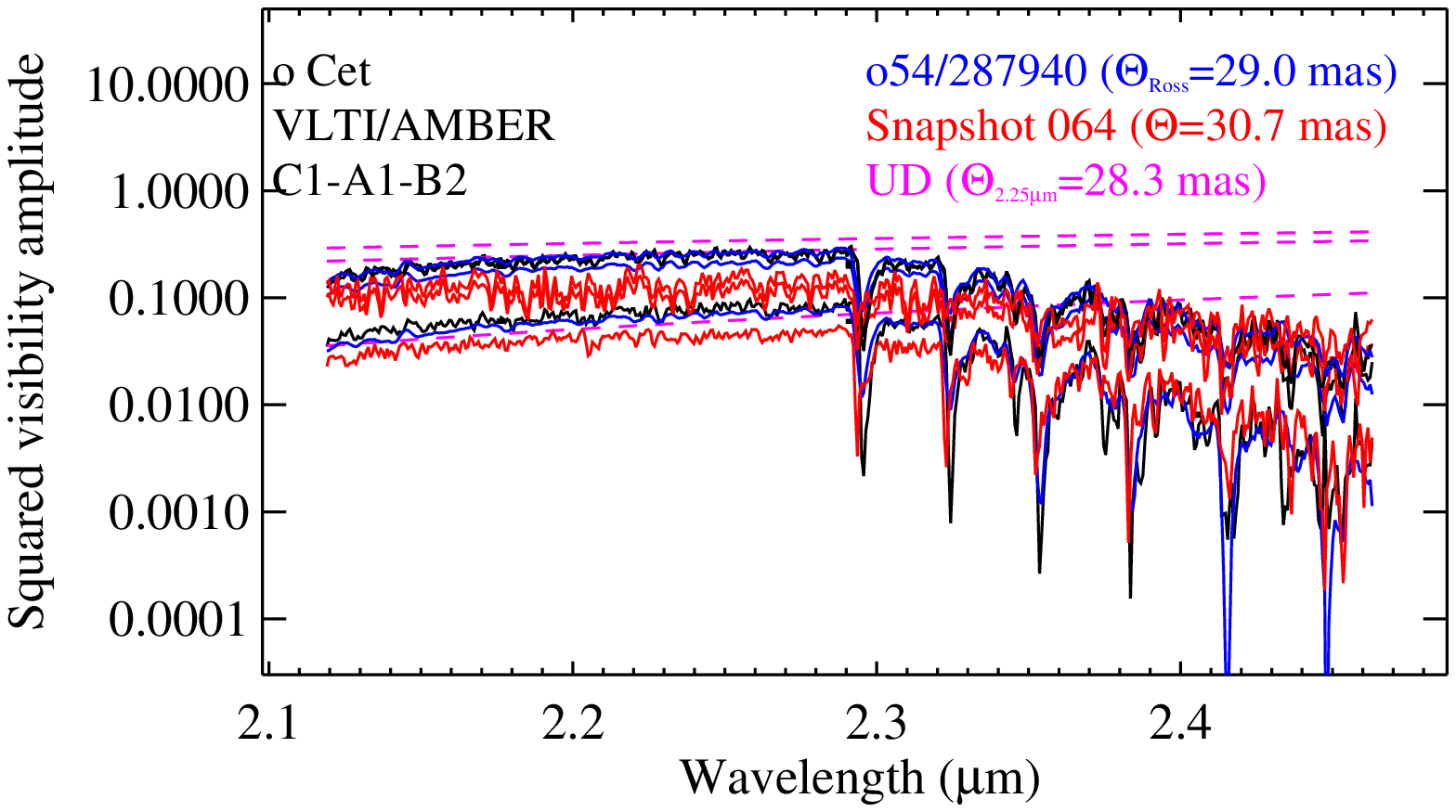}
  \includegraphics[width=0.32\textwidth]{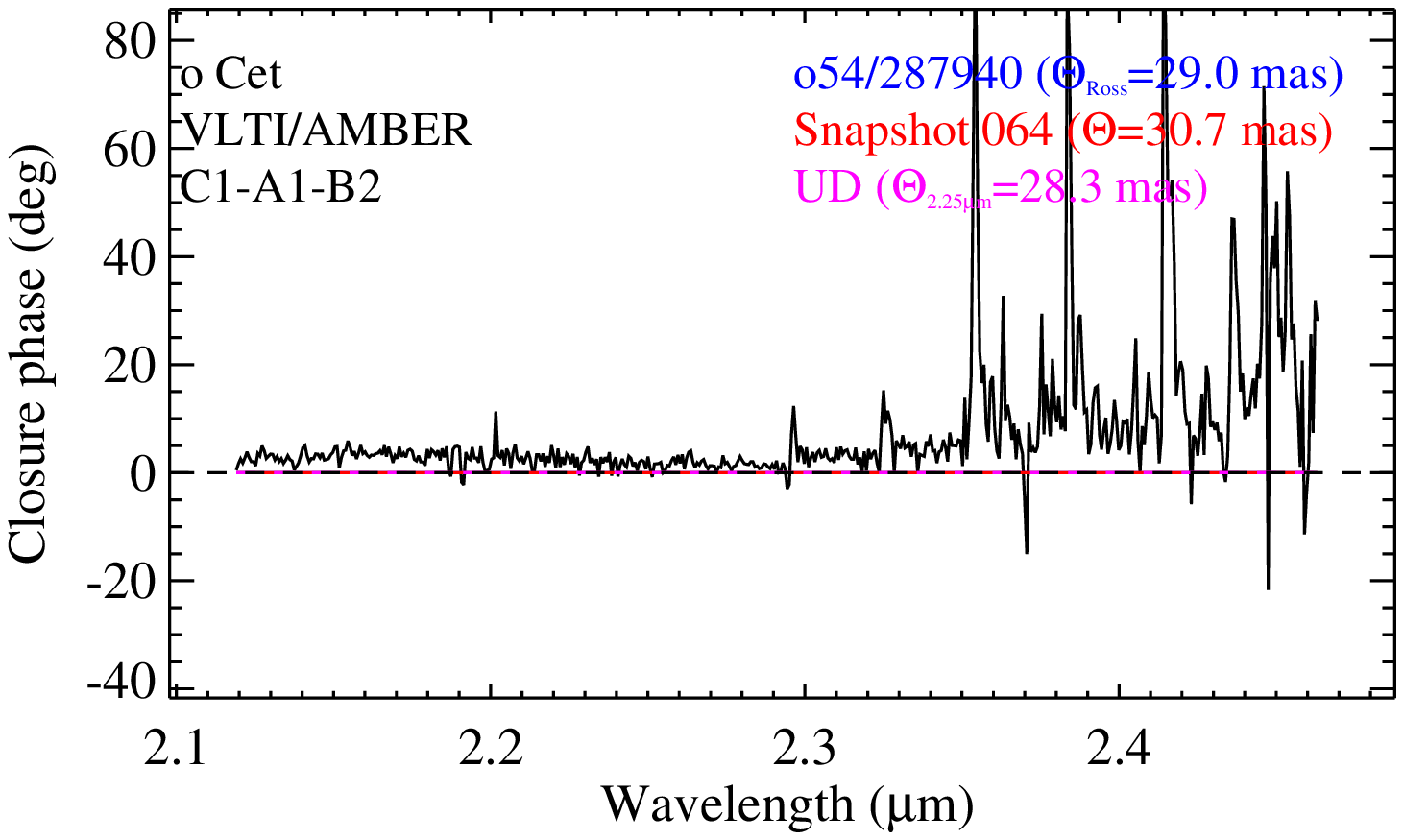}

\caption{As Fig.~\protect\ref{fig:rcnc}, but for o Cet, data sets 14--17
from Table~\protect\ref{tab:obs}.}
\label{fig:omicet}
\end{figure*}
}
\onlfig{
\begin{figure*}[p]
\centering
  \includegraphics[width=0.32\textwidth]{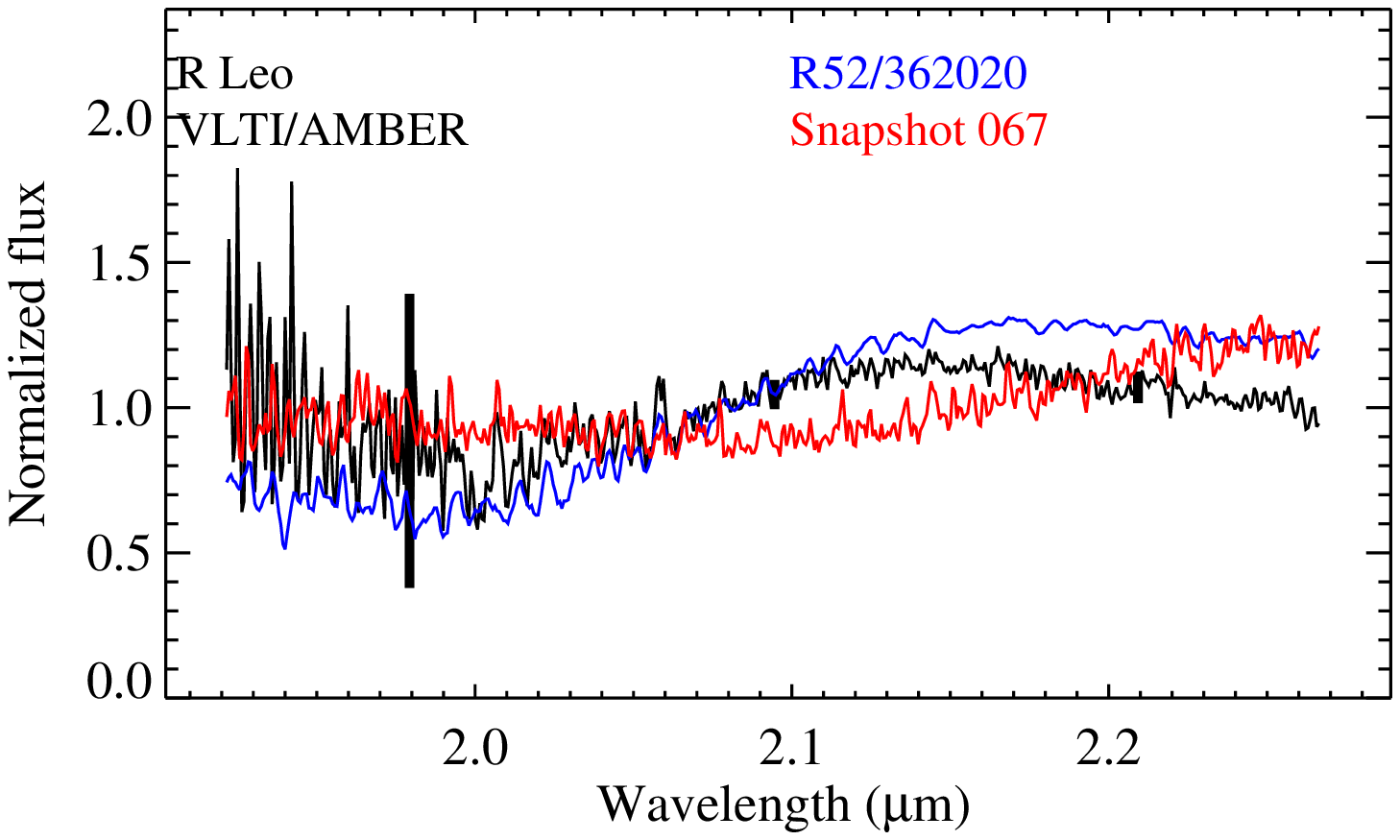}
  \includegraphics[width=0.32\textwidth]{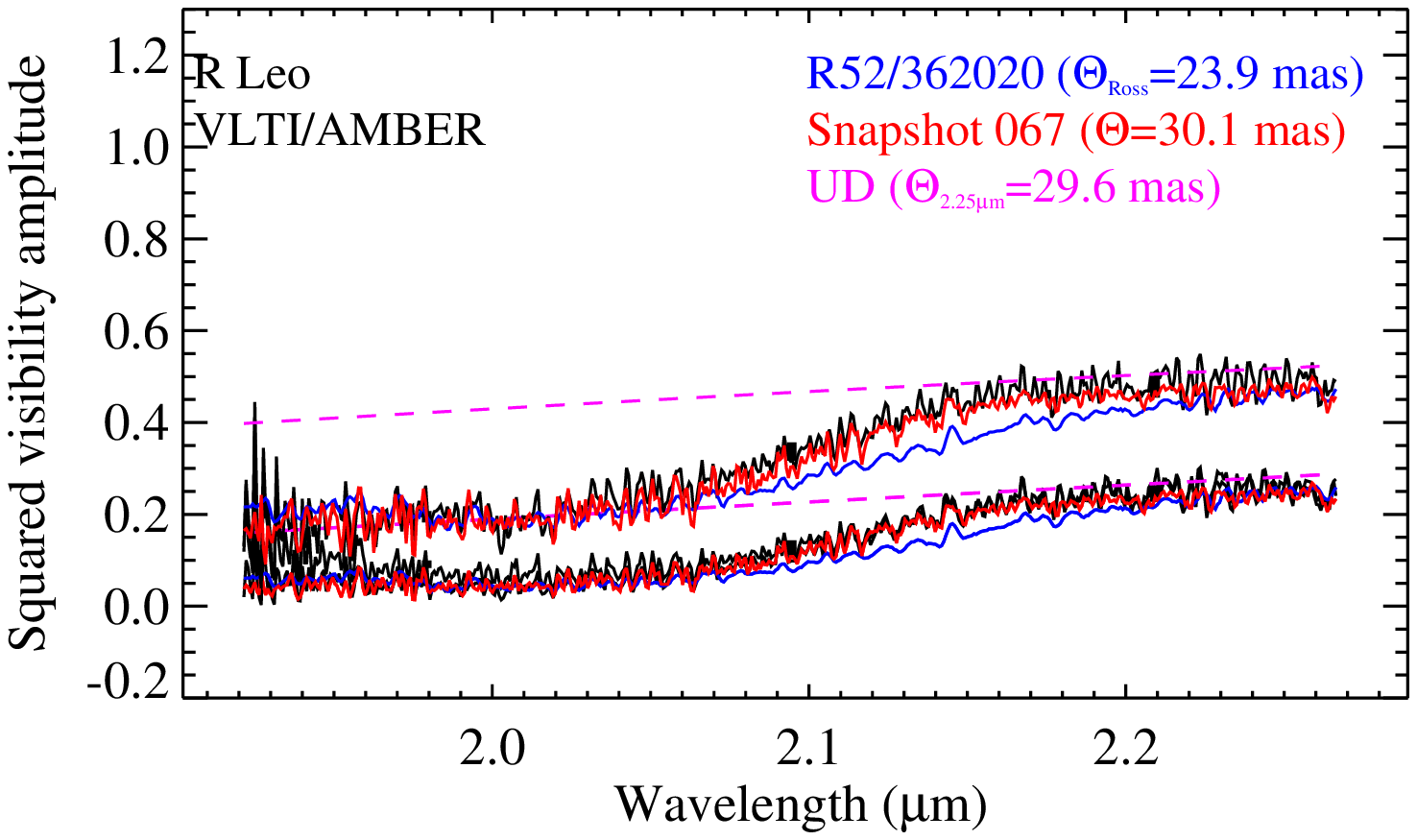}
  \includegraphics[width=0.32\textwidth]{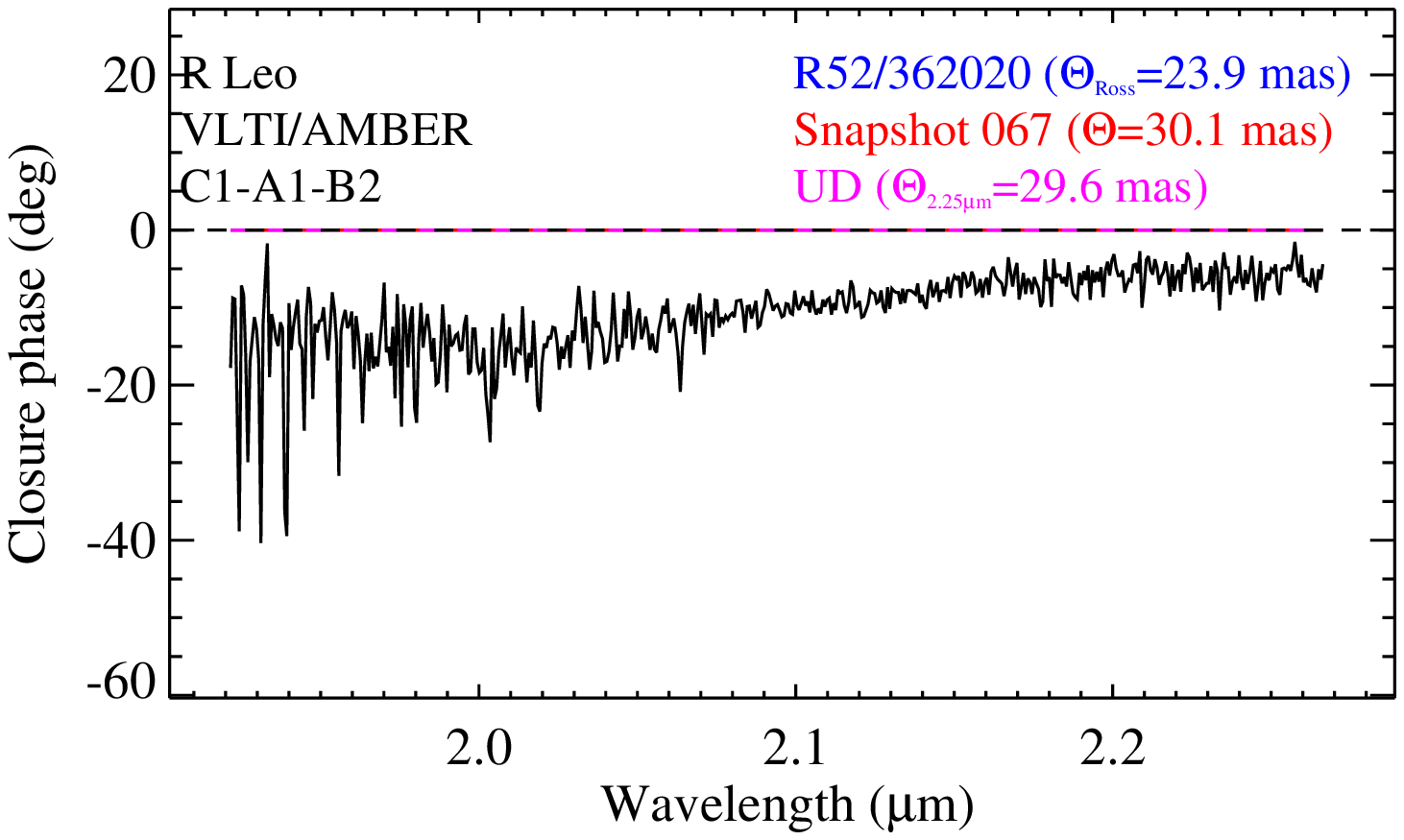}

  \includegraphics[width=0.32\textwidth]{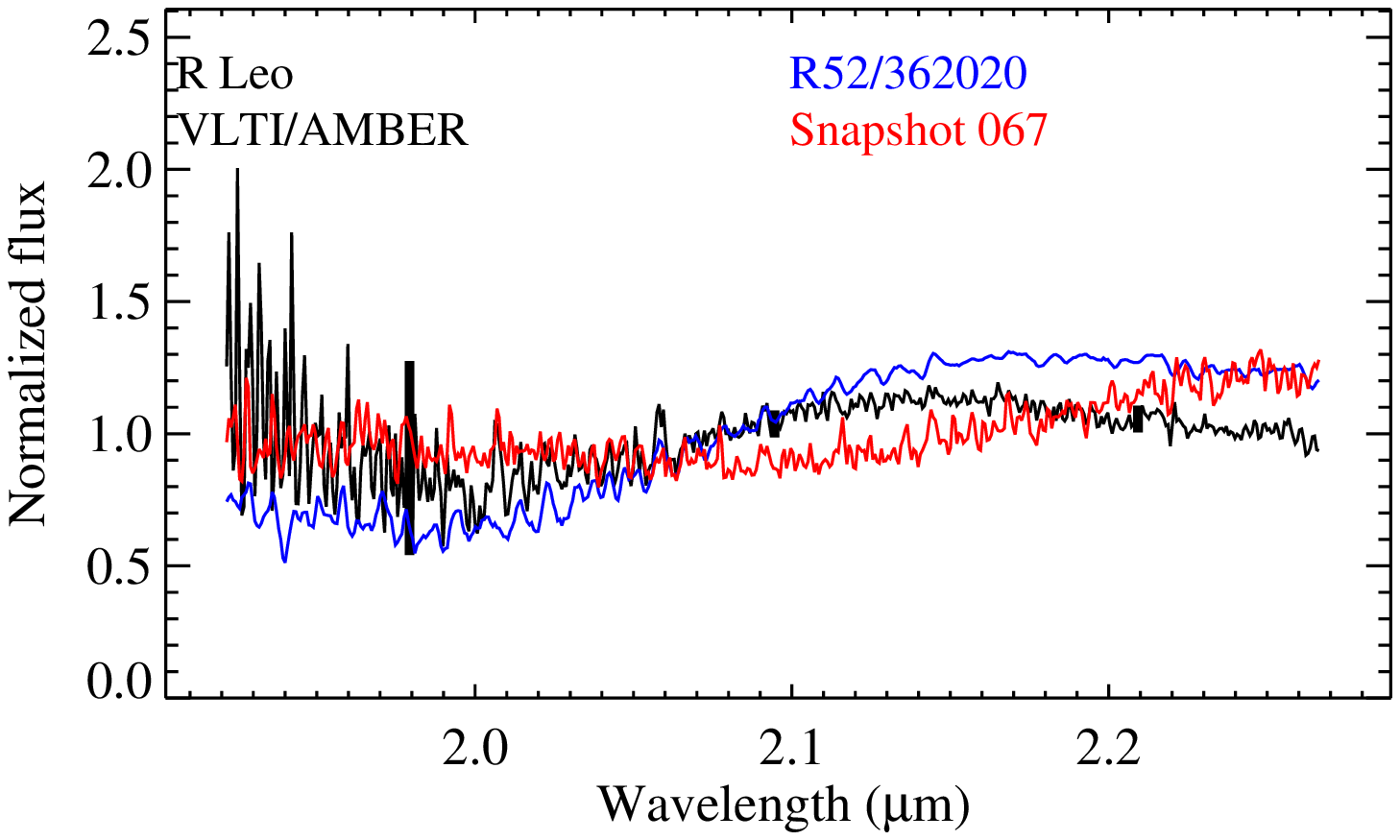}
  \includegraphics[width=0.32\textwidth]{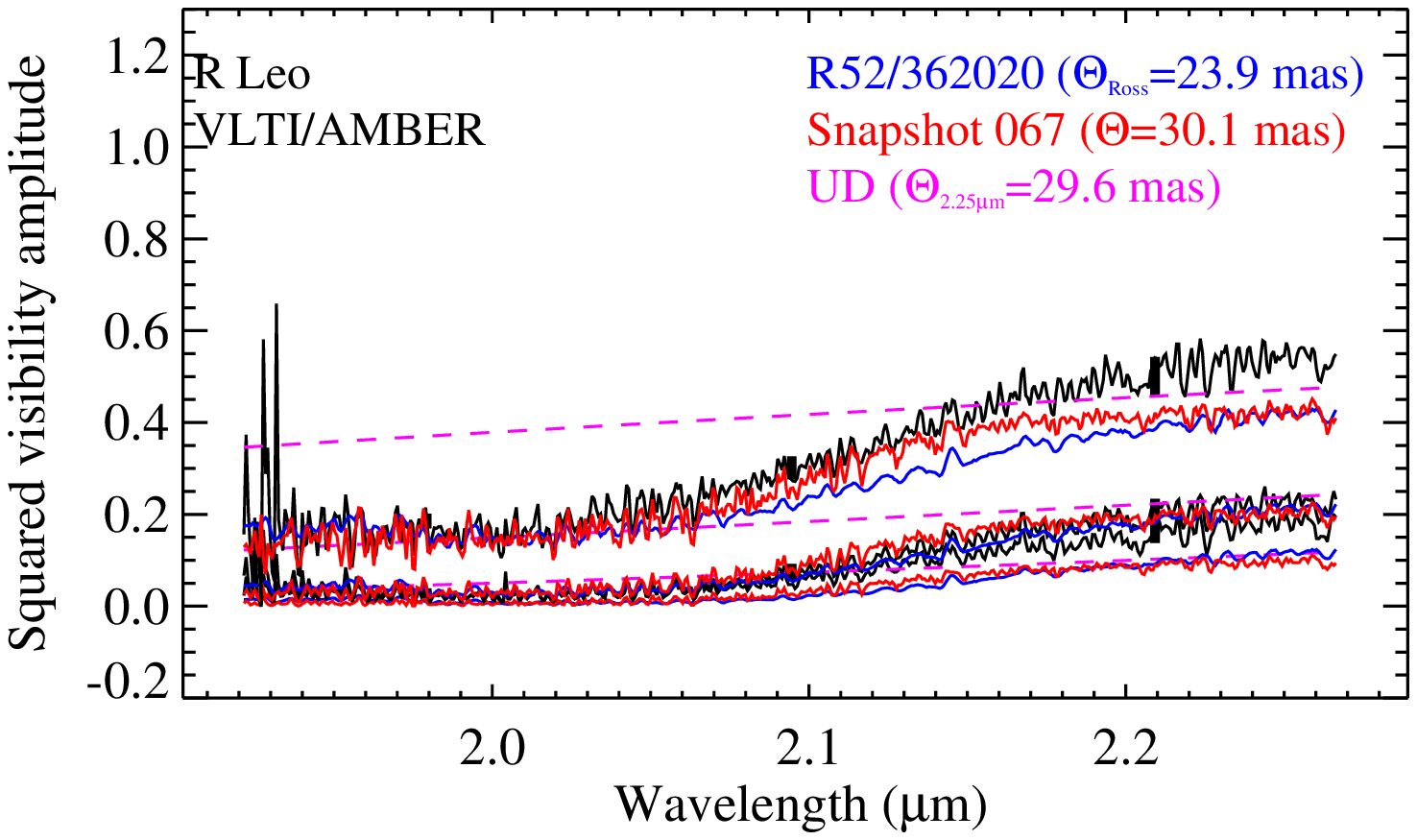}
  \includegraphics[width=0.32\textwidth]{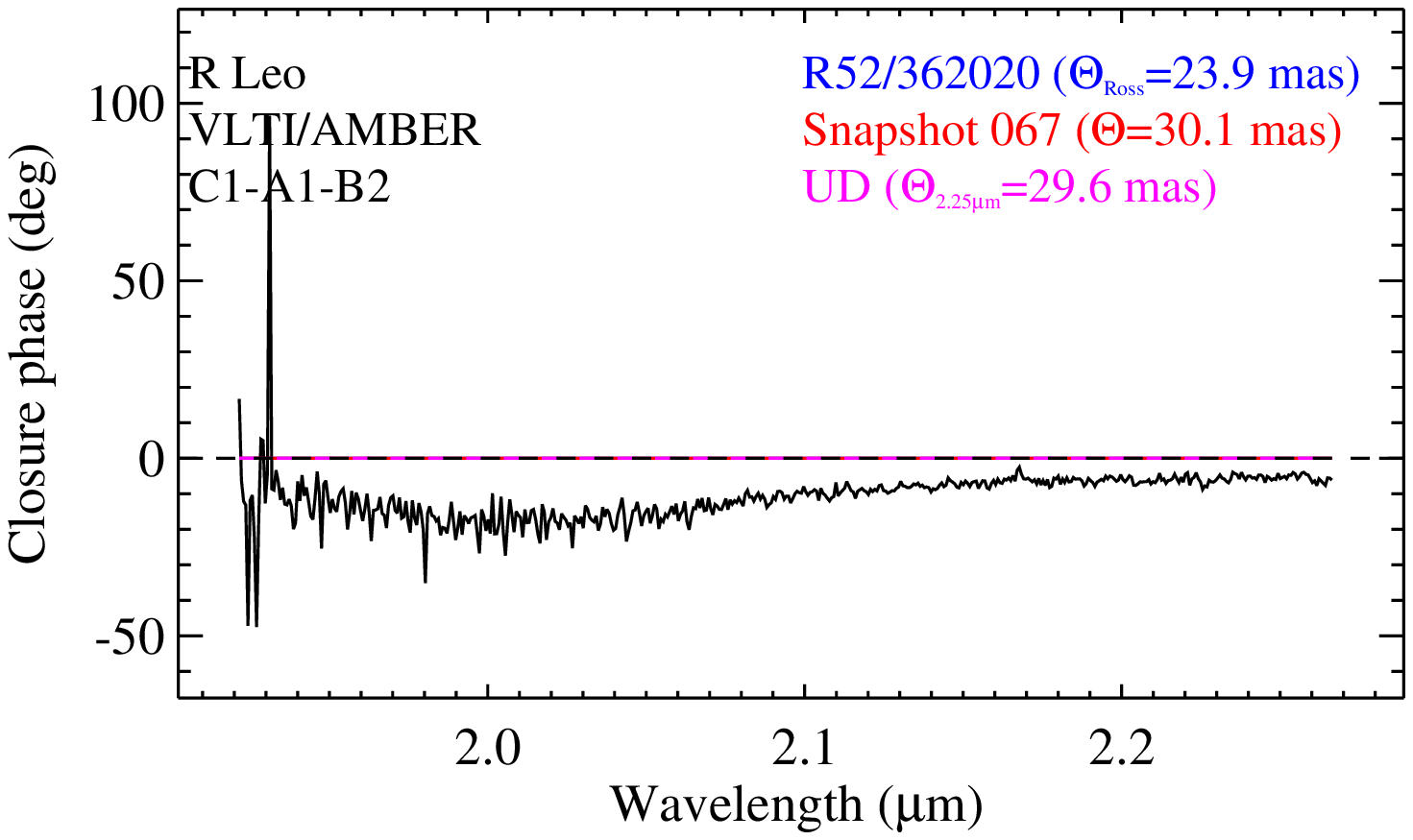}

  \includegraphics[width=0.32\textwidth]{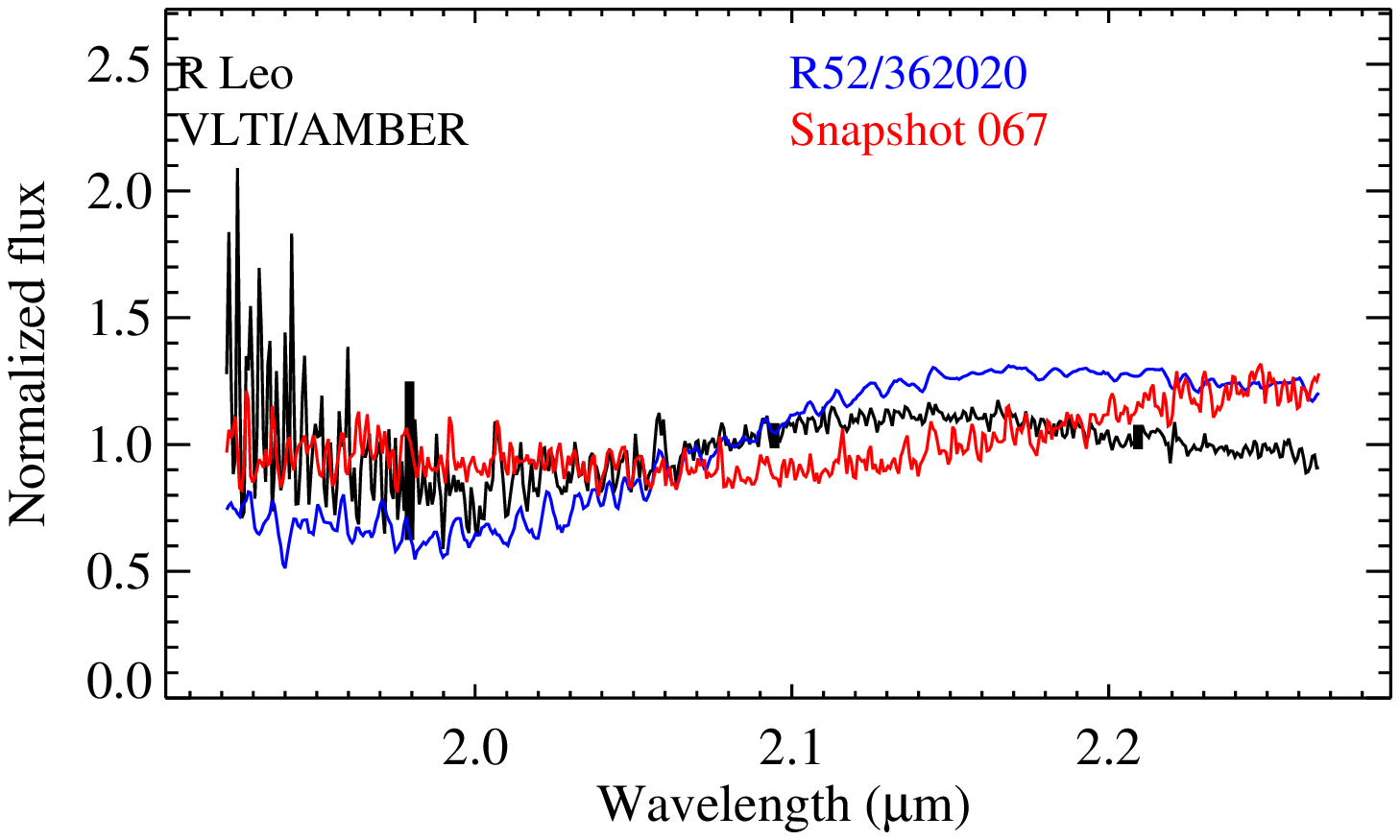}
  \includegraphics[width=0.32\textwidth]{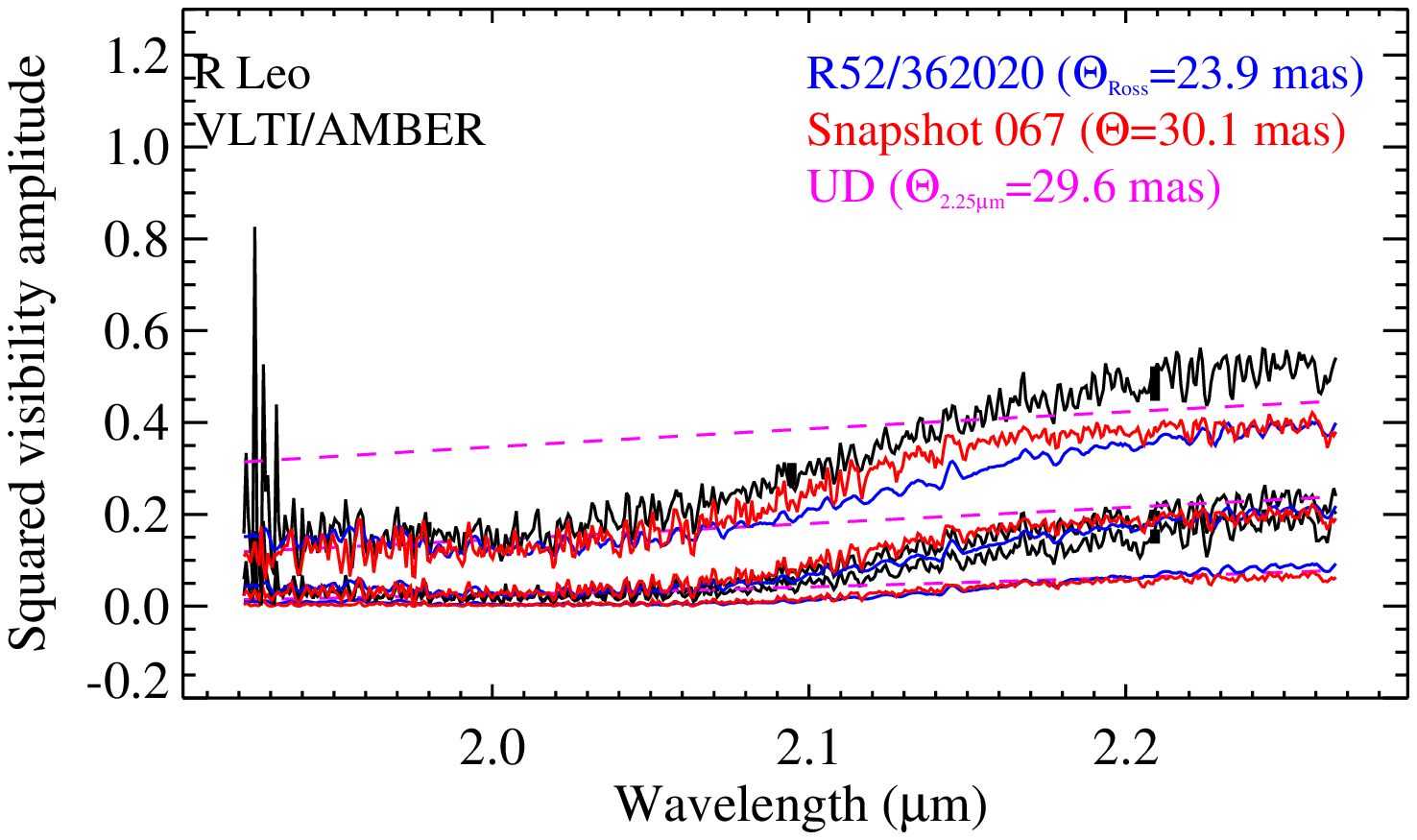}
  \includegraphics[width=0.32\textwidth]{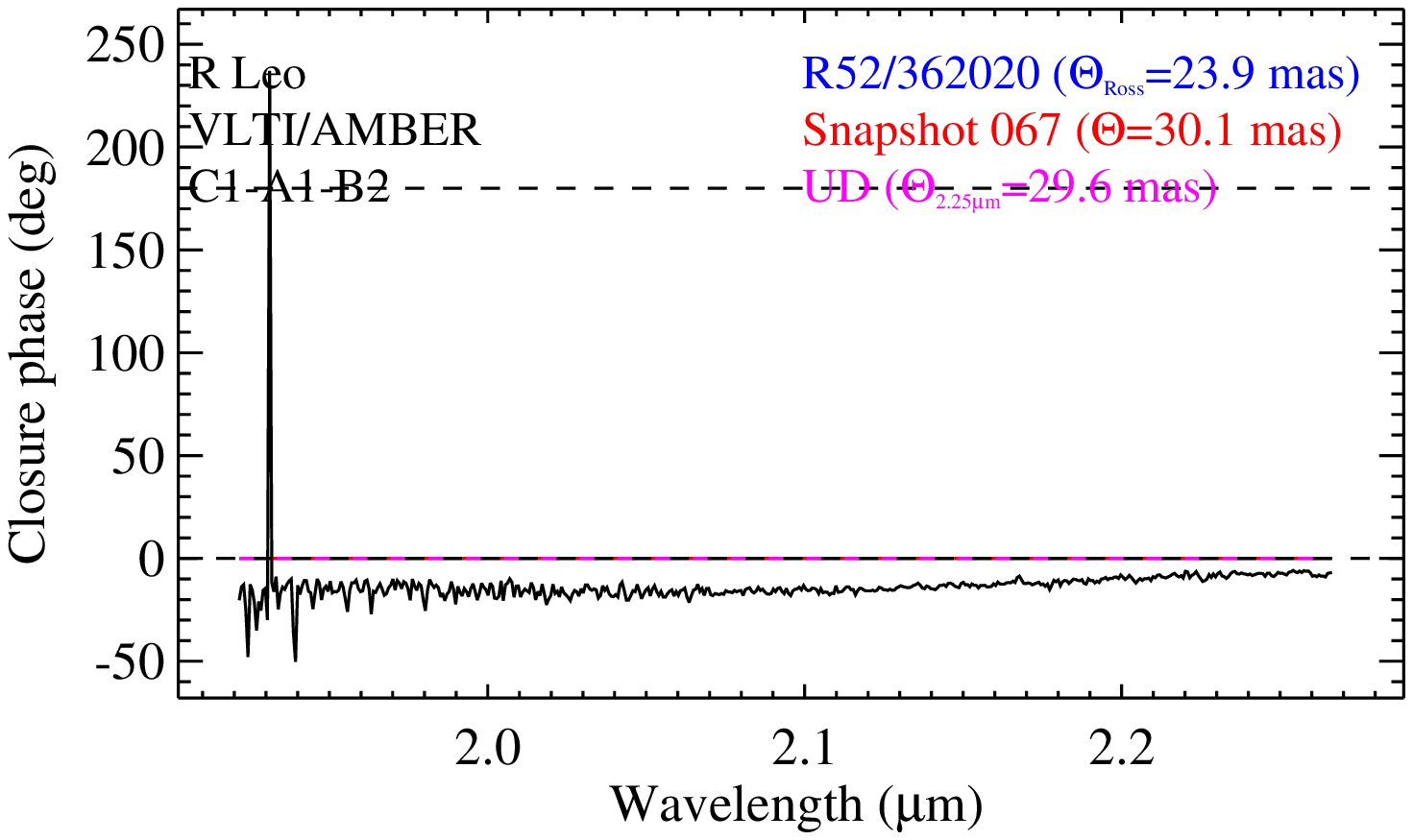}

\caption{As Fig.~\protect\ref{fig:rcnc}, but for R Leo, data sets no. 18--20
from Table~\protect\ref{tab:obs}.}
\label{fig:rleo}
\end{figure*}
}
\begin{table}
\centering
\caption{SAAO photometry\label{tab:saao}}
\begin{tabular}{lrrrrrrr}
\hline\hline
JD\tablefootmark{a}   &     $J$   &  $H$     &  $K$     & $L$      & $A_V$   & $m_\mathrm{bol}$ \\
     & (mag)     & (mag)    & (mag)    & (mag)    & (mag)   & (mag) \\\hline
R Cnc\\
4825 &   0.815 & -0.215 & -0.688 & -1.119 &  0.07 &  2.50 \\
4825 &   0.806 & -0.212 & -0.676 & -1.095 &  0.07 &  2.50 \\
4827 &   0.838 & -0.205 & -0.665 & -1.112 &  0.07 &  2.52 \\
4831 &   0.866 & -0.155 & -0.652 & -1.063 &  0.07 &  2.55 \\
4892 &   1.280 &  0.225 & -0.328 & -0.89  &  0.07 &  2.94 \\
4893 &   1.444 &  0.468 & -0.213 & -0.909 &  0.07 &  3.11 \\
W Vel\\
4825 &   2.066 &  1.048 &  0.553 & -0.074 &  0.42 &  3.70 \\
4831 &   2.002 &  0.987 &  0.503 & -0.097 &  0.42 &  3.66 \\
4892 &   1.797 &  0.759 &  0.303 & -0.23  &  0.42 &  3.44 \\
4893 &   1.780 &  0.719 &  0.250 & -0.278 &  0.42 &  3.39 \\
X Hya\\
4831 &   2.375 &  1.476 &  0.982 &  0.397 &  0.18 &  4.13 \\
4892 &   2.125 &  1.219 &  0.814 &  0.33  &  0.18 &  3.87 \\
4893 &   2.152 &  1.236 &  0.832 &  0.331 &  0.18 &  3.90 \\\hline
\end{tabular}
\tablefoot{
\tablefoottext{a}{JD-2450000}}
\end{table}

We also obtained concurrent near-infrared $JHKL$ photometry of R~Cnc, 
X~Hya, and W~Vel at the South African Astronomical Observatory (SAAO)
Mk~II instrument. Table~\ref{tab:saao} lists the 
obtained photometry and the estimated bolometric magnitudes using the 
procedures outlined by \citet{Whitelock2008}. 
We used $A_V$ values from \citet{Whitelock2000}. For o~Cet, R, Leo,
and R~Aqr, we estimated bolometric magnitudes at the phases of our
observations based on the mean $m_\mathrm{bol}$ values and the 
amplitudes $\Delta m_\mathrm{bol}$ from \citet{Whitelock2000}
(o Cet: $m_\mathrm{bol}$ 0.56 and $\Delta m_\mathrm{bol}$ 1.01; 
R Leo: $m_\mathrm{bol}$ 0.65 and $\Delta m_\mathrm{bol}$ 0.63;
R Aqr: $m_\mathrm{bol}$ 2.23 and $\Delta m_\mathrm{bol}$ 0.91).
We estimated $m_\mathrm{bol}$=1.04 for o Cet at phase 0.1,
$m_\mathrm{bol}$=0.75 for R Leo at phase 0.6, and
$m_\mathrm{bol}$=2.37 for R Aqr at phase 0.6. We adopted errors of 0.05 mag
for the concurrent measurements of R Cnc, X Hya, and W Vel,
and 0.1 mag for the estimates for o Cet, R Leo, and R Aqr.

\section{Visibility results and variability}
\begin{figure*}
\centering
 \includegraphics[width=0.45\textwidth]{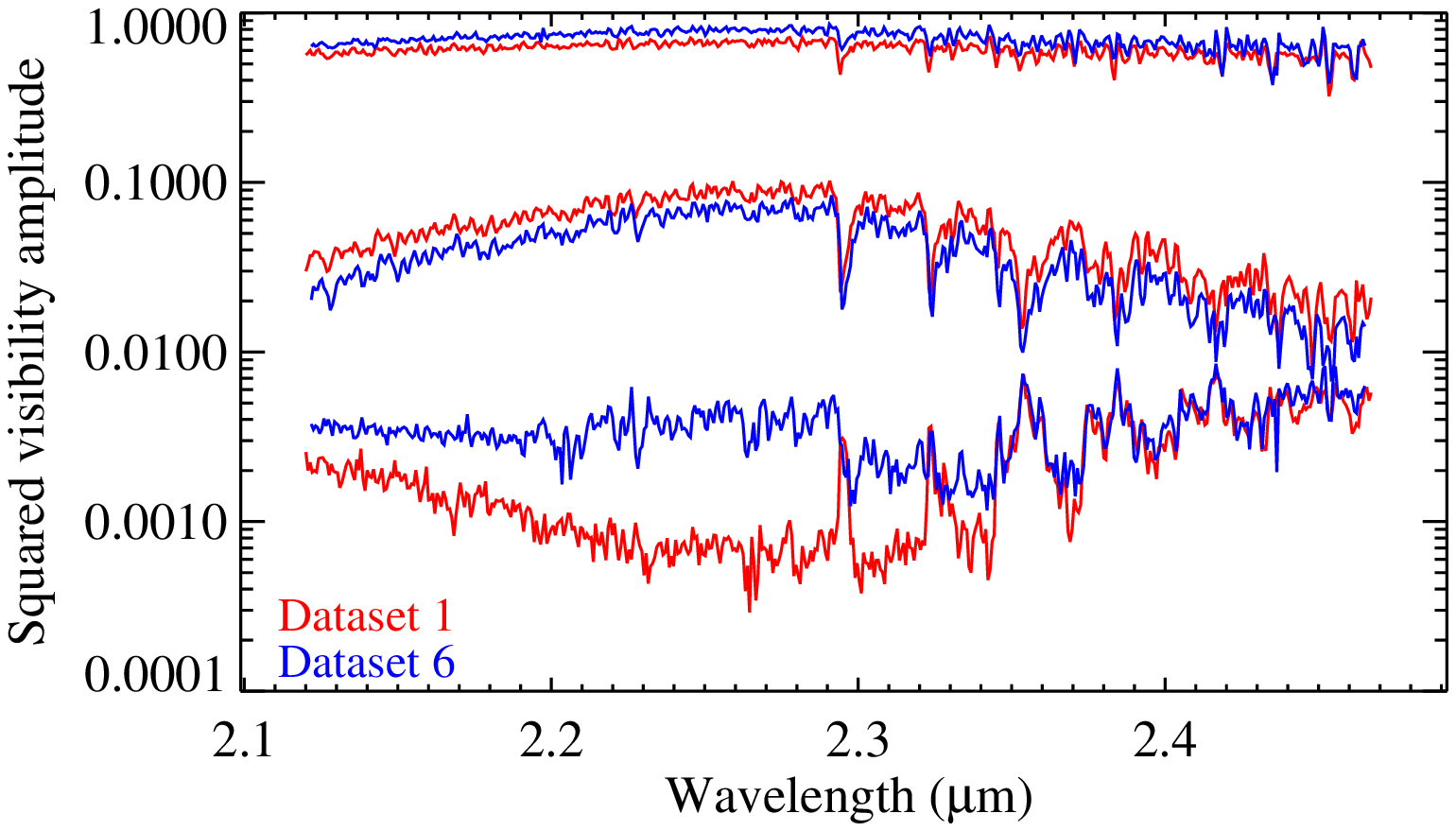}
 \includegraphics[width=0.45\textwidth]{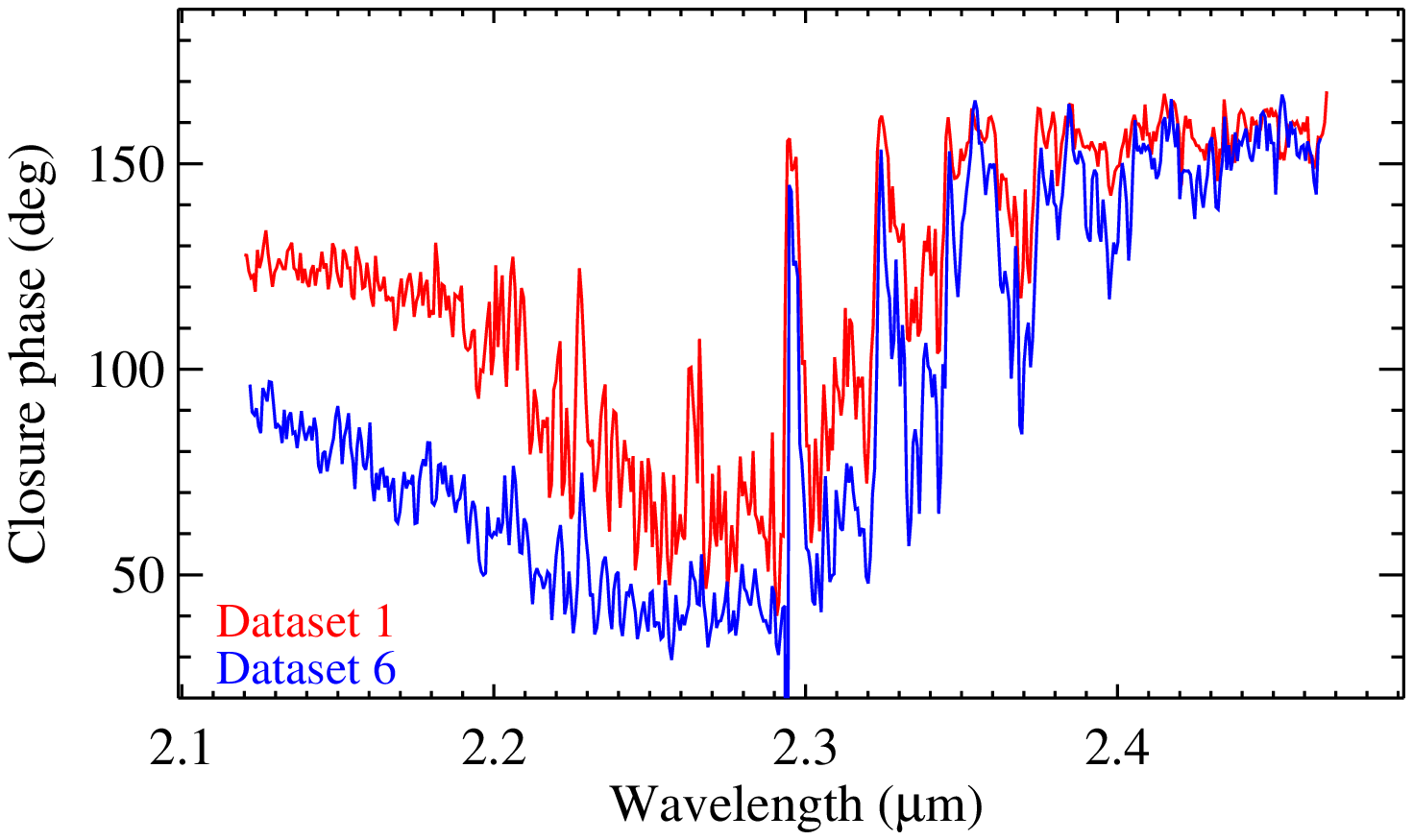}

 \includegraphics[width=0.45\textwidth]{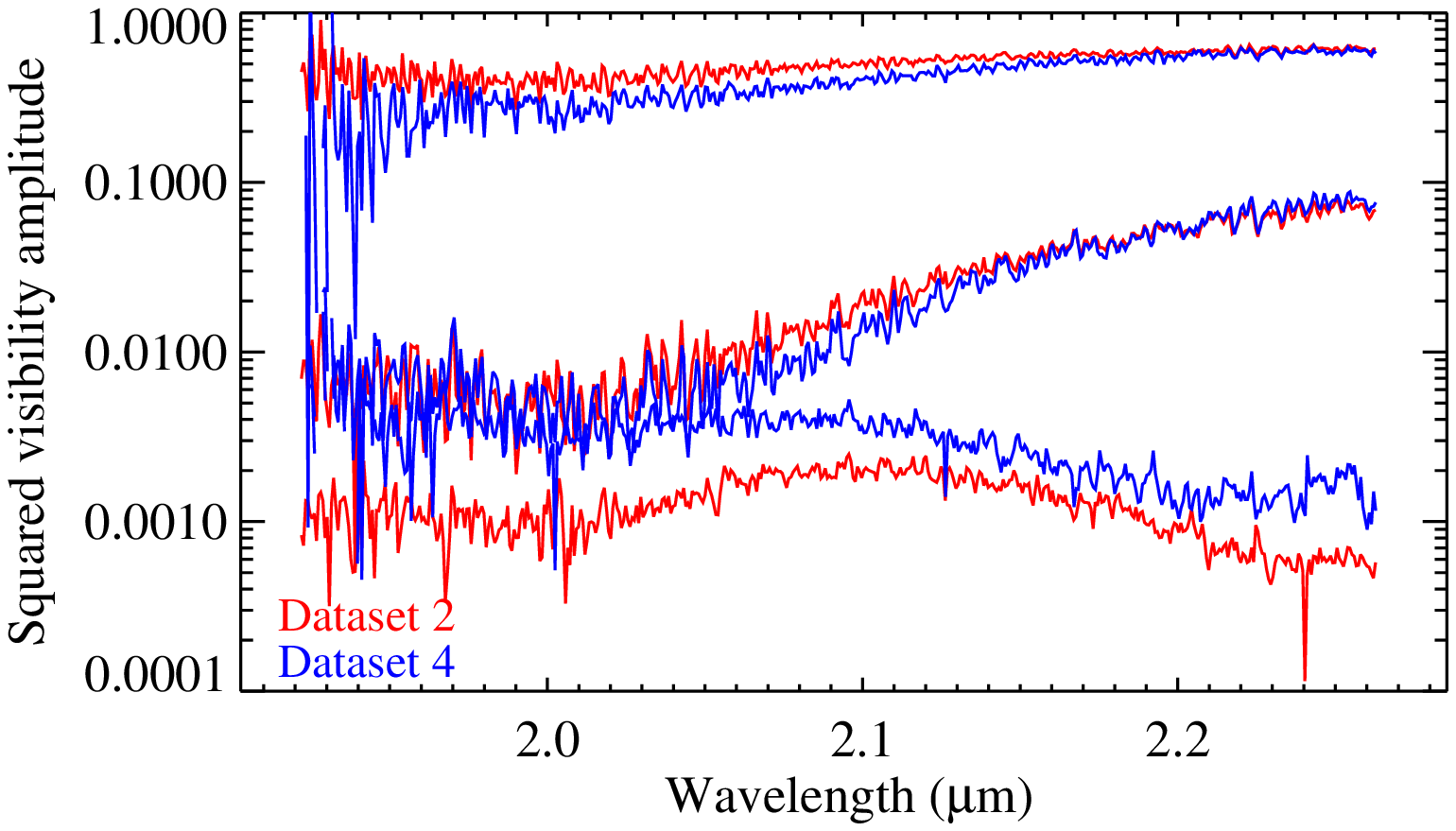}
 \includegraphics[width=0.45\textwidth]{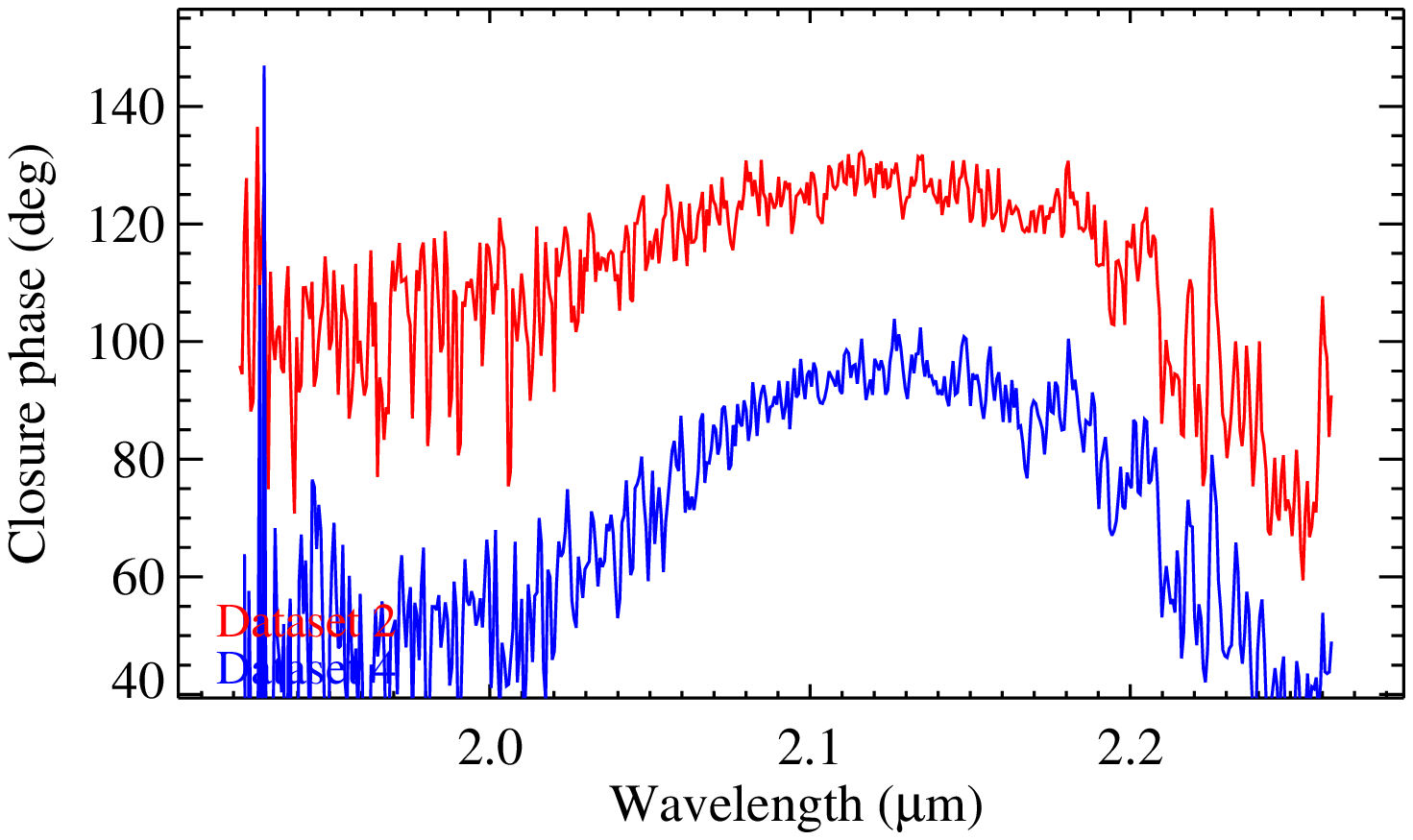}

 \includegraphics[width=0.45\textwidth]{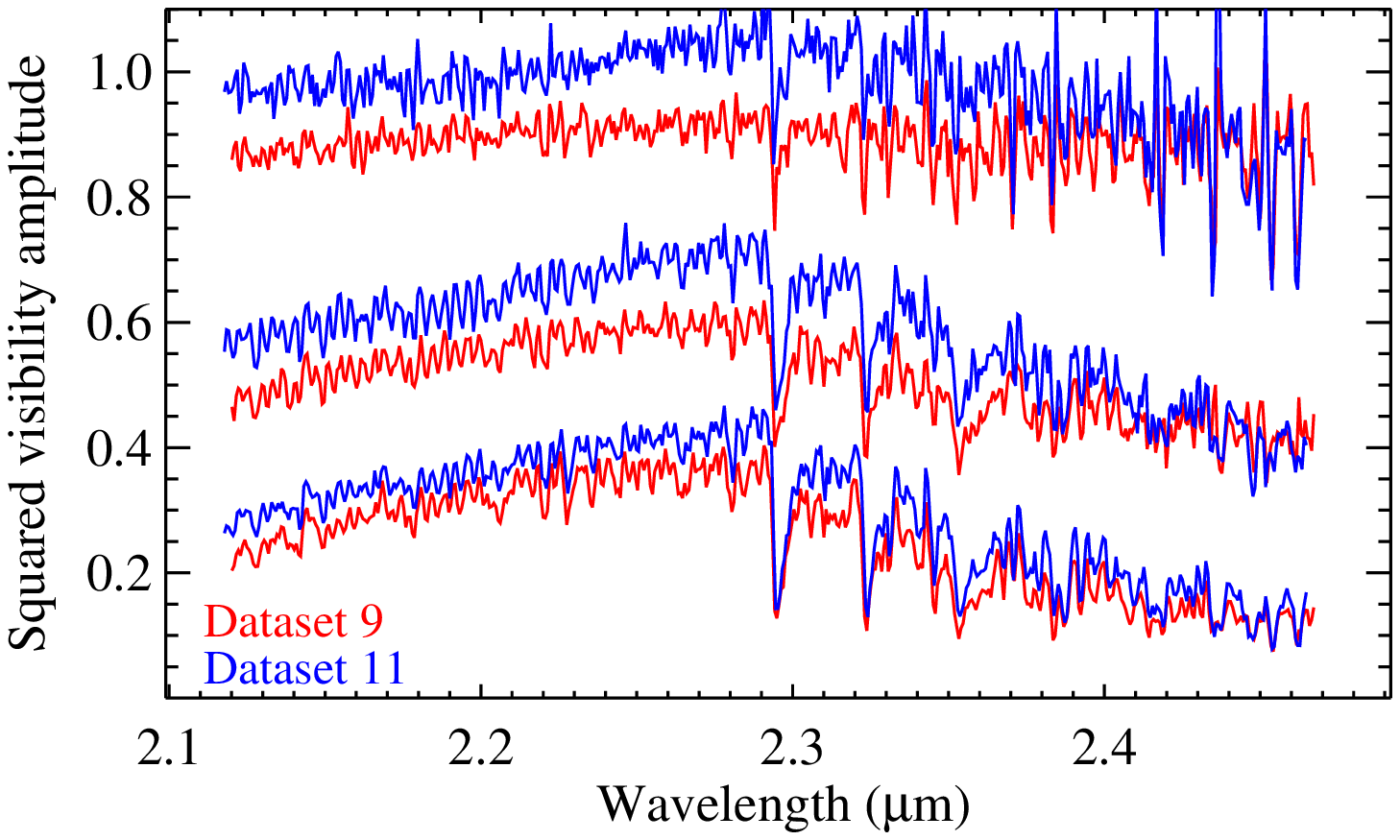}
 \includegraphics[width=0.45\textwidth]{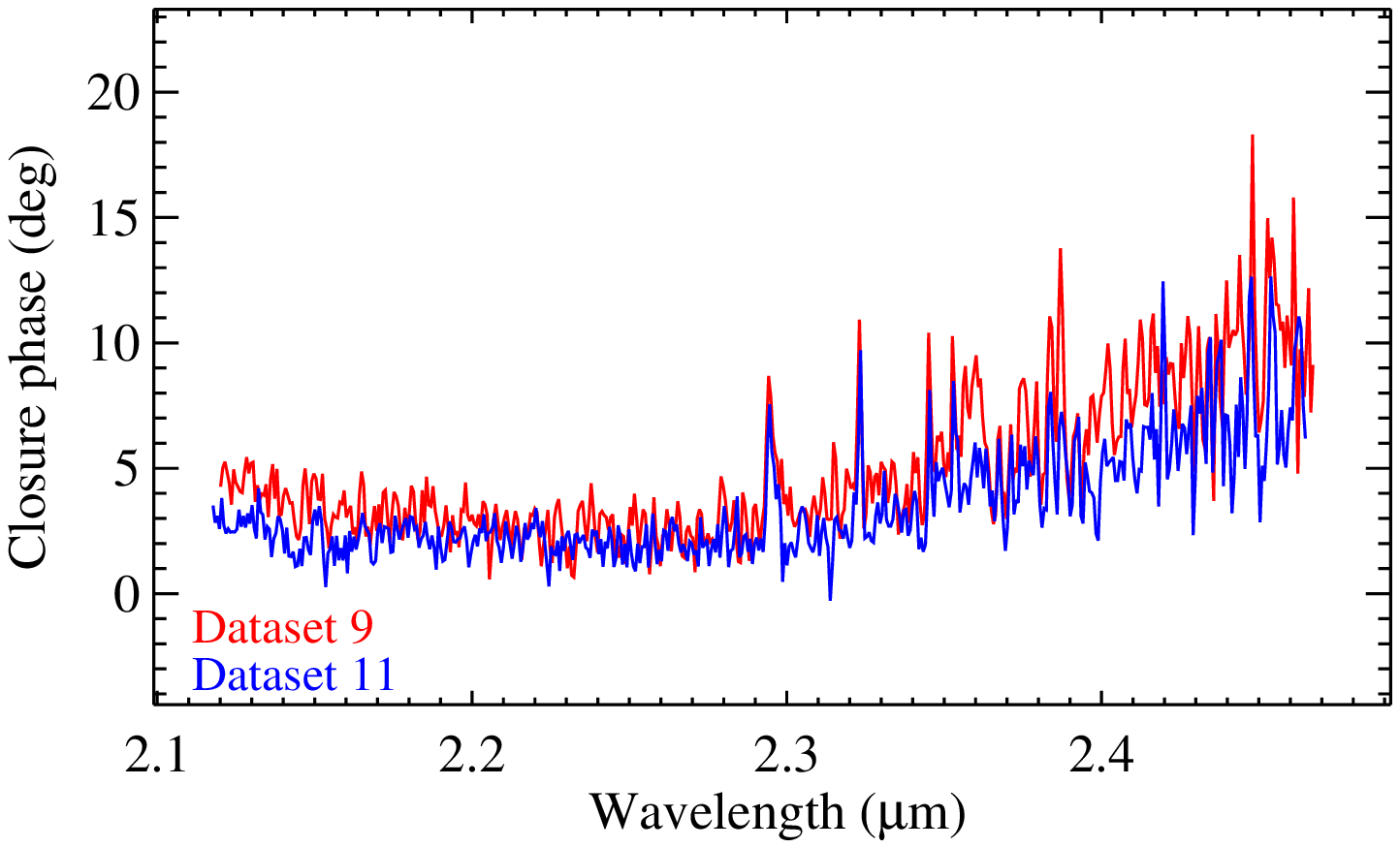}
\caption{Direct comparison of (left) the visibility function and 
(right) the
closure phases of data obtained of the same source at virtually 
the same baseline length and angle but at different epochs. 
From top to bottom:
data sets 1\&6 of R Cnc, 2\&4 of R Cnc, 9\&11 
of X Hya, cf. Table \protect\ref{tab:obs}.
Errors of the squared visibility amplitudes are in the range
of 5--15\%; errors of the closure phases are in the range
of 5--15$\deg$.
\label{fig:comp}}
\end{figure*}

The AMBER medium spectral resolution visibility results shown in 
Figs.~\ref{fig:rcnc}--\ref{fig:rleo} exhibit the same characteristic 
shape as a function of wavelength as previously observed 
by \citet{Wittkowski2011}.
The squared visibility amplitudes within 
the $K$-band show a maximum near 2.25\,$\mu$m and decrease both toward
shorter and longer wavelengths with pronounced drops at the positions
of the CO bandheads. This is interpreted as an indication of 
extended atmospheric molecular layers of -- most importantly -- H$_2$O 
(centered at $\sim$2.0\,$\mu$m) and CO (longward of 2.29\,$\mu$m).
The closure phases as a function of wavelength show a characteristic shape
as well, where deviations from point symmetry are observed at all 
wavelengths, depending on how well the target is resolved, with most
pronounced deviations from symmetry at the positions of H$_2$O and CO 
bands. \citet{Wittkowski2011} interpreted this closure phase signal
as a signature of large-scale inhomogeneities or clumps
in the extended molecular layers contributing a few percent of the total
flux. This is consistent with imaging studies of the Mira star X~Hya
by \citet{Haubois2015}.
Their reconstructed images at two epochs separated by about two months
also revealed a variability of the asymmetric structures.

Our data sets of R Cnc and X Hya include data obtained at different epochs 
using the same mode with virtually the same projected baseline lengths, 
so that we can attempt to directly (i.e. model independently) probe for 
visibility variability. In particular, data set numbers 1 and 6 of R Cnc taken
in mode MR23, data set numbers 2 and 4 of R Cnc taken in mode MR21, and data
set numbers 9 and 11 of X Hya were taken with projected baseline lengths
that differ by only 0.1\% (cf. Table~\ref{tab:obs}) and projected baseline
angles that differ by only 1--3 $\deg$. Figure~\ref{fig:comp} shows
a comparison of the visibility and closure phase data for these three
pairs. The visibility data of R Cnc at the short and intermediate baseline
differ within the error bars, so that we cannot conclude on a 
variation in the overall angular diameter of the star within a phase 
difference of 0.2. However, the visibility on the longest baselines
and the closure phases do show significant changes within a phase
difference 0.2. Both the low visibilities on the third baseline,
which fully resolves the stellar disk, and the closure phases are affected
by the inhomogeneities or clumps mentioned above. This indicates
a variability in the detailed structure of these clumps for phase
differences as small as 0.2 (cf. data sets no. 2 \& 4), again consistent 
with the imaging study by \citet{Haubois2015}. Hereby, 
data sets no. 1 and 6 were taken
during consecutive cycles, so that for this comparison we cannot
distinguish between intracycle and cycle-to-cycle variabilities.
For the comparison of the two datasets of X Hya,
we cannot conclude on any variability within two months, owing to the
lower spatial resolution that we obtained for this star.

\section{Comparison to model atmosphere predictions}
\label{sec:modelcomp}
\begin{sidewaystable*}
\centering
\caption{Fit results\label{tab:fitresults} }
\begin{tabular}{|llrr|rr|rrrrrr|rrrrr|}
\hline\hline
Target  & Epoch         & JD & $\Phi_\mathrm{\small Obs}$ & \multicolumn{2}{c|}{UD fit} &\multicolumn{6}{c|}{Best-fit 1D CODEX}  & \multicolumn{5}{c|}{Best-fit 3D CO5BOLD}  \\
        &               &           &     &  $\Theta_\mathrm{UD}^{\mathrm{2.25}\mu\mathrm{m}}$ & $\chi^2_\nu$   & Model & $\Phi^\mathrm{\small Mod.}$ & $T_\mathrm{eff}^\mathrm{\small Mod.}$ & $L^\mathrm{\small Mod.}$ & $\Theta_\mathrm{Ross}$ & $\chi^2_\nu$ & Snaps. & $T_\mathrm{eff}^\mathrm{\small Snaps.}$ & $L^\mathrm{\small Snaps.}$ & $\Theta$ & $\chi^2_\nu$ \\
        &               &         &   & (mas) &                  &       &                       & (K)                             & ($L_\odot$)        & (mas)                  &              &          & K                                  & ($L_\odot$)           & (mas)    &              \\\hline

R Cnc   & Dec. 2008     & 4831.0 & 0.3 & 13.2 $\pm$ 0.3 & 2.3 & C50/377760 & 0.4 & 2618 & 5465 & 12.3 $\pm$ 0.4 & 4.5 & 70 & 2253 & 5661 & 12.9 $\pm$ 0.2 & 9.2 \\
R Cnc   & Mar. 2009     & 4893.2 & 0.5 & 13.3 $\pm$ 0.2 & 0.8 & C50/377880 & 0.7 & 2606 & 2883 & 11.5 $\pm$ 0.8 &15.8 & 68 & 2234 & 5775 & 13.4 $\pm$ 0.3 &12.5 \\
R Cnc   & Feb. 2010     & 5239.2 & 1.5 & 13.1 $\pm$ 0.4 & 3.2 & C50/377760 & 0.4 & 2618 & 5465 & 12.5 $\pm$ 0.3 & 3.7 & 70 & 2253 & 5661 & 12.9 $\pm$ 0.2 & 4.6 \\[1ex]
W Vel   & Dec. 2008     & 4830.4 & 0.9 &  7.4 $\pm$ 0.2 & 4.4 & R52/390300 & 0.9 & 3482 & 4674 &  5.2 $\pm$ 0.9 & 2.9 & 61 & 2416 & 7662 &  8.5 $\pm$ 0.4 & 5.4 \\
W Vel   & Mar. 2009     & 4892.2 & 1.1 &  7.4 $\pm$ 0.1 & 1.5 & R52/363900 & 0.7 & 3225 & 3910 &  7.1 $\pm$ 1.1 & 2.5 & 61 & 2416 & 7662 &  8.4 $\pm$ 0.4 & 4.0 \\[1ex]
X Hya   & Dec. 2008     & 4831.0 & 0.7 &  6.2 $\pm$ 0.1 & 0.8 & o54/250640 & 0.7 & 2990 & 3305 &  4.1 $\pm$ 0.8 & 2.2 & 67 & 2242 & 5995 &  6.1 $\pm$ 0.1 & 2.7 \\
X Hya   & Mar. 2009     & 4892.2 & 0.9 &  5.7 $\pm$ 0.1 & 0.7 & o54/286100 & 0.5 & 2614 & 3845 &  5.7 $\pm$ 1.0 & 1.4 & 67 & 2242 & 5995 &  5.8 $\pm$ 0.3 & 2.5 \\[1ex]
R Aqr   & Oct. 2012     & 6219.1 & 0.6 & 18.4 $\pm$ 0.4 & 1.8 & o54/288140 & 0.7 & 2984 & 3342 & 16.2 $\pm$ 3.5 & 3.4 & 66 & 2263 & 6211 & 15.8 $\pm$ 1.8 & 6.3 \\[1ex]
o Cet   & Oct. 2012     & 6219.2 & 0.1 & 28.6 $\pm$ 1.0 & 5.1 & o54/286100 & 0.5 & 2614 & 3845 & 31.1 $\pm$ 3.7 & 5.8 & 68 & 2234 & 5775 & 28.0 $\pm$ 0.5 & 4.9 \\
o Cet   & Sep. 2013     & 6549.3 & 0.1 & 28.3 $\pm$ 1.5 &12.4 & o54/287940 & 0.5 & 2629 & 3820 & 29.0 $\pm$ 3.7 &15.9 & 64 & 2271 & 6364 & 30.7 $\pm$ 1.8 &56.6 \\[1ex]
R Leo   & Dec. 2012     & 6272.3 & 0.6 & 29.6 $\pm$ 1.3 & 8.5 & R52/362020 & 0.6 & 2851 & 2901 & 23.9 $\pm$ 2.2 &23.4 & 67 & 2242 & 5995 & 30.2 $\pm$ 0.7 &20.0 \\\hline

\end{tabular}
\end{sidewaystable*}
As a first approach to characterizing the angular size of the targets,
we fit a uniform disk (UD) model to the squared visibility amplitudes
obtained at a near-continuum wavelength band between 
2.22\,$\mu$m and 2.28\,$\mu$m. 
Previous studies have shown that this
wavelength band is not contaminated much by molecular bands and that it
provides a good estimate of the angular diameter of continuum-forming layers.
We merged data of the same targets
that were obtained at the same epoch with separations of up to a few days.
Our previous studies \citep{Wittkowski2011} indicated surface 
inhomogeneities at flux levels of a few percent that affected low
visibility values, while higher visibility values within the first
lobe were not affected by inhomogeneities and were consistent with 
spherically symmetric models.
We thus excluded squared visibility amplitudes 
below 0.01 in our fitting procedure, and these correspond to flux levels 
below 10\%.
The resulting UD angular diameters 
$\Theta_\mathrm{UD}^{\mathrm{2.25}\mu\mathrm{m}}$ obtained at a wavelength of
2.25\,$\mu$m, together with the corresponding reduced $\chi^2_\nu$ values
are listed in Table~\ref{tab:fitresults}, together with the fit
results of the 1D and 3D model atmospheres described below.
Reduced $\chi^2_\nu$ values are partly above unity. The most likely
cause of the comparably higher $\chi^2_\nu$ values is that we 
underestimated the systematic error of the interferometric transfer 
function. Our error was estimated by 
the variability of the transfer functions obtained before and after
the science target observations.
There may be systematic effects between the transfer function of
(almost) unresolved calibrator stars and (often fully resolved) 
science targets that could not be included in this error, 
even though we made sure that the FINITO fringe-tracking performance
(phase rms values) was similar (cf. Sect. \ref{sec:obs}).
We note that the statistical error based on the individual scans and 
that of the uncertainty of the adopted calibration star diameters are smaller
than for the transfer function stability.

We obtained $\Theta_\mathrm{UD}^{\mathrm{2.25}\mu\mathrm{m}}$ values
for R~Cnc between 13.1\,mas and 13.3\,mas at phases between 0.3 and 0.5, 
for W~Vel of 7.4\,mas at phases between 0.9 and 1.1, for X~Hya between
6.2\,mas and 5.7\,mas at phases between 0.7 and 0.9, for R~Aqr of 
18.4\,mas at phase 0.6, for o~Cet between 28.6\,mas and 28.3\,mas
at phase 0.1, and for R~Leo of 29.6\,mas at phase 0.6. Errors range
between 1\% and 3\%. For those targets for which we obtained data
at different phases (R~Cnc, W Vel, X Hya) separated by ~0.2, there
are no significant differences of 
$\Theta_\mathrm{UD}^{\mathrm{2.25}\mu\mathrm{m}}$. These values
are broadly consistent with earlier estimates of the angular diameters
of these targets. For example, \citet{Mennesson2002} obtained
broad-band $K$ angular diameters of 16.9$\pm$0.6\,mas for R Aqr at
phase 0.4, for o Cet between 24.4$\pm$0.1\,mas and 28.8$\pm$0.1\,mas
at phases 0.9--0.0, and for R Leo between 28.2$\pm$0.1\,mas and 
30.7$\pm$0.1\,mas at phases between 0.2 and 0.3. \citet{Woodruff2004}
estimated a Rosseland angular diameter of o~Cet between 28.9$\pm$0.3\,mas
and 34.9$\pm$0.4\,mas at phases of 0.1 and 0.4, respectively.
\citet{Fedele2005} estimated a Rosseland angular diameter of R~Leo
between 28.5$\pm$0.4\,mas and 26.2$\pm$0.8\,mas at phases of 0.1 and 0.0,
respectively.

The visibility curves of Figs.~\ref{fig:rcnc}--\ref{fig:rleo} show
that the UD curves provide good fits to the measured visibilities
in the near-continuum bandpass around 2.25\,$\mu$m at the different
baselines. The measured visibilities drop significantly below the 
UD curves toward shorter and larger wavelengths, which are affected 
by molecular bands of H$_2$O and CO. This behavior is consistent
with our earlier results \citep{Wittkowski2011} and has been interpreted 
as an indication that extended molecular layers are located above the 
continuum-forming layers.

Closure phases are wavelength-dependent resembling the locations
of the water vapor band around 2.0\,$\mu$m and the CO band and 
bandheads longward of 2.29\,$\mu$m. As in \citet{Wittkowski2011}
we interpret the closure phase signal as an indication of a few
surface spots across the extended molecular atmosphere. As
spherical model atmospheres (cf. Sect.~\ref{sec:codex}) are generally 
consistent with squared visibility amplitudes above $\sim$ 0.01,
and deviations occur at lower values, we can estimate that these
inhomogeneities contribute up to $\sim$\,10\% of the total flux.

\subsection{Comparison to 1D CODEX model atmospheres}
\label{sec:codex}
As a next step, we compared our data to 1D CODEX model atmospheres
of Mira variables by \citet{Ireland2008,Ireland2011} as in 
\citet{Wittkowski2011}. These
model atmospheres are based on self-excited pulsation models and
use the opacity sampling method to calculate model intensity
profiles. CODEX models are available in four different series
that correspond to different stellar parameters of the underlying 
hypothetical non-pulsating parent star. The parameters of the
CODEX model series are listed in Table~\ref{tab:codex}.
\begin{table}
\centering
\label{tab:codex}
\caption{Parameters of the CODEX model series}
\begin{tabular}{llllll}
\hline\hline
Series & Mass & Lum. & Radius & Eff. temp. & Period \\\hline
o54 & 1.1\,M$_\odot$ & 5400\,L$_\odot$ & 216\,R$_\odot$ & 3370\,K & 330\,d \\
R52 & 1.10\,M$_\odot$& 5200\,L$_\odot$ & 209\,R$_\odot$ & 3400\,K & 307\,d \\
C50 & 1.35\,M$_\odot$& 5050\,L$_\odot$ & 291\,R$_\odot$ & 2860\,K & 430\,d \\
C81 & 1.35\,M$_\odot$& 8160\,L$_\odot$ & 278\,R$_\odot$ & 3290\,K & 427\,d \\\hline
\end{tabular}
\end{table}
The parameters of the o54 series are designed to match those
of the Mira variable o~Cet, those of the R52 series
to match the Mira variable R~Leo, and those of the C50/C81
series to match a typical longer-period Mira star such as R Cas.
C81 is a higher luminosity series compared to C50.
Each series consists of a number of models that correspond to 
different phases with a separation of $\sim$0.1 and covering a few
cycles. The o54 series deliberately includes cycles where the atmosphere
is particularly compact and particularly extended, in addition to normal
cycles.
The CODEX models contain a simple description of dust, which only moderately
affects the equation of state, the spectrum, and the
effective temperature and which does not lead to a wind 
\citep{Ireland2008,Ireland2011}. We note that the nature of real dust
in Miras is still rather controversial regarding the layers, the 
composition, and the grain sizes 
\citep[e.g.,][]{Bladh2012,Goumans2012,Gail2013,Karovicova2013,
Bladh2015,Gobrecht2016}.

We tabulated the {\tt CODEX} intensity profiles in
steps of 0.0005\,$\mu$m ($\lambda/\Delta\lambda=40000$ at 
$\lambda=2$\,$\mu$m) for wavelengths between 1.4\,$\mu$m and 2.5\,$\mu$m,
covering the AMBER medium resolution modes at a sufficient spectral resolution
to average more than 20 monochromatic intensity profiles for each individual 
AMBER spectral channel. Synthetic visibility values were computed
using the Hankel transform and averaging monochromatic squared visibility
amplitudes over the AMBER spectral channels.

We fit the visibility data of each of our targets and epochs 
separately to every available CODEX model, where the Rosseland angular 
diameter $\theta_\mathrm{Ross}$ was the only free parameter, and we
selected the models with the best reduced $\chi^2_\nu$ values. 
The Rosseland angular diameter is the diameter that corresponds to the
model layer where the Rosseland optical depth equals unity 
($\tau_\mathrm{Ross}=1$)\footnote{Gray pulsation models typically
use a $\tau_\mathrm{Ross}=2/3$ radius definition. In practice, the radius positions
of these deep 2/3 and 1 layers are  not very different in almost any model. Our discussion
is not affected by this definition detail.}.
We then selected a series
that provided good fits to all epochs of a given target and used the 
best-fit model of only this series. For o~Cet and R~Leo, we restricted 
the fits to the o54 and R52 series, respectively, which are designed to 
match the parameters of these sources. As for the UD fits, we excluded
squared visibility values below 0.01, which correspond to flux levels below
10\% because these may be affected by surface inhomogeneities.
The error of $\theta_\mathrm{Ross}$
is dominated by the choice of the model, reflecting
the uncertainty of defining a radius of a star with a very extended
atmosphere. We used the standard deviation of the $\theta_\mathrm{Ross}$
values based on the ten best-fit models of the adopted series
as the error of $\theta_\mathrm{Ross}$. 

The resulting best-fit models, together with their model phase, effective
temperature, luminosity, and the corresponding $\theta_\mathrm{Ross}$
and $\chi^2_\nu$ values are listed in the middle panel of 
Table~\ref{tab:fitresults} for each of our sources and epochs.
The model phases of the best-fit CODEX models are generally consistent 
with the visual phases within 0.1--0.2, which is considered the 
uncertainty of phase assignments \citep{Ireland2011}. Exceptions
are the near-maximum observation of W Vel at phase 0.1, of X Hya at phase 0.9,
and of o Cet at phase 0.1, for which the molecular features can be  better
described by near-minimum models. Discrepancies between observed and 
model phases
in particular for near-maximum observations has already been noted 
by \citet{Hillen2012}. This might be caused by an absence of a wind in the 
CODEX models. \citet{Bladh2015} show that models with and without a wind
have different colors because molecular features are related to the 
structure in the wind acceleration region.

The Rosseland angular diameters are broadly consistent (within 1--3$\sigma$)
with the $\Theta_\mathrm{UD}^{\mathrm{2.25}\mu\mathrm{m}}$
and with previous observations listed above.
This is a satisfactory agreement, considering that the UD model is not expected
to be a very good model even for the near-continuum bandpass, that the estimate 
of the Rosseland angular diameter is model-dependent, and that 
in general a radius definition of a star with a very extended atmosphere is
problematic \citep[cf.][]{Scholz2001}. As for the 
$\Theta_\mathrm{UD}^{\mathrm{2.25}\mu\mathrm{m}}$ values, the  
$\theta_\mathrm{Ross}$ values also do not indicate an intracycle or 
cycle-to-cycle
variability within the limited separations ($\sim$ 2 months, corresponding to 
phase differences of $\sim$0.2) and within our uncertainties.

Figures~\ref{fig:rcnc} to \ref{fig:rleo} show the synthetic flux, squared
visibility amplitudes, and closure phases as predicted by the best-fit 
CODEX models compared to the observed quantities.
The flux spectra predicted by the CODEX model are consistent with our
observed AMBER spectra within the error bars. The observed spectra
taken with the AMBER MR-$K$ 2.3\,$\mu$m mode are systematically higher than
the model-predicted spectra in the wavelength range between about 2.1\,$\mu$m
and 2.2\,$\mu$m. However, the same wavelength range is consistent for data
taken with the AMBER MR-$K$ 2.1\,$\mu$m mode. This discrepancy is thus 
most likely to be caused by a systematic calibration effect of the 
AMBER MR-$K$ 2.3\,$\mu$m mode.
Although some spectral calibration can be performed, we note that the 
AMBER instrument is optimized for interferometry and that it is not a 
good spectrograph with its single-mode fibers.

The synthetic visibility
curves as predicted by the best-fit CODEX models are consistent with the
measured visibility curves. In particular, they match the decrease in the
visibility function well at the location of the H$_2$O band around 2.0\,$\mu$m
and at the location of the CO bandheads longward of 2.29\,$\mu$m. This shows
that the CODEX models predict an extended molecular atmosphere and that
the model-predicted extensions of the atmosphere are consistent with 
our observations. Deviations between observed and model-predicted visibility
amplitudes below $\sim$0.1 and the wavelength-dependent closure phases showing
strongest deviations from point symmetry at the locations of the water vapor
and CO bands indicate a few inhomogeneities within the molecular atmosphere
within an overall spherical geometry. This is consistent with similar
results for oxygen-rich Mira variables 
by, for example, \citet{Lebouquin2009}, \citet{Wittkowski2011}, 
\citet{Monnier2014}, and \citet{Haubois2015}.

\subsection{Comparison to 3D CO5BOLD model atmospheres}
\label{sec:rhd}
\begin{figure*}
\centering
  \includegraphics[width=0.53\textwidth]{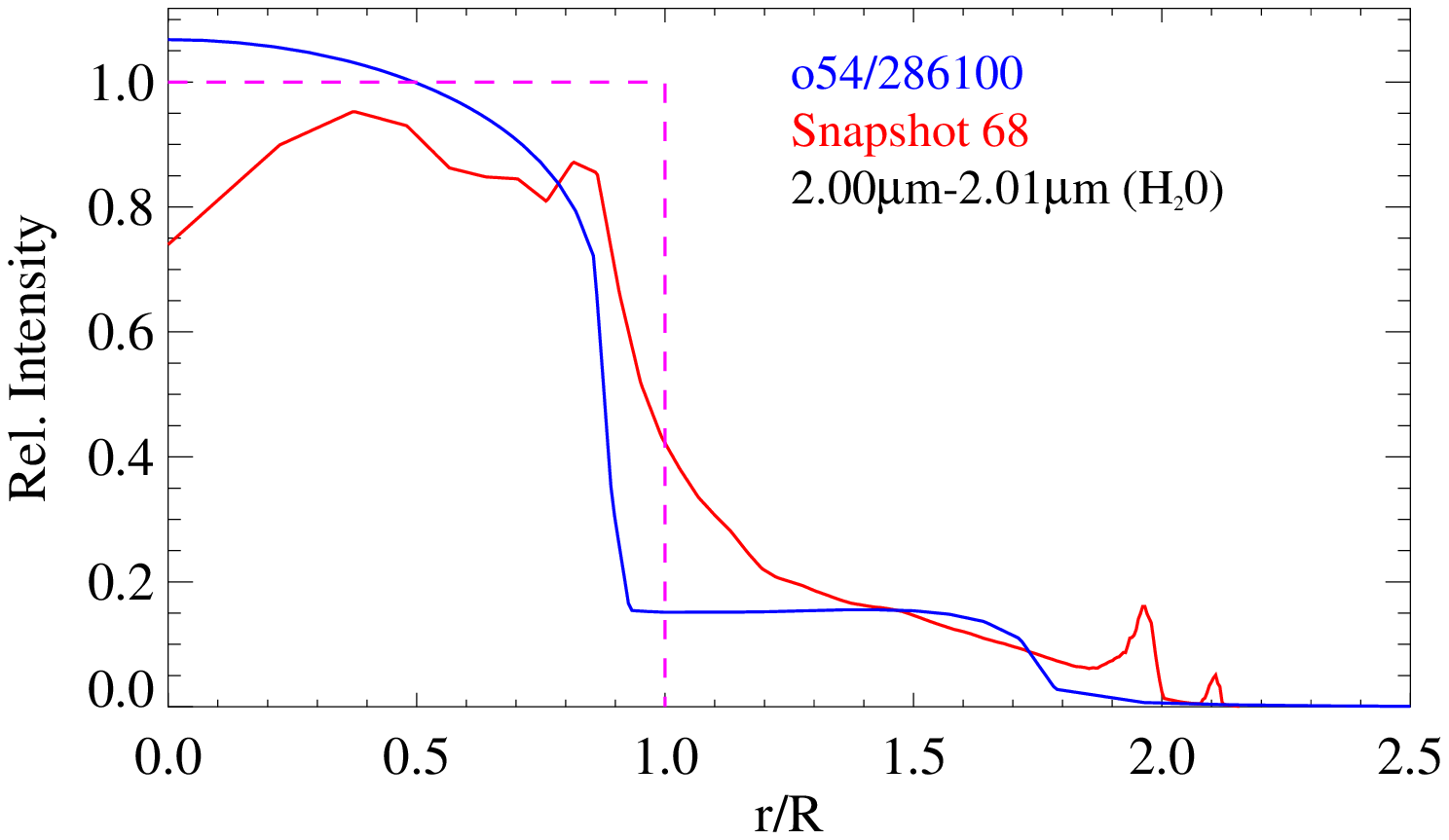}
  \includegraphics[width=0.30\textwidth]{image_parambf070_20000.00-20100.00_sqrt_scale.epsi}

  \includegraphics[width=0.53\textwidth]{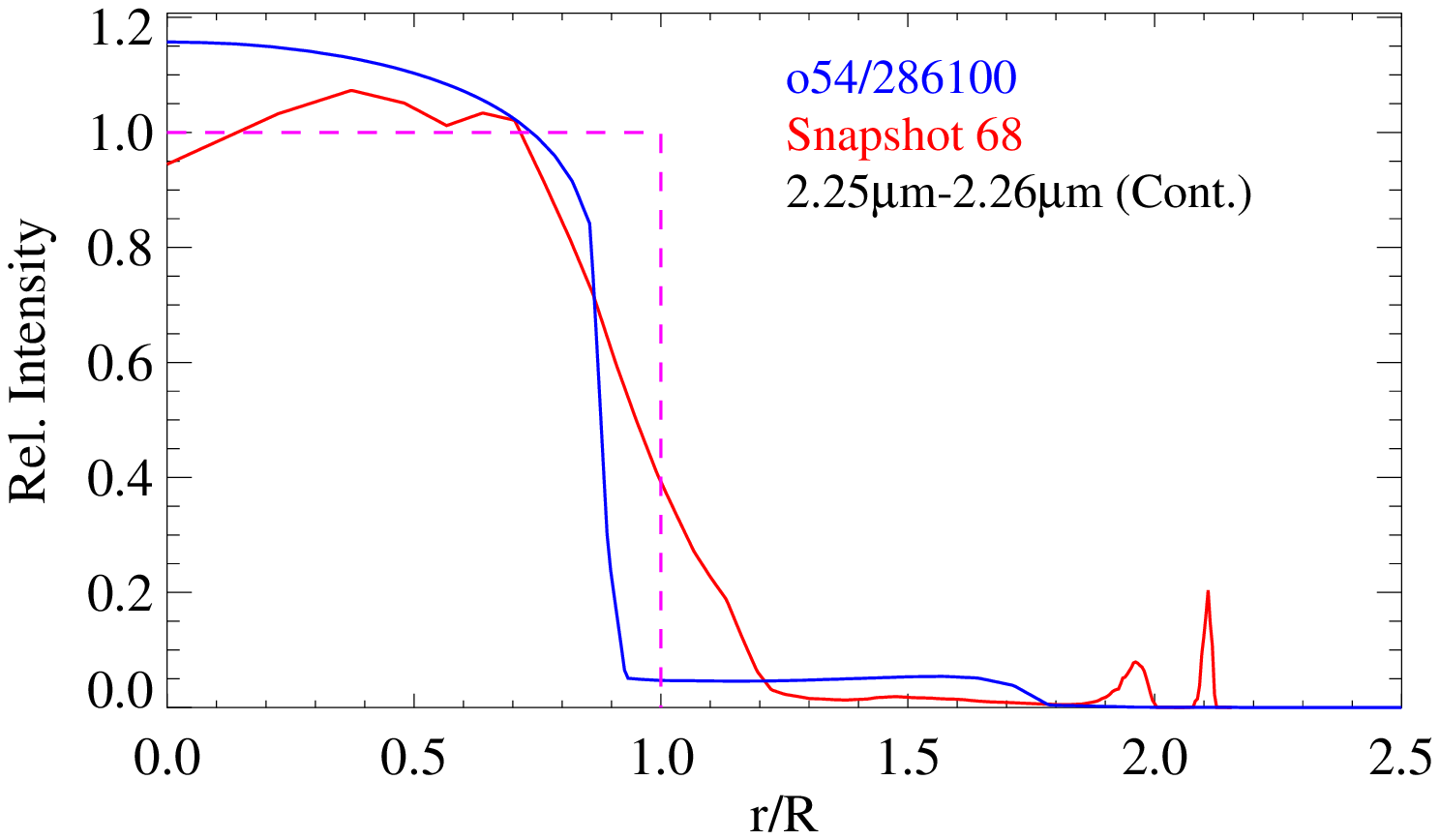}
  \includegraphics[width=0.30\textwidth]{image_parambf070_22500.00-22600.00_sqrt_scale.epsi}

  \includegraphics[width=0.53\textwidth]{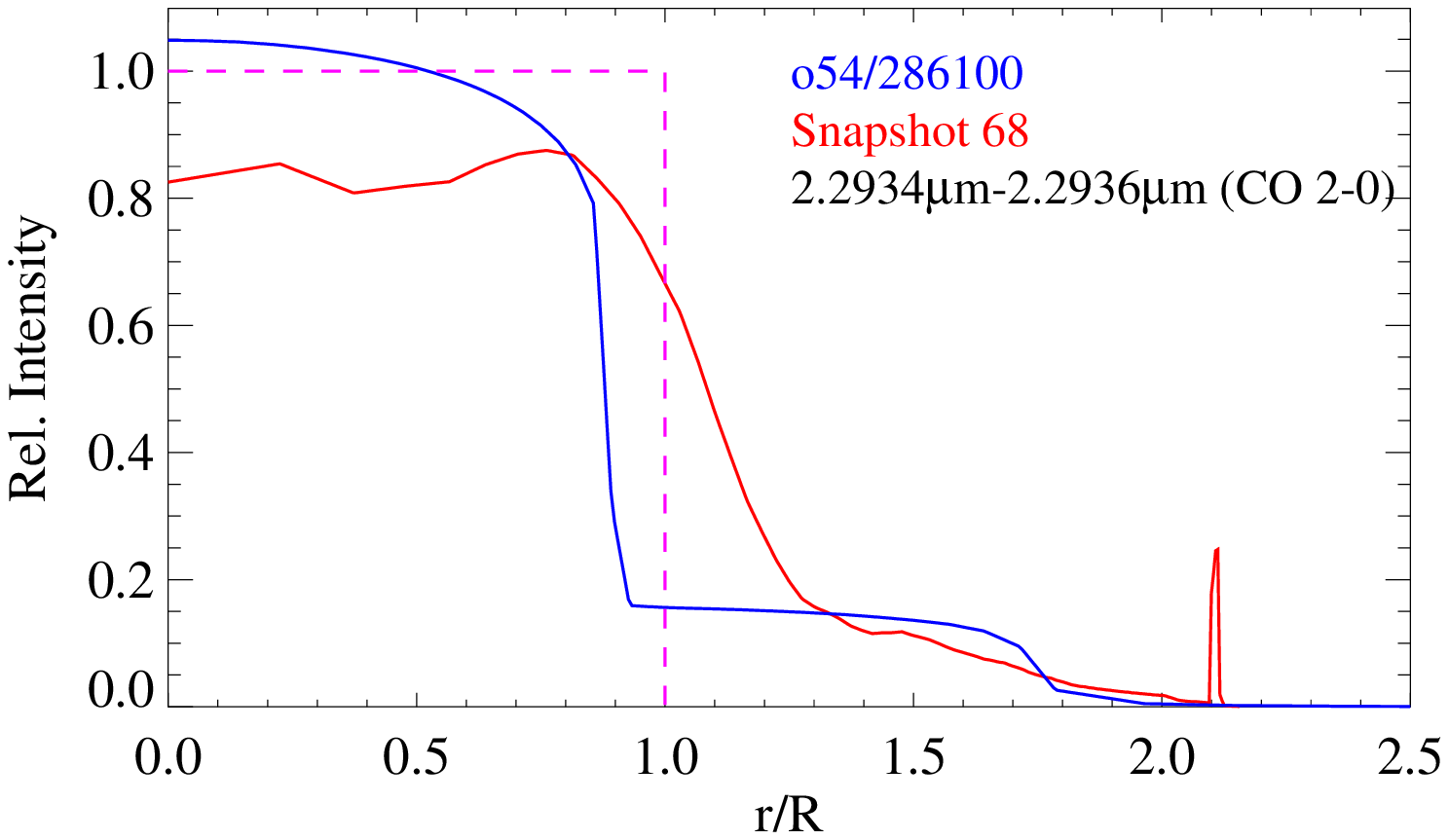}
  \includegraphics[width=0.30\textwidth]{image_parambf070_22934.00-22936.00_sqrt_scale.epsi}

  \includegraphics[width=0.53\textwidth]{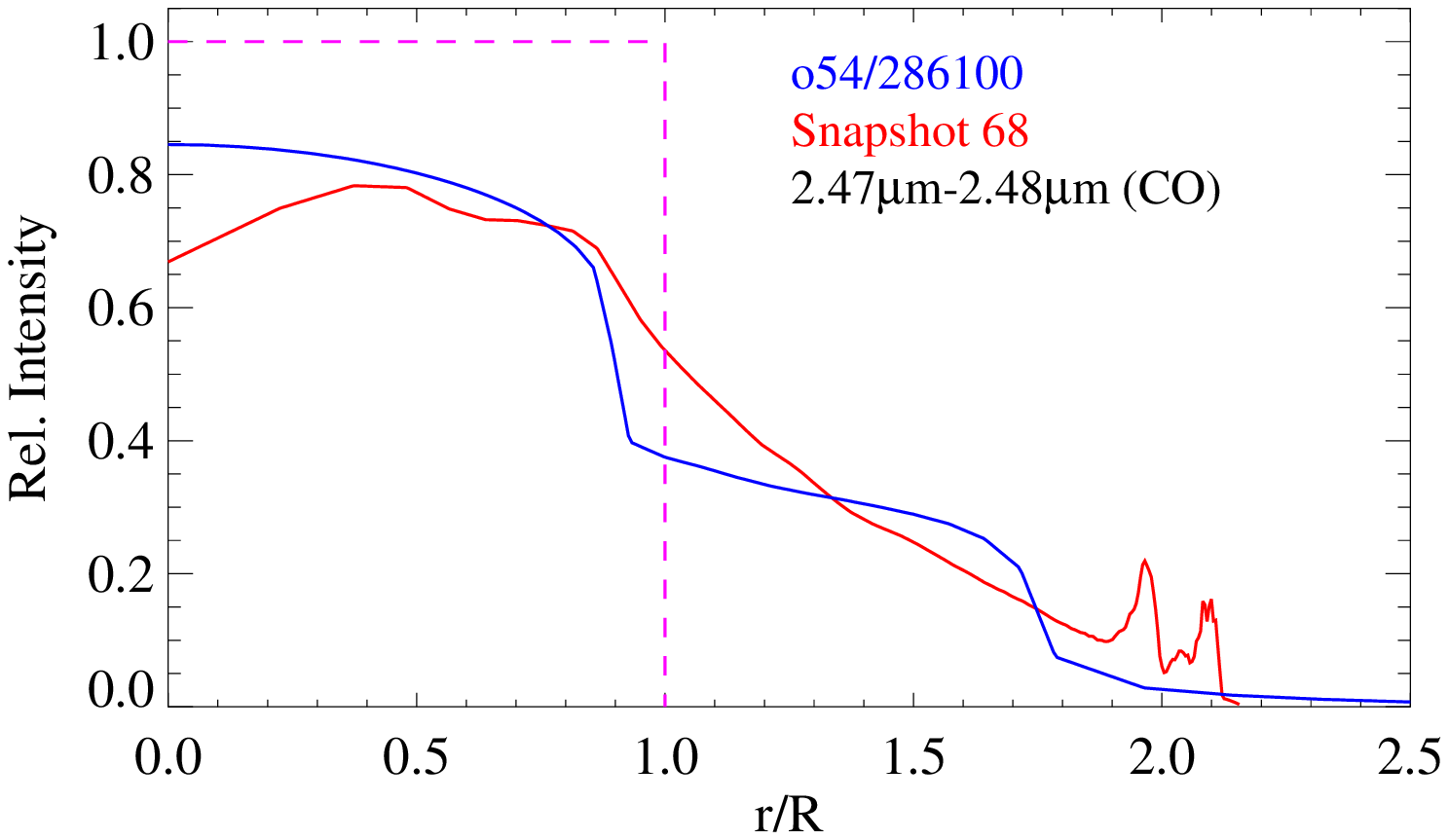}
  \includegraphics[width=0.30\textwidth]{image_parambf070_24700.00-24800.00_sqrt_scale.epsi}
\caption{Left panels: Intensity profiles of (blue) the 1D CODEX model 
o54/286100 compared to (red) the azimuthally averages intensity profiles
of snapshot 68 of the RHD simulation {\tt st28gm06n06} at
bandpasses (from top to bottom) 
2.00\,$\mu$m--2.01\,$\mu$m (water vapor), 
2.25\,$\mu$m--2.26\,$\mu$m (continuum), 
2.2934\,$\mu$m--2.2936\,$\mu$m (CO 2-1), and 
2.47\,$\mu$m--2.48\,$\mu$m (CO). Both models provide a satisfactory
fit to our visibility data of o Cet obtained in October 2012.
The purple dashed line
indicates a UD as reference. The intensities are normalized
so that the integrated flux of each curve is unity.
Right panels: Images of the 3D intensity maps of snapshot 68
at the same bandpasses. A squareroot color scale is used to enhance
the low intensities of the extended molecular atmosphere.}
\label{fig:modelcomp}
\end{figure*}
As a next step we compared our visibility data to predictions by
3D CO5BOLD dynamic model atmospheres including pulsation and convection 
of an AGB star by \citet{Freytag2008}, which represents
the first direct comparison of interferometric data of an AGB star 
to this model.
\citet{Freytag2008} used the RHD code CO5BOLD \citep{Freytag2012} to produce 
time-dependent
3D models of the outer convective layers and the atmosphere of an AGB star.
In short, the stellar parameters were chosen to closely resemble those of a 
1D model studied by \citet{Hoefner2003}. The 3D simulations started with
a hydrostatic, spherically symmetric structure, mapped onto a 3D grid.

Small-amplitude random velocities were added. In a sequence of models,
the physical parameters of the dynamical model and the size of the computational
box were adjusted until the flow properties stabilized. Dust formation was
added as a last step without feeding the dust opacities back into the RHD
equations. 
Models use gray (Rosseland) opacities, computed from
detailed continuum and line opacities from atoms and molecules."
In fact, the opacities are merged from Rosseland opacity tables
from PHOENIX and the OPAL project for the atmosphere and the interior,
respectively.
Here, we used their final
3D model {\tt st28gm06n06}. The basic parameters of this model are
listed in Table~\ref{tab:3d}.
\begin{table*}
\centering
\caption{Parameters of the 3D model {\tt st28gm06n06}.}
\label{tab:3d}
\begin{tabular}{llllllll}
\hline\hline
Grid & Box & $M_\mathrm{pot}$ & $M_\mathrm{env}$ & $L$ & $R$ & $T_\mathrm{eff}$ & $\log g$\\\hline 
171$^3$ & 1674 $R_\odot$ & 1 $M_\odot$ & 0.186 $M_\odot$ & 6935 $L_\odot$ & 429 $R_\odot$ & 2542 K & -0.83\\\hline
\end{tabular}
\end{table*}
Compared to the CODEX series, this is an AGB star model with relatively high
luminosity, relatively large radius, and relatively low effective temperature.
Indeed, the initial 1D model from \citet{Hoefner2003} was designed to 
match a C-rich AGB star, but a O-rich atmosphere was used for the 3D modeling.
Unlike the CODEX models, which include a simple description
of dust, the 3D model is dust-free.
Thus far this is the only 3D dynamic model atmosphere of an AGB star 
that is 
currently readily available. New AGB models, including gray models with
the current version of CO5BOLD and a much better resolution, as well
as first non-gray models, are being prepared \citep[cf.][]{Freytag2015}
and may become available within the next years.

We used the pure-LTE radiative transfer Optim3D \citep{Chiavassa2009} to 
compute intensity maps from 12 snapshots of the RHD simulation 
{\tt st28gm06n06}. 
These snapshots cover about 8.5 months with separations of about 0.7 months 
and thus
cover varying stellar parameters (luminosities between 5418 L$_\odot$ and
8170 L$_\odot$, radii between 489 $R_\odot$ and 519 $R_\odot$, and effective 
temperatures between 2190\,K and 2469\,K).
Optim3D takes the 
Doppler shifts occurring due to convective motions into account. The radiative 
transfer equation is solved monochromatically using pretabulated extinction 
coefficients as a function of temperature, density, and wavelength. The lookup 
tables are computed 
using the same extensive atomic and molecular continuum and line opacity data 
as the latest generation of MARCS models \citep{Gustafsson2008}. 
Again, a description of dust has not been added.
We computed intensity maps in the wavelength range from 1.90\,$\mu$m to
2.60\,$\mu$m with a constant spectral resolution of about 20 000. Moreover, 
for every wavelength, a top-hat filter including five wavelengths close by has 
been considered. In total, about 70\,000 images for every simulation snapshot 
were computed to cover the wavelength range of our AMBER observations 
with a spectral simulation so that we can average more than ten wavelengths 
for each AMBER spectral channel in a similar manner to what was performed for 
the CODEX models in Sect.~\ref{sec:codex} above. 

Instead of a direct comparison to the intensity maps, we chose to compare our 
data to azimuthally averaged intensity profiles obtained from the 
intensity maps for a number of reasons: (1) The detailed
surface structure has a random origin and is unlikely to match
the observations at a given epoch exactly. (2) \citet{Freytag2008} discuss
that the convection cells in the 3D simulations are so large that the
associated shock fronts appear almost spherical and  that they are 
consistent with pulsating 1D models. (3) Our interferometric
data already indicated that the visibilities above about a flux level
of 10\% are consistent with spherically symmetric models
(see Sect.~\ref{sec:codex}). (4) We intend to compare our data to
the results from the 3D model atmospheres in the same way as for the
1D CODEX models and to compare the intensity profiles of these 
modeling attempts.\\
We obtained the azimuthally averaged intensity profiles using the method
explained in detail by \citet{Chiavassa2009}. 
They were constructed using rings that are regularly spaced in $\mu$ 
(where $\mu=\cos(\theta)$ 
with $\theta$ being the angle between the line of sight and the 
radial direction). 
Finally, we estimated synthetic squared visibility amplitudes from the 
azimuthally averaged intensity profiles using the Hankel transform and averaging
over the bandpass of each AMBER spectral channel as we did for the CODEX models.
These procedures follow those described by \citet{Arroyo2015} for
RHD simulations of red supergiants.

In the same way as for the CODEX models in Sect.~\ref{sec:codex} we fit
our visibility data of each target and epoch to each of the available
12 snapshots of the RHD simulation and chose the snapshot with the best
$\chi^2_\nu$ value. The fit results including the best-fit radius and
the corresponding $\chi^2_\nu$ value are listed in the righthand panel of 
Table~\ref{tab:fitresults}, together with the effective temperature
and luminosity of the best-fit snapshot. Here, the radius corresponds
to the model radius, which is calculated from the luminosity and
effective temperature of the snapshot. A layer corresponding to 
a Rosseland optical depth of unity as used for the CODEX models 
is more difficult to define for a 3D model with a different model 
structure in different directions. However, the radius defined in 
this way should be comparable to the Rosseland radius used with 
the CODEX models.

The synthetic flux and squared visibility spectra of the best-fit
RHD snapshot are shown again in Figs.~\ref{fig:rcnc}--\ref{fig:rleo}
compared to the observed data and to the predictions by the
best-fit CODEX models. 

The flux spectra of the RHD simulation have a systematically
different shape than the observations and are compared to the
CODEX flux spectra, in particular in the wavelength range
shortward of the near-continuum bandpass at $\sim$2.25\,$\mu$m
down to about 2.0\,$\mu$m, a range that is dominated by
water vapor bands. The reason for this difference is not yet
known.

The synthetic squared visibility amplitudes
based on the RHD simulation are consistent with our 
measurements and with the predictions by the CODEX models.
They predict the decrease in the visibility function at the locations
of the water vapor band centered at about 2.0\,$\mu$m and at the
locations of the CO bandheads as well as the CODEX models.
This means that the 3D CO5BOLD dynamic model atmospheres also lead to an 
extended molecular atmosphere that is consistent 
with observations. Likewise, the best-fit angular diameters
and the $\chi^2_\nu$ values between the fits to the CODEX models
and to the RHD simulation are comparable, considering the caveats
already mentioned in Sect.~\ref{sec:codex} and considering that the
stellar parameters of the RHD simulation do not match those
of our sources well.

In Figure~\ref{fig:modelcomp} we compare the model
predicted intensity profiles at certain bandpasses for the example
of the 1D CODEX model o54/286100 and the 3D model snapshot 68,
which both provide a good fit to our visibility data of o Cet
obtained in October 2012. We chose four bandpasses, including a 
water vapor band at 2.0\,$\mu$m, a near-continuum band at 2.25\,$\mu$m,
a band centered on the CO 2-0 line at 2.2935\,$\mu$m, and a
band including CO and possibly water vapor toward the edge of the
$K$-band at 2.47\,$\mu$m. We also show 2D images of the intensity
distribution based on the 3D model. These images use
a squareroot color scale to visually enhance the low intensities 
of the
extended molecular layers.
Both model attempts predict a compact peak intensity
up to about 1\,$R_*$ and an extended envelope up to about 2\,$R_*$.
The peaks in the azimuthal average of the 3D models at about two stellar
radii are a signature of density peaks at the position of shock fronts.
These peaks do not significantly affect our results.
The ratio of the extended component relative to the central compact
component varies as a function of wavelength. In other words, our  
visibility variation as a function of wavelength more precisely
indicates different flux contributions of the extended atmosphere
rather than different spatial extensions. The exact intensity profiles
of the two modeling attempts differ, but they agree in the overall
spatial extent of the extended component and in the wavelength-dependent
flux contributions of the extended component. 
High-fidelity interferometric imaging campaigns 
at high spectral resolution are needed to constrain the exact
shape of the intensity profiles as a function of wavelength
and to differentiate further between different model atmospheres
of AGB stars. The images of the 3D model atmosphere indicate 
an inhomogeneous distribution of the extended molecular layers,
which in this example are more elongated toward the south than
toward the north. This is qualitatively consistent with the
observed inhomogeneities or clumps discussed above and by
\citet{Wittkowski2011} and \citet{Haubois2015} based on AMBER data.
In this context, we also note that for R~Cnc, W~Vel, and X~ Hya, 
all our current observations are taken along similar position angles, 
thus providing limited information on the global geometry 
of the photospheres of these objects, while observations of R Aqr,
o Cet, and R Leo cover a wider range of position angles.

\subsection{Comparison of the broad-band photometry to model atmosphere 
predictions}
\begin{table}
\centering
\caption{Observed and model-predicted colors}
\label{tab:colors}
\begin{tabular}{lrrr}
\hline\hline
                                        & $J-H$ & $J-K$ & $J-L$ \\
                                        & (mag) & (mag) & (mag) \\\hline
\multicolumn{4}{l}{R Cnc, JD 2454831, $\Phi_\mathrm{Obs}=0.3$}  \\
Observed                                & 1.01  & 1.51  & 1.91  \\
Best-fit CODEX                          & 1.15  & 1.56  & 1.91  \\
Best-fit 3D                             & 0.87  & 1.30  & 0.05  \\[1ex]
\multicolumn{4}{l}{R Cnc, JD 2454893, $\Phi_\mathrm{Obs}=0.5$} \\
Observed                                & 0.97  & 1.65  & 2.33  \\
Best-fit CODEX                          & 1.27  & 1.77  & 2.28  \\
Best-fit 3D                             & 0.81  & 1.31  & 0.22  \\[1ex]
\multicolumn{4}{l}{W Vel, JD 2454831, $\Phi_\mathrm{Obs}=0.9$}  \\
Observed                                & 0.97  & 1.43  & 1.98  \\
Best-fit CODEX                          & 0.98  & 1.27  & 1.69  \\
Best-fit 3D                             & 0.89  & 1.28  & 0.02  \\[1ex]
\multicolumn{4}{l}{W Vel, JD 2454892, $\Phi_\mathrm{Obs}=1.1$}  \\
Observed                                & 1.00  & 1.42  & 1.91  \\
Best-fit CODEX                          & 1.08  & 1.38  & 1.83  \\
Best-fit 3D                             & 0.89  & 1.28  & 0.02  \\[1ex]
\multicolumn{4}{l}{X Hya, JD 2454831, $\Phi_\mathrm{Obs}=0.7$}  \\ 
Observed                                & 0.88  & 1.36  & 1.93  \\
Best-fit CODEX                          & 1.15  & 1.65  & 2.24  \\
Best-fit 3D                             & 0.78  & 1.24  & 0.10  \\[1ex]
\multicolumn{4}{l}{X Hya, JD 2454892, $\Phi_\mathrm{Obs}=0.9$}  \\
Observed                                & 0.89  & 1.28  & 1.74  \\
Best-fit CODEX                          & 1.14  & 1.57  & 1.96  \\
Best-fit 3D                             & 0.78  & 1.24  & 0.10  \\\hline
\end{tabular}
\end{table}
We collected simultaneous $JHKL$ broad-band photometry at the SAAO
for two epochs of three of our six targets as described in Sect.~\ref{sec:obs}
and as listed in Table~\ref{tab:saao}. The primary goal of these observations
was to obtain simultaneous bolometric fluxes and other
fundamental stellar parameters (Sect.~\ref{sec:fundpar}). Here, we compare
the broad-band photometry to synthetic values predicted by the model
atmospheres to constrain them further.
For each of these targets and epochs, Table~\ref{tab:colors} shows the observed $J-H$, $J-K$, and
$J-L$ colors derived from Table~\ref{tab:saao} taking the
extinction as listed there into account, together with the synthetic values 
from the best-fit CODEX model and the best-fit 3D snapshot 
(cf. Table~\ref{tab:fitresults}). To calculate the synthetic values,
we used the $H,K,L$ filter curves from \citet{Glass1973} and the
$J$ filter curve from \citet{Glass1993}. 
The predicted $J-H$, $J-K$, and $J-L$ broad-band colors of the
best-fit CODEX models generally agree with the observations to about
0.0--0.3\,mag. The predictions of the best-fit 3D snapshots in general
agree similarly well for the $J-H$ and $J-K$ colors, 
while they predict much bluer $J-L$ colors near 0\,mag. The latter
discrepancy can most likely be explained because the 3D models do not
include any dust description while the CODEX models do include a simple
description of dust (cf. Sects.~\ref{sec:codex} and \ref{sec:rhd}).
Indeed, using the dust-free predecessors
of the CODEX models, the P/M model series by \citet{Ireland2004a,Ireland2004b},
we also obtain typical $J-L$ colors near 0\,mag (for example, model 
P20 gives $J-H=0.41$, $J-K=0.28$, $J-L=0.03$).
This means that dust is transparent for our sources at $J$, $H$, and $K$ bands 
and starts to contribute to broad-band colors from the $L$-band onward.
Hereby, the simple dust description of the CODEX models is sufficient to 
explain the broad-band $J-L$ color. These findings are consistent with dust
grains being transparent at near-infrared wavelengths as discussed by
\citet{Ireland2005}, \citet{Norris2012}, and \citet{Bladh2015}.
\section{Fundamental stellar parameters}
\label{sec:fundpar}
\begin{table*}
\centering
\caption{Fundamental stellar parameters\label{tab:fund}}
\begin{tabular}{lrrrrrr}
\hline\hline
                                                        & o Cet            & R Leo            & R Aqr          & R Cnc          & X Hya          & W Vel \\\hline
Phase                                                   & 0.1              & 0.6              & 0.6            & 0.3--0.5       & 0.7--0.9       & 0.9--1.1\\[1ex]
$\Theta_\mathrm{UD}^{\mathrm{2.25}\mu\mathrm{m}}$ (mas)\tablefootmark{a} & 28.5 $\pm$ 1.5   & 29.6 $\pm$ 1.3   & 18.4 $\pm$ 0.4 & 13.2 $\pm$ 0.3 & 6.0 $\pm$ 0.3  & 7.4 $\pm$ 0.2 \\
$m_\mathrm{bol}$ (mag)\tablefootmark{b}                                 & 1.04 $\pm$ 0.1   & 0.75 $\pm$ 0.1   & 2.37 $\pm$ 0.1 & 2.83 $\pm$ 0.4 & 4.00 $\pm$ 0.2 & 3.55 $\pm$ 0.2 \\ 
Parallax (mas)                                          & 9.1 $\pm$ 1.4\tablefootmark{c} & 10.0 $\pm$ 1.5\tablefootmark{c} & 4.59 $\pm$ 0.24\tablefootmark{d}& 3.6 $\pm$ 0.5\tablefootmark{c}& 2.3 $\pm$ 0.3\tablefootmark{c}& 2.0 $\pm$ 0.3\tablefootmark{c}\\[1ex]
R ($R_\odot$)                                           & 340 $\pm$ 80     & 320 $\pm$ 70     & 431 $\pm$ 33   & 400 $\pm$ 70   & 280 $\pm$ 60   & 400 $\pm$ 80 \\
$T_\mathrm{eff}$ (K)                                    & 2450 $\pm$ 120   & 2570 $\pm$ 120   & 2250 $\pm$ 76  & 2390 $\pm$ 260 & 2700 $\pm$ 195 & 2700 $\pm$ 160\\
log $L/L_\odot$                                         & 3.57 $\pm$ 0.27  & 3.60 $\pm$ 0.25  & 3.63 $\pm$ 0.12& 3.66 $\pm$ 0.35& 3.58 $\pm$ 0.28& 3.88 $\pm$ 0.26\\\hline
\end{tabular}
\tablefoot{
\tablefoottext{a}{This work, see Table~\protect\ref{tab:fitresults}.}
\tablefoottext{b}{Based on SAAO photometry, see Sect.~\ref{sec:obs}.}
\tablefoottext{c}{$P-L$ distance from \protect\citet{Whitelock2008}. We adopt an error of 15\% in distance.}
\tablefoottext{d}{\protect\citet{Min2014}.}
}
\end{table*}
\begin{figure}
\centering
  \includegraphics[width=0.5\textwidth]{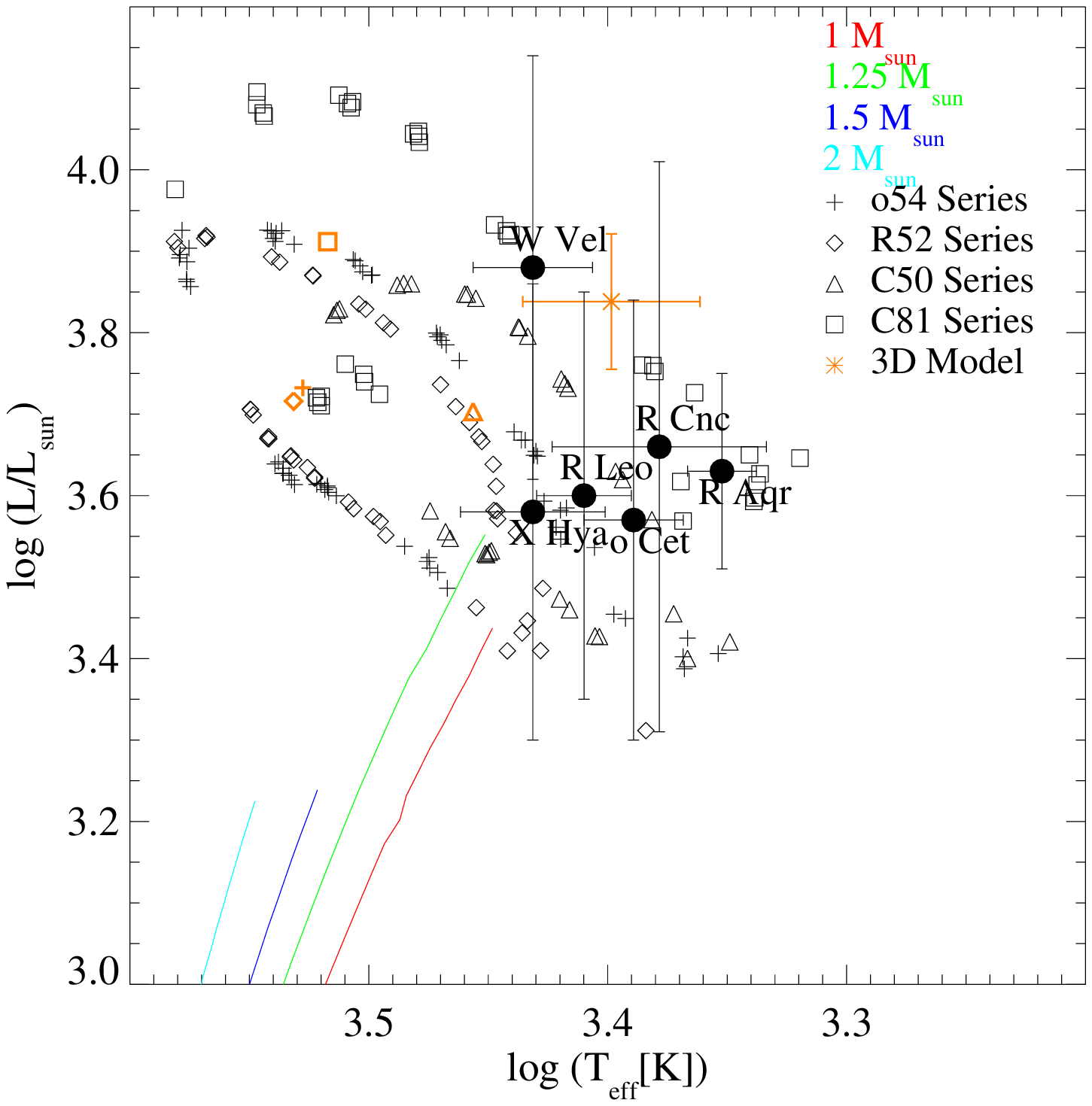}
\caption{Location of our sources in the HR diagram as 
listed in Table~\protect\ref{tab:fund},
together with those of the CODEX models and the 3D model st28gm06n06.
Shown are the positions of individual CODEX models with different symbols 
for each of the four series, where the orange symbols represent the locations 
of the hypothetical non-pulsating parent star. The orange symbol and error 
bars of the 3D model denote the average and standard deviation of different 
snapshots of the 3D model. Individual snapshots of the 3D model are 
not shown.  Also shown are stellar evolutionary
tracks of the early AGB from \citet{Lagarde2012} for 1.0, 1.25, 
1.5, and 2.0 solar masses and solar chemical composition.
\label{fig:hrdiagram}}
\end{figure}
We used the UD diameters at a near-continuum wavelength of 2.25\,$\mu$m
$\Theta_\mathrm{UD}^{\mathrm{2.25}\mu\mathrm{m}}$ from Tab.~\ref{tab:fitresults}
as a model-independent estimate of the stellar angular continuum radius. 
These are generally consistent with model-dependent Rosseland angular diameters (cf. the discussion above in Sect.~\ref{sec:modelcomp}). Together with apparent 
bolometric magnitudes based on SAAO photometry (Sect.~\ref{sec:obs}) and with
distance estimates from the literature, we computed linear radii, effective
temperatures, and luminosities of our sources at the respective phase.
The resulting values are listed in Table~\ref{tab:fund}. 
R Aqr has a recent high-precision VLBI parallax measurement \citep{Min2014},
which we used. For our other sources, 
we used the period-luminosity ($P-L$) distances from \citet{Whitelock2008}.
It is difficult to give a precise error on the $P-L$ distance making us 
conservatively adopt an error of 15\%.

Compared to the best-fit CODEX models,
the measured $T_\mathrm{eff}$ values are typically lower than those
of the CODEX models. The luminosities of the best-fit CODEX models
are generally consistent with the observed luminosities within the 
error bars, but tend to be systematically higher within the uncertainty.
Compared to the best-fit 3D snapshots, the measured $T_\mathrm{eff}$ values
are higher than those of the models except for R Cnc. The luminosities
of the 3D snapshots tend to be higher than the measured values,
although they are consistent within the errors for R Cnc, W Vel, and X Hya.
However, we note that, similar to the discussion of the radius in 
Sect.~\ref{sec:codex}, it is also difficult to define an effective 
temperature for a star with a very extended atmosphere. The definition of 
the effective temperature depends on the model-dependent definition of the 
stellar radius. 

Figure~\ref{fig:hrdiagram} illustrates the positions of our sources in the 
HR diagram
compared to the positions of the individual CODEX models and the average of
the 3D CO5BOLD model. Also shown are the stellar evolutionary tracks
of the early AGB by \citet{Lagarde2012} for masses between 1.0\,$M_\odot$ 
and 2.0\,$M_\odot$ and solar chemical composition.  The observed positions 
are consistent with the ranges of the model atmospheres
and are located shortly above the early AGB evolution.

\section{Summary, discussion, conclusions}
We obtained AMBER near-infrared $K$-band (1.9--2.5\,$\mu$m)
spectro-interferometric (spectral resolution
$R\sim$1500) observations of a total of six Mira variable AGB stars.
For three of the sources,
observations were obtained at two epochs separated by about two months using 
virtually the same projected baseline lengths and angles.

The squared visibility amplitudes as a function of wavelength within
the $K$-band show a maximum near 2.25\,$\mu$m and decrease both toward
shorter and longer wavelengths with pronounced drops at the positions
of the CO bandheads. This is interpreted as an indication of
extended atmospheric molecular layers of -- most importantly -- H$_2$O
(centered at $\sim$2.0\,$\mu$m) and CO (longward of 2.29\,$\mu$m) as 
previously observed for Mira stars with AMBER 
by \citet{Wittkowski2008}, \citet{Wittkowski2011}, \citet{Haubois2015}.
The closure phases as a function of wavelength show a characteristic shape
as well, where deviations from point symmetry are observed at all
wavelengths, depending on how well the target is resolved, with most
pronounced deviations from symmetry at the positions of H$_2$O and CO
bands. This closure-phase signal is interpreted as a signature of 
large-scale inhomogeneities or clumps at the photospheric layers
and within the extended molecular layers. These inhomogeneities or clumps
contribute a few percent of the total flux. Observations taken
at two epochs separated by about two months indicate a variability of the
detailed structure of the inhomogeneities, while a variability
of the overall radius could not be detected within our calibration
uncertainties.

We compared our data with the latest 1D dynamic model atmospheres based 
on self-excited pulsation models (CODEX models) and for the first time 
also with 3D radiation hydrodynamic (RHD) simulations of the convective 
interiors and the atmospheres of AGB stars (CO5BOLD code). For the latter
we chose to use azimuthally averaged intensity profiles because our main 
interest
in this current investigation concerns the overall extension of 
the atmospheres and because we aim at comparing the 3D models to the 
results by 1D models. In addition, we have previously shown that
visibilities above a flux level of about 10\% are consistent with 
spherically symmetric models so that the detailed 3D structure
mainly concerns only relatively low flux levels.
Both model attempts provide satisfactory fits to our data and, in particular
reproduce the wavelength dependence of the visibility data.
This means that both models include a spatially extended molecular atmosphere
that is consistent with our observations. A comparison between the intensity
profiles of the best-fit 1D and 3D models show that indeed both model attempts
include an extended atmosphere up to roughly two to three stellar radii and that the
wavelength-dependent flux contributions of the extension are consistent 
with observations. This agreement of both the different 1D and 3D modeling 
attempts with our data supports the theoretical result by \citet{Freytag2008} 
that shocks induced by convection and pulsation in 3D models are roughly 
spherically expanding and of similar nature to those of 1D self-excited 
pulsations.

Moreover, our results show that indeed both modeling attempts, 
1D pulsation models and 3D convection simulations, can levitate
the atmosphere of Mira stars to at least two to three stellar radii, which 
is consistent
with our observations. We note that, on the contrary the same modeling
attempts failed to extend the atmospheres of red supergiant stars to similarly
observed extensions \citep{Arroyo2015}.

In addition, observed and model-predicted broad-band $J-H$ and $J-K$ are
consistent for both the 1D CODEX and 3D CO5BOLD models, while the $J-L$
color can only be re-produced by the 1D CODEX, which includes a simple dust
description, and not by the dust-free 3D CO5BOLD models, indicating that
dust starts to contribute to broad-band colors from the $L$-band onward.

The detailed intensity profile across the stellar disk
could not be constrained by our current snapshot observations. This would
require future imaging campaigns. 
More detailed comparisons between 1D and 3D models
and between observations and 3D models along different angles are planned 
for a forthcoming publication. However, devising a proper strategy 
for comparing 
variable objects with dynamical models and, in particular, 3D models is a
non-trivial task. For more detailed comparisons, one should try to 
disentangle (or at least discuss) whether discrepancies are due to bugs in 
the code, the omission of essential physics (such as dust, magnetic fields, 
rotation), simplifications (such as gray opacities, finite numerical grid),
or inappropriate stellar parameters. Moreover, for a dynamic and 
non-spherical structure, it is a fundamental problem to select the
correct snapshot in time and the correct viewing angle.

Based on the model-independent UD diameters at a near-continuum bandpass 
around 2.25\,$\mu$m, on SAAO photometry to estimate the bolometric flux, 
and on distances available in the literature, we estimated absolute radii, 
effective temperatures, and luminosities of our sources at the respective 
variability phases. The UD diameters at 2.25\,$\mu$m are generally consistent
with Rosseland angular diameters that are derived from the comparison with 
model atmospheres.
In general, the CODEX models tend to have
higher effective temperatures than observed. The 3D RHD simulation
tends to have lower effective temperatures and higher luminosities
than observed for our sample. Observational variability
phases are mostly consistent with those of the best-fit CODEX models, except
for near-maximum phases, where data are better described by 
near-minimum models. 
The availability of more complete model grids 
of different stellar parameters would be important for better matching 
observed stellar parameters.

\begin{acknowledgements}
This paper uses observations made at the South African Astronomical 
Observatory (SAAO). We are grateful to Francois van Wyk for doing the 
infrared photometry from SAAO.
PAW thanks the South African National Research Foundation (NRF) for a 
research grant.
AC acknowledges support from the ESO Scientific Visitor Program.
SH and BF acknowledge support from the Swedish Research Council.
The 3D simulations were performed on resources provided by SNIC
through Uppsala Multidisciplinary Center for Advanced Computational 
Science (UPPMAX) under Projects p2003019 and p2013234.
We acknowledge with thanks the variable star observations from the AAVSO 
International Database contributed by observers worldwide and used in this 
research. This research has made use of the SIMBAD and AFOEV databases,
operated at the CDS, France. This research made use of NASA's 
Astrophysics Data System."
\end{acknowledgements}
\bibliographystyle{aa}
\bibliography{27614}
\end{document}